\documentclass[
amsmath,amssymb,
aps,
reprint,
nofootinbib,
]{revtex4-2}

\usepackage{graphicx}
\usepackage{dcolumn}
\usepackage{bm}
\usepackage{color}
\usepackage{hyperref}

\newcommand{\bea}{\begin{eqnarray}}
\newcommand{\eea}{\end{eqnarray}}
\newcommand{\be}{\begin{equation}}
\newcommand{\ee}{\end{equation}}
\newcommand{\ba}{\begin{align}}
\newcommand{\ea}{\end{align}}
\newcommand{\im}{\mathfrak{Im}\,}

\newcommand{\conid}{C_{\text{in}, \text{down}}}
\newcommand{\coniu}{C_{\text{in}, \text{up}}}
\newcommand{\conou}{C_{\text{out}, \text{up}}}
\newcommand{\conod}{C_{\text{out}, \text{down}}}
\newcommand{\cnin}{C_{\text{n.}, \text{in}}}
\newcommand{\cnout}{C_{\text{n.}, \text{out}}}
\newcommand{\cnnin}{C_{\text{n.n.}, \text{in}}}
\newcommand{\cnnout}{C_{\text{n.n.}, \text{out}}}
\newcommand{\res}{\text{res}}

\begin{document}

\title{
Beyond quasinormal modes:\\ a complete mode decomposition of black hole perturbations
}

\author{Paolo Arnaudo}
 \email{p.arnaudo@soton.ac.uk}
\author{Javier Carballo}
 \email{j.carballo@soton.ac.uk}
\author{Benjamin Withers}
 \email{b.s.withers@soton.ac.uk}
\affiliation{Mathematical Sciences and STAG Research Centre,\\ University of Southampton, Highfield,\\ Southampton SO17 1BJ, UK}%

\date{2025}

\begin{abstract}
We show that retarded Green's functions of black hole spacetimes can be expressed as a convergent mode sum everywhere in spacetime. At late times a quasinormal mode sum converges, while at early times a Matsubara (or, Euclidean) mode sum converges. The two regions are separated by a lightcone which scatters from the black hole potential. The Matsubara sum is a Fourier series on the Euclidean thermal circle associated to the early time region.  We illustrate our results for P\"oschl-Teller, BTZ, and Schwarzschild. In the case of Schwarzschild, we express the branch cut contribution as a convergent sum of de Sitter quasinormal modes as $\Lambda\to 0^+$, and exploit recent exact solutions to the Heun connection problem. In each case we analytically show convergence by studying the asymptotics of residue sums and also provide numerical demonstrations.
\end{abstract}

\maketitle

\tableofcontents

\section{Introduction}

Gravitational wave astronomy is a rapidly advancing field backed by the development of increasingly sensitive detectors \cite{LIGOScientific:2016aoc, GWcollab}, which now boasts a catalogue of hundreds of compact binary events \cite{LIGOScientific:2025slb}. This constitutes an unparalleled opportunity to test the foundations of gravitational physics through precision predictions of binary black hole merger waveforms.

A core theoretical tool in this context is black hole perturbation theory, which emerged over six decades ago as a framework to test the stability of black hole solutions to Einstein's equations under external disturbances \cite{Regge_Wheeler,Zerilli}. By perturbing a black hole, one can study its response and ultimate ringdown back to its equilibrium state, inviting the term: black hole spectroscopy \cite{Berti:2025hly}. Recently, through measurements of the loudest gravitational wave event observed to date, it was possible to place impressive bounds on the first few characteristic spectral frequencies and decay times of a black hole \cite{LIGOScientific:2025obp}.

In light of the increasing volume of gravitational wave data, it is evident that a complete theoretical understanding of black hole perturbations is now required. Such a treatment is still lacking because black holes are dissipative systems and this has created certain technical obstructions. Specifically, perturbations to black holes are governed by non-self adjoint operators, and there is no associated spectral theorem that guarantees orthogonality and completeness of perturbations.

In this paper, we address the issue of completeness in black hole perturbation theory. Our central new result is that black hole spacetimes are partitioned into distinct regions by certain lightrays, and in each region the black hole Green's function is represented by a different convergent mode sum. A quasinormal mode (QNM) sum provides a convergent representation at late times, while a new ingredient in this context, a Matsubara mode (MM) sum, provides a convergent representation at early times. 

The results of our work are directly applicable to gravitational wave physics and black hole spectroscopy as well as a class of strongly coupled quantum many-body systems through the AdS/CFT correspondence. 

The retarded Green's function $G(t,t',r,r')$ is the solution to the following PDE,
\be
\left(-\partial_t^2 + \partial_{r_*}^2 - V(r)\right)G(t,t',r,r') = -\delta(t-t')\delta(r_*-r_*'), \label{GreensPDE}
\ee
which obeys causal response, vanishing for $t<t'$. Here, $r_*$ is a tortoise coordinate, we have decomposed the perturbation into transverse spherical harmonics\footnote{For simplicity we focus on static, spherically symmetric black holes.}, and $V$ is the potential whose details depend on the spacetime, species of perturbation, and the orbital quantum number, $\ell$. It is convenient to move to frequency space, 
\be
G(t,t',r,r') = \int_{-\infty + i\epsilon}^{\infty + i\epsilon} \frac{d\omega}{2\pi} \widetilde{G}(\omega,r,r') e^{-i \omega (t-t')}, \label{intro:FT}
\ee
since then $\widetilde{G}(\omega,r,r')$ obeys an ODE instead, solvable by standard techniques.
The parameter $\epsilon$ in \eqref{intro:FT} is chosen as an infinitesimal positive real number.
With $\widetilde{G}(\omega,r,r')$ obtained, the remaining task is to compute the integral \eqref{intro:FT}, and this work is focussed on expressing this integral as a convergent mode sum.

\begin{figure}[h]
\centering
\includegraphics[width=0.6\columnwidth]{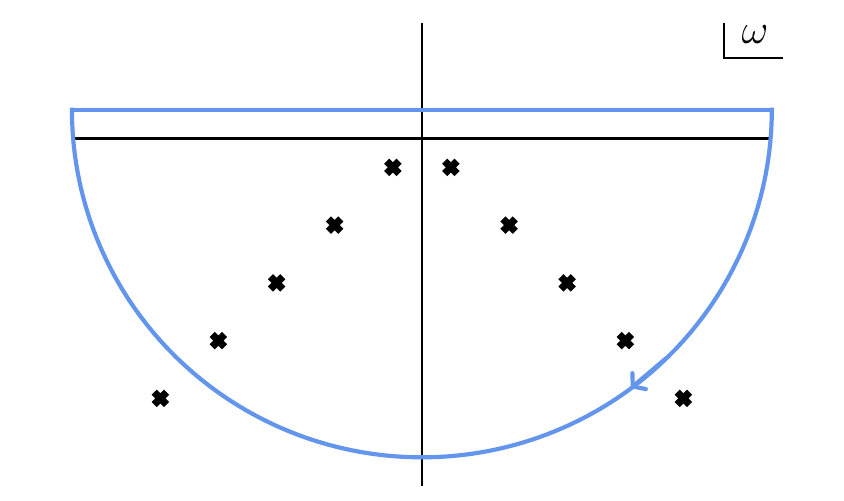}
\caption{Computation of a retarded black hole Green's function $G(t,t',r,r')$ by Fourier transforming the frequency-space expression $\widetilde{G}(\omega,r,r')$ using a contour integral. The analytic structure of $\widetilde{G}(\omega,r,r')$ consists of poles corresponding to QNM modes, and, depending on the black hole, branch cut contributions (not shown here). Closing the contour in the LHP leads to a QNM mode sum which diverges in some spacetime regions. Our work addresses this issue.}
\label{fig:lore}
\end{figure}

\begin{figure*}[p]
\centering
\includegraphics[width=1.9\columnwidth]{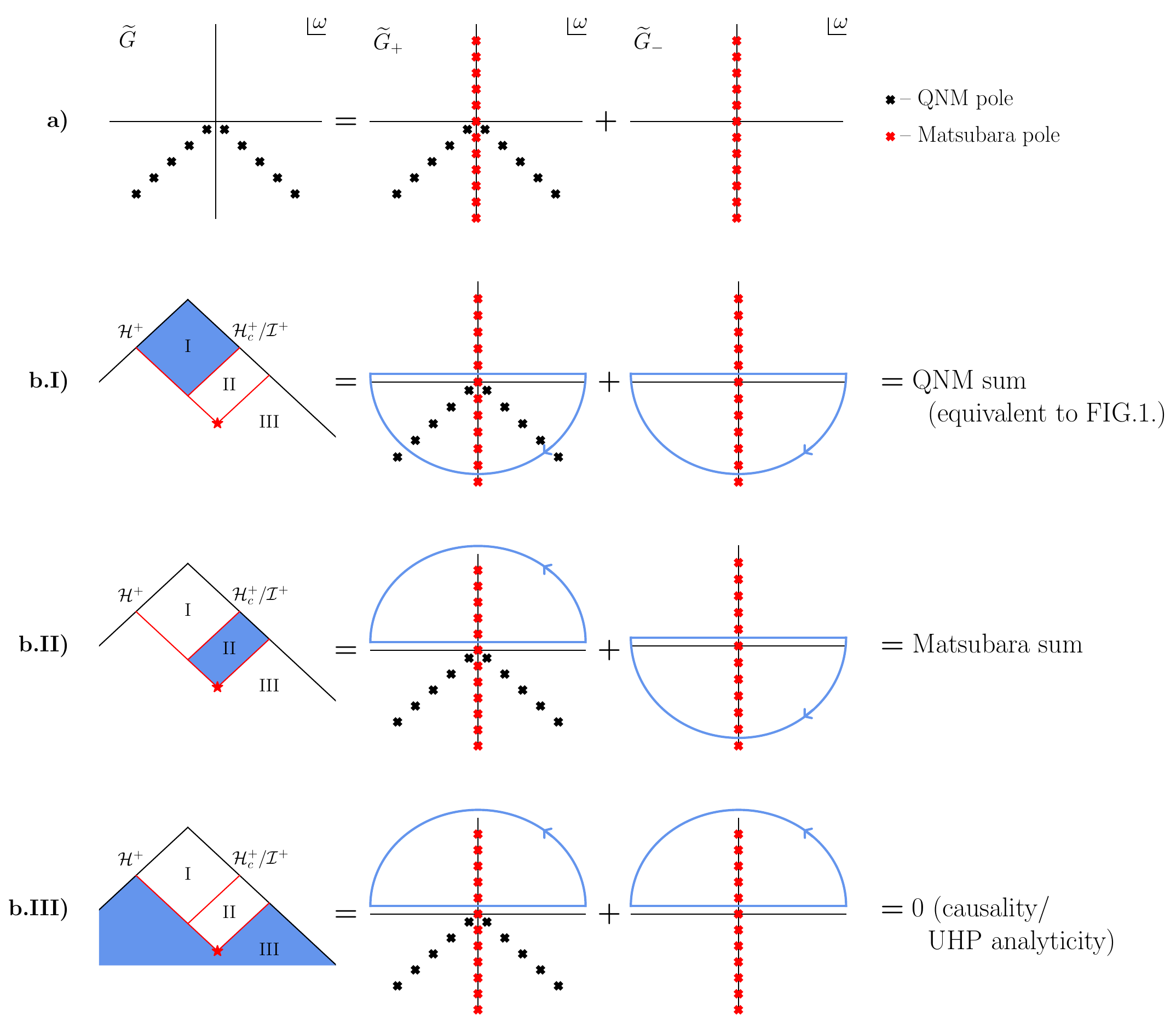}
\caption{The anatomy of a retarded black hole Green's function. \\\\\textbf{a)} The split of the frequency space Green's function $\widetilde{G}(\omega, r, r')$ into two terms $\widetilde{G}_+$ and $\widetilde{G}_-$ based on connection formulae. Both terms contain poles at Matsubara frequencies (red points) with opposite sign for their residues, so that these poles cancel in $\widetilde{G}(\omega, r, r')$. Only $\widetilde{G}_+$ contains QNM poles (black points). \\\\\textbf{b)} Computing the Fourier transform of $\widetilde{G}(\omega, r, r')$ in different spacetime regions. Different regions require different contour closure choices so that arc contributions vanish, leading to different convergent residue (mode) sums in each case. The leftmost plot shows a portion of the conformal diagram, with the star corresponding to the delta function location, and the red lines are the lightrays which delineate different regions. In region \textbf{I)} both terms must be closed in the LHP, leading to a cancellation of the MM contributions. Thus  $G(t,t',r,r')$ is given by a convergent sum of QNMs only. In region \textbf{II)} $\widetilde{G}_+$ requires UHP closure and $\widetilde{G}_-$ LHP closure. There are no QNM contributions and $G(t,t',r,r')$ is given by a convergent sum of MMs only. In region \textbf{III)} both terms must be closed in the UHP, the residue contributions cancel leading to a zero as required by causality. \\\\An analogous construction appears in asymptotically AdS spacetimes, where lightrays reflect off the timelike boundary instead (see FIG.\ref{fig:btzregions}).} 
\label{fig:introimage}
\end{figure*}

\begin{figure*}[t]
\centering
\includegraphics[width=2\columnwidth]{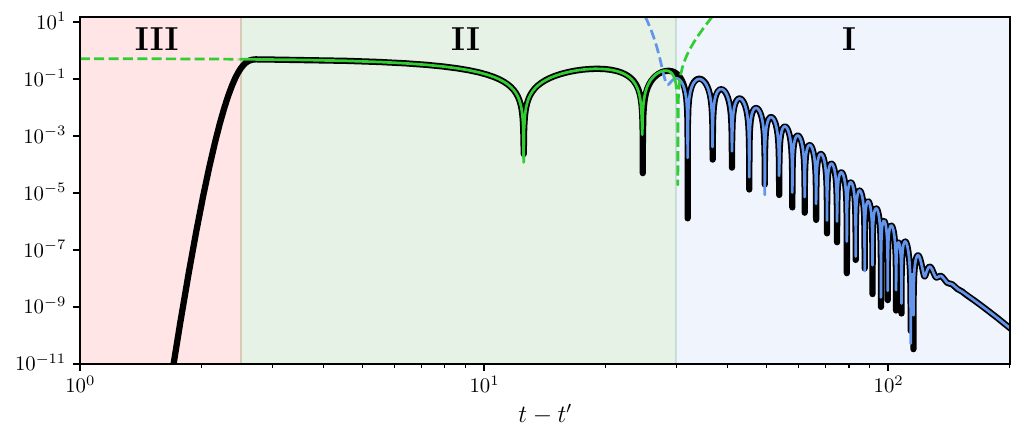}
\caption{The retarded Green's function $|G(t,t',r,r')|$ for small $\Lambda$ Schwarzschild-de Sitter as a convergent mode sum at fixed $r,r'$. There is no fitting, this is directly evaluated by summing residues of $\widetilde{G}_\pm$. In \textbf{region I} (blue) the Green's function is a convergent sum of QNMs. Here we have included the first 6 off-axis QNMs and the first 6 on-axis QNMs in the sum. In \textbf{region II} (green) it is a convergent sum of MMs. Here we have included 30 UHP MMs, 30 LHP MMs and the zero mode residue. In \textbf{region III} (red) it is zero. The lines are drawn as solid in the regions where they apply, and also extended to regions where they do not as dashed lines.
For comparison, we show a numerical solution to the Green's function PDE, \eqref{GreensPDE} (black), where the delta function is approximated by a Gaussian of width $\sigma$ (note that the finite width of the Gaussian gives rise to the non-zero support of the black curve in region III). Parameters used are $s=2$, $\ell=2$, $R_h = 1$, $\Lambda = 10^{-3}$, $r_* = 16$, $r_*' = 13.5$, and $\sigma = 1/10$.}
\label{fig:SdSpartial}
\end{figure*}

First, a note on branch cuts. When there are branch cut discontinuities of $\widetilde{G}$ in the complex $\omega$ plane, we choose to resolve them into a set of poles. For example, in the case of Schwarzschild we resolve its cut into a line of de Sitter QNM poles by adding a small cosmological constant, $\Lambda > 0$ (i.e. using Schwarzschild-de Sitter). Further details of this procedure are given in \cite{Arnaudo:2025kit}.\footnote{This is analogous to the finite temperature resolution of the cut for charged scalar correlators in RNAdS$_5$ \cite{Arnaudo:2024sen} (both cases are a confluence limit of the Heun equation). A mode resolution of cut contributions was also made in \cite{Besson:2024adi}.} In other words, for all intents and purposes, the contributions coming from poles in this paper may be regarded as also including contributions from cut discontinuity integrals. Such cut resolutions are easier to work with and do not affect our conclusions.

In general, the QNMs are a set of complex discrete frequencies which are singled out by imposing two suitable boundary conditions to the perturbation field. When the black hole background is stable against the perturbations, the QNMs lie in the complex lower half-plane. Moreover, being the frequencies at which both boundary conditions are satisfied, they correspond to poles of the retarded Green's function $\widetilde{G}(\omega,r,r')$.

The prevailing lore in the literature is that $G(t,t',r,r')$ is decomposed as a sum of QNMs plus arc contributions from large $|\omega|$ \cite{Leaver, PhysRevD.45.2617, AnderssonBookSection, Andersson:1996cm, Dolan:2011fh}.\footnote{See the previous paragraph for a discussion on our treatment of cuts.} This arises from treating \eqref{intro:FT} as a complex contour integral, closed in the lower-half $\omega$-plane, picking up residues from poles corresponding to QNMs. See FIG. \ref{fig:lore}. However, in some regions of spacetime, the semicircular arc contribution diverges (see \cite{Andersson:1996cm} for a detailed discussion) and accordingly, the residue sum corresponding to the contribution of QNM poles diverges too (so that the net result is finite). Therefore, a QNM sum does not provide an appropriate description of the spacetime in such regions, leading one to ask the question: what does?

It is also well-known that QNMs do not provide a complete set of modes everywhere in spacetime, even in the absence of branch cuts. This is sharply illustrated by the `QNM completeness counterexample' \cite{Warnick:2022hnc}. This again leaves the same question, where a QNM sum does not work, is there something else that does?

In this work, we resolve these questions by showing that an additional set of modes which appear at Matsubara frequencies associated to the horizon temperature, which we call Matsubara modes (MMs), provide a convergent sum in regions where QNMs do not. 

The Matsubara frequencies typically arise in Euclidean signature as the set of frequencies that appear in the Fourier mode decomposition of a bosonic field satisfying periodic boundary conditions along the compactified time direction. 
More explicitly, in a thermal quantum field theory, after performing a Wick rotation $t\to -i\tau$, the theory is formulated in Euclidean spacetime with imaginary time compactified on a circle of circumference $\beta=1/T$, where $T$ is the temperature of the system.
Every bosonic field $\phi$ satisfies periodic boundary conditions $\phi(\tau)=\phi(\tau+\beta)$, and its Fourier mode decomposition reads
\begin{equation}
\phi(\tau)=\sum_{n\in\mathbb{Z}}\widetilde{\phi}_n\,e^{\frac{2\pi i n}{\beta}\tau}.
\end{equation}
In a black hole background, the temperature associated to the horizon is defined as $T_{\text{hor}}=\kappa_{\text{hor}}/(2\pi)$, where $\kappa_{\text{hor}}$ is the surface gravity of the horizon.

We will illustrate our findings using retarded Green's function calculations for several black hole spacetimes. The common steps to reach these conclusions are as follows, and are also illustrated in FIG. \ref{fig:introimage}.\footnote{The details also differ slightly for asymptotically AdS spacetimes, and we discuss these in section \ref{sec:BTZ} and also Appendix \ref{sec:AdS2}.}
\begin{enumerate}
    \item The solution to \eqref{GreensPDE} in frequency space is given by standard techniques as follows
    \be
    \widetilde{G}(\omega, r, r') = \frac{1}{\mathcal{W}} \times
\begin{cases}
\phi_\text{in}(r) \, \phi_\text{up}(r'), & r < r' \\
\phi_\text{in}(r') \, \phi_\text{up}(r), & r > r',
\end{cases} \label{Gformal}
    \ee
    where the Wronskian $\mathcal{W} = \phi_\text{up} \partial_{r_\ast}\phi_\text{in} - \phi_\text{in} \partial_{r_\ast}\phi_\text{up}$, which is an $r_*$-independent quantity and has zeros at the QNM frequencies. The fields $\phi_\text{in}$ and $\phi_\text{up}$ are the homogeneous solutions ingoing through $\mathcal{H}^+$ and outgoing through $\mathcal{H}_c^+/\mathcal{I}^+$.  Similarly, for later use, we denote the outgoing black hole homogeneous solution as $\phi_\text{out}$ and the incoming solution from $\mathcal{H}_c^-/\mathcal{I}^-$ as $\phi_\text{down}$.\footnote{In asymptotically AdS spacetimes instead of $\phi_\text{up}$ and $\phi_\text{down}$ we have $\phi_\text{n.n.}$ and $\phi_\text{n.}$ for non-normalisable and normalisable perturbations respectively.} In formulae, the homogeneous solutions in \eqref{Gformal} satisfy $\phi_{\text{in}}\sim e^{-i\omega r_*}$ as $r_*\to-\infty$ and $\phi_{\text{up}}\sim e^{i\omega r_*}$ as $r_*\to+\infty$.
   
    \item Next, $\widetilde{G}$ is split into two terms using connection formulae. For example, when $r'$ is to the right of the peak of $V(r)$ and we are interested in the response at $r> r'$ (as in FIG. \ref{fig:introimage}) we use
    \be
    \phi_\text{in}(r') = \coniu\, \phi_\text{up}(r') + \conid\, \phi_\text{down}(r')
    \ee
    which splits \eqref{Gformal} into two terms,
    \bea
    \widetilde{G}(\omega, r, r') &=& \widetilde{G}_+(\omega, r, r')  + \widetilde{G}_-(\omega, r, r'), \label{introsplit}\\
    \widetilde{G}_+ &=& \frac{\coniu \phi_\text{up}(r')\phi_\text{up}(r)}{\mathcal{W}}, \label{introG+}\\
    \widetilde{G}_- &=& \frac{\conid \phi_\text{down}(r')\phi_\text{up}(r)}{\mathcal{W}}. \label{introG-}
    \eea
    Each term has different analytic properties and asymptotics in the $\omega$-plane. $\widetilde{G}_+$ contains poles at QNM frequencies, at the zeros of $\mathcal{W}$, while $\widetilde{G}_-$ does not, because $\conid \propto \mathcal{W}$. A crucial observation is that both $\widetilde{G}_+$ and $\widetilde{G}_-$ contain poles at Matsubara frequencies, which cancel when summed to form $\widetilde{G}$. 
    \item The Fourier integral \eqref{intro:FT} can be evaluated term-by-term using a semicircle contour, but the choice of UHP or LHP closure differs term-by-term, depending on the spacetime point, $t,r$. This choice is dictated by demanding a vanishing arc contribution, which implies a convergent QNM or MM sum. These choices partition the spacetime into different regions by lightrays, which we label I, II, and III as shown in FIG. \ref{fig:introimage}. The lightrays are those emanating from $t',r'$ in both directions, as well as a third ray which reflects from the black hole potential, leading to the three regions of interest. Region II could be described as the `direct transmission' component. In region I, both $\widetilde{G}_+$  and $\widetilde{G}_-$ are closed in the LHP, the MM residues cancel, leaving a convergent QNM sum. This is equivalent to the standard calculation illustrated in FIG. \ref{fig:lore}. In region II,  $\widetilde{G}_+$ is closed in the UHP while $\widetilde{G}_-$ is closed in the LHP. Thus, $G$ is expressed as a convergent sum over only MMs in this region. In region III, both $\widetilde{G}_+$ and $\widetilde{G}_-$ are closed in the UHP, the MM residues cancel, and one is left with the usual result that $G$ vanishes due to analyticity in the UHP.
    As an additional technical note, we remark that the parameter $\epsilon$ in \eqref{intro:FT} will be chosen as an infinitesimal positive real number such that the integration line does not cross any singularity of $\widetilde{G}_+$, and so that all Matsubara frequencies lie above the integration line.
\end{enumerate}
The final result is thus a convergent mode expansion for black hole Green's function in all regions, using QNMs and a new ingredient, MMs.

As an illustration of the efficacy of this mode decomposition in practice, we draw the reader's attention to FIG. \ref{fig:SdSpartial} in the case of the Schwarzschild-de Sitter spacetime at small $\Lambda$, the analysis of which is given in section \ref{sec:SdS}.

Of course, it is well known that, in some cases, a QNM sum converges at sufficiently late times \cite{SunPrice88, Bachelot1993, Ching:1993gt, Ching:1995rt, SaBarretoZworski1997, Beyer:1998nu, Szpak:2004sf, Berti:2006wq, bony2008decay, Dyatlov:2010hq, Ansorg:2016ztf, PanossoMacedo:2018hab, Chen:2023hra, Besson:2024adi} and such work is consistent with our region I results. The study of QNMs is a rich and varied active research subject and we defer to reviews for a more complete treatment \cite{HansPeterNollert_1999, Kokkotas:1999bd, Berti:2009kk, Konoplya:2011qq, Berti:2025hly}. In particular, it is worth highlighting that despite a lack of a spectral theorem, there are known orthogonality relations among QNMs \cite{Jafferis:2013qia, Green:2022htq, London:2023aeo, London:2023idh, Arnaudo:2025bnm}.\footnote{See also \cite{Finster:2000jz,Finster:2003fu}.}

Our main contribution to this story is what happens at earlier times, in region II (see FIG. \ref{fig:introimage}), where we find that the Matubara mode sum converges. These are Fourier modes of the Euclidean thermal circle associated to the region. For example, in FIG. \ref{fig:introimage}, region II intersects the right horizon, and the MMs are those with the temperature of that horizon. In the case of asymptotically flat spacetimes with region II intersecting null infinity, the relevant temperature is zero, and the line of MMs forms a cut instead, and the Matsubara contribution takes the form of cut discontinuity integrals.

One may question how $G$ can be expressed as an MM sum, since QNMs are the only perturbations that obey all appropriate boundary conditions in the problem. The key insight is that the MM sum only applies in region II which intersects only one boundary. In the case discussed above, region II intersects $\mathcal{H}_c^+/\mathcal{I}^+$, and so the mode sum in this region need not respect the ingoing black hole boundary conditions. Similarly, when $r'$ is inside the peak of $V(r)$, region II intersects $\mathcal{H}^+$, and need not exhibit outgoing behaviour through $\mathcal{H}_c^+/\mathcal{I}^+$.

The inclusion of MMs becomes crucial for the correct analysis of gravitational wave signals, as for generic initial data the linear gravitational perturbation receives contributions from these modes, as we discuss in section \ref{sec:initialdata}.

We note that Lorentzian perturbations at Matsubara frequencies have appeared in previous work in various contexts, sometimes called `redshift modes' or `horizon modes' \cite{Mino:2008at,Zimmerman:2011dx,DeAmicis:2025xuh,Oshita:2025qmn}. For example, they appear in the near-horizon region for a plunging particle in a black hole \cite{DeAmicis:2025xuh}. Here the Matsubara modes appear in a systematic treatment of the spectral decomposition of the black hole Green's function, and as far as we are aware this is the first such treatment. We also note that the appearance of different regions has also arisen, \cite{Andersson:1996cm, Berti:2006wq, Okuzumi:2008ej, Lagos:2022otp, Chen:2023hra, Chavda:2024awq, Ling:2025wfv}. Perhaps the closest existing work in the literature is \cite{Andersson:1996cm} in which different requirements for closing contour integrals in different regions of spacetime is mentioned, though not pursued, as well as \cite{Chen:2023hra} in which the QNM region for scalar correlators in the Rindler patch of AdS$_2$ was identified (this example is discussed in Appendix \ref{sec:AdS2}).
\\\\
The outline of the paper is as follows. In section \ref{sec:PT} we use the P\"oschl-Teller potential as a toy model to analytically demonstrate the decomposition of retarded Green's functions. Next, in section \ref{sec:BTZ} we turn to another mostly-analytic example, the BTZ black hole. We address Schwarzschild and Schwarzschild-de Sitter in section \ref{sec:SdS}. A vastly simpler example, scalar correlators in Rindler-sliced AdS$_2$, can be found in Appendix \ref{sec:AdS2}. We finish with a discussion in section \ref{sec:discussion}.

\section{P\"oschl-Teller} \label{sec:PT}
As a first example of mode decomposition, we analyse the P\"oschl-Teller model. It can be regarded as a toy model of perturbations to a black hole, with the benefit that our analysis is entirely analytic. It most closely resembles a black hole such as Schwarzschild-de Sitter rather than an asymptotically flat spacetime, as the points $r_\ast \to \pm \infty$ (the `horizons') are both regular singular points. The model takes the form of \eqref{GreensPDE} where
\be
V = -\left(\nu-\frac{1}{2}\right)\left(\nu + \frac{1}{2}\right)\text{sech}^2(r_*),
\ee
where $\nu$ is a parameter of the model, chosen so that $V$ has a maximum $V_0 > 0$ at $r_* = 0$, i.e. $\nu= \pm\frac{1}{2}\sqrt{1-4 V_0}$. It is convenient to work with a new coordinate $z = r = \tanh{r_*}$. The homogeneous form of \eqref{GreensPDE} in frequency space is given by
\be
\left(\partial_{r_*}^2 + \omega^2 - V\right)\psi(z) = 0, \label{PThom}
\ee
whose solutions are associated Legendre functions,
\bea
\psi_\text{down}(z) &=& c_d \, P_{\nu-\frac{1}{2}}^{-i \omega}(z), \label{PThom1}\\
\psi_\text{up}(z) &=& c_u \, P_{\nu-\frac{1}{2}}^{i \omega}(z), \\
\psi_\text{in}(z) &=& c_i^+ \, P_{\nu - \tfrac{1}{2}}^{i \omega}(z) + c_i^-\, P_{\nu - \tfrac{1}{2}}^{-i \omega}( z), \\
\psi_\text{out}(z) &=& c_o^+\, P_{\nu - \tfrac{1}{2}}^{i \omega}(z) + c_o^-\, P_{\nu - \tfrac{1}{2}}^{-i \omega}( z),\label{PThom4}
\eea
where the coefficients are functions of $\omega$ found in \eqref{PTmodescd}-\eqref{PTmodescom}. These are normalised such that $\psi_\text{down}(z) \sim e^{-i \omega r_*}$, $\psi_\text{up}(z) \sim e^{i \omega r_*}$ as $r_*\to \infty$, and $\psi_\text{in}(z) \sim e^{-i \omega r_*}$, $\psi_\text{out}(z) \sim e^{i \omega r_*}$ as $r_*\to -\infty$. As \eqref{PThom} is a second order ODE there are only two independent solutions, and so there are two connection formulae, 
\bea
\psi_\text{in}(z) &=& \coniu \psi_\text{up}(z) + \conid \psi_\text{down}(z), \label{PTcon1}\\
\psi_\text{out}(z) &=& \conou \psi_\text{up}(z) + \conod \psi_\text{down}(z), \label{PTcon2}
\eea
with connection coefficients given in \eqref{PTconid}-\eqref{PTconou}. In particular, $\conid = i\mathcal{W}/(2\omega)$.
Next, we construct $\widetilde{G}(\omega, z, z')$ according to \eqref{Gformal},
\be
\widetilde{G}(\omega, z, z') = \frac{\psi_\text{in}(z_<)\psi_\text{up}(z_>)}{\mathcal{W}}
\ee
where we have introduced the notation $z_< \equiv \min(z,z')$ and $z_> \equiv \max(z,z')$, and where the Wronskian is given by
\be
\mathcal{W} =\frac{2\Gamma(1-i\omega)^2}{\Gamma(\frac{1}{2}-\nu - i\omega)\Gamma(\frac{1}{2}+\nu - i\omega)}.
\ee
Since $V$ is symmetric under $z \to -z$, without loss of generality, we restrict our attention to the case $z' > 0$. In this case, as shall become clear shortly, it is beneficial to express $\widetilde{G}(\omega, z, z')$ in terms of the functions defined at infinity, $\psi_\text{up}$ and $\psi_\text{down}$, using connection formula \eqref{PTcon1}. As described in the introduction, this gives rise to a two-term split of $\widetilde{G}(\omega, z, z')$, as in \eqref{introsplit} with terms \eqref{introG+} and \eqref{introG-}.

The benefit of this split can be seen by turning to the large $r_*, r_*'$ asymptotics of each term,
\be
\widetilde{G}_\pm (\omega, z, z') \sim \alpha_\pm e^{i\omega \left| r_{*} \pm r_{*}'\right|} \qquad \text{as}\quad r_*,r_*'\to\infty, \label{PTasy}
\ee
where $\alpha_\pm$ are given in \eqref{PTalphaplus} and \eqref{PTalphaminus}.
Inserting this into the Fourier integral \eqref{intro:FT} gives an integrand $\sim\frac{\alpha_\pm}{2\pi} e^{-i\omega\left(t-t'-\left| r_{*} \pm r_{*}'\right|\right)}$.  This exponential behaviour in $\omega$ allows us to guarantee vanishing arc contributions by closing the integration contour in the UHP or LHP depending on the sign of $t-t'-\left| r_{*} \pm r_{*}'\right|$ (see section \ref{sec:PTasyarcs} for full details). If $t-t' < \left| r_{*} \pm r_{*}'\right|$ the contour should be closed in the UHP, while if $t-t'>\left| r_{*} \pm r_{*}'\right|$ it should be closed in the LHP. This motivates the following definition of three regions of spacetime, shown in FIG.\ref{fig:introimage},
\bea
\text{region I:}\quad && t-t' > \left| r_{*} + r_{*}'\right| 
\;\cap\; t-t' > \left| r_{*} - r_{*}'\right|\quad\label{PTreg1def}\\
\text{region II:}\quad && t-t' < \left| r_{*} + r_{*}'\right| 
\;\cap\; t-t' > \left| r_{*} - r_{*}'\right|\\
\text{region III:}\quad && t-t' < \left| r_{*} - r_{*}'\right|
\eea
$t-t' = |r_{*} - r_{*}'|$ is the forward lightcone from the insertion, while $t-t' = |r_{*} + r_{*}'|$ is the  reflection of this lightcone from the peak of the potential at $z=r=r_\ast = 0$. Thus, in region I we close both the $\widetilde{G}_+$ and $\widetilde{G}_- $ integrals in the LHP, in region II we close $\widetilde{G}_+$ in the UHP and $\widetilde{G}_-$ in the LHP, in region III we close both integrals in the UHP. This is illustrated in FIG.\ref{fig:introimage}. These choices give us vanishing arc contributions, and therefore an expression for the Fourier transform as a \emph{convergent} mode sum in each region. This split of terms and the associated regions also apply to finite $r_*, r_*'$, which we verify a posteriori using convergence tests below.

Let us turn to analytic structure for the integrand of \eqref{intro:FT} separated into the two terms corresponding to the $\widetilde{G}_+$ integral and $\widetilde{G}_-$ integral respectively,
\bea
I_+ &=& -\frac{i}{4\pi}\text{csch}(\pi\omega)\cos(\pi\nu)\Gamma\left(\frac{1}{2}-\nu-i\omega\right)\label{PTIplus}\\
&&\times\;\Gamma\left(\frac{1}{2}+\nu-i\omega\right)P_{\nu-\frac{1}{2}}^{i\omega}(z_<)P_{\nu-\frac{1}{2}}^{i\omega}(z_>)e^{-i\omega(t-t')},\nonumber\\
I_- &=& \frac{i}{4}\text{csch}(\pi\omega)P_{\nu-\frac{1}{2}}^{-i\omega}(z_<)P_{\nu-\frac{1}{2}}^{i\omega}(z_>)e^{-i\omega(t-t')}.\label{PTIminus}
\eea
$I_+$ has poles at QNM frequencies, these are the poles of the gamma functions, 
\be
\omega^\pm_n = -i\left(\frac{1}{2}+ n\right) \pm i \nu\qquad n\in \mathbb{Z}_{\geq 0}. \label{PTQNMfreq}
\ee
They are QNMs because they are zeros of $\mathcal{W}$ and thus are both ingoing at $\mathcal{H}^+$ and outgoing at $\mathcal{H}_c^+/\mathcal{I}^+$.

In addition, there are a set of poles at Matsubara frequencies in both $I_+$ and $I_-$ due to the common $\text{csch}(\pi\omega)$ prefactor, i.e. at $\omega = i n$ for $n \in \mathbb{Z}$. These are the Fourier modes for the Euclidean thermal circle associated to the `horizon' in the P\"oschl-Teller model. In other words, these poles are located at
\be
\omega_n = \frac{2n\pi}{\beta}i, \qquad n \in \mathbb{Z},
\ee
with inverse temperature $\beta = 2\pi$. This connection to horizon temperature will be more direct in other models in this paper, but it can also be seen in this toy model by studying the near horizon limit of \eqref{PThom}. We take $z = 1-\alpha \rho^2$ then at small $\rho$ one has $(\rho^2\partial_\rho^2 + \rho\partial_\rho +\omega^2)\psi = 0$ at leading order, which is the wave equation on Rindler spacetime, $-\rho^2 dt^2 + d\rho^2$, which indeed, has a Euclidean thermal period $\beta = 2\pi$.

All associated residues are straightforward to obtain from $I_\pm$ \eqref{PTIplus}, \eqref{PTIminus}, and we will not quote them here directly, except to point out that the residues at Matsubara frequencies obey
\be
\res\left(I_+,i n\right) =- \res\left(I_-,i n\right), \qquad n \in \mathbb{Z}. \label{PTsignprop}
\ee
This had to be the case, since $\widetilde{G}$ itself has poles only at QNM frequencies. 

Finally, we can compute the Green's function in each region as a convergent mode sum.

\subsection{Region I} \label{sec:PTreg1}
See row \textbf{b.I} in FIG.\ref{fig:introimage} for reference.
In this region we close both $\widetilde{G}_+$ and $\widetilde{G}_- $ integrals in the LHP. The sum over Matsubara residues cancels due to the property \eqref{PTsignprop}, leaving only the sum over QNMs which arises from the $\widetilde{G}_+$ integral. Thus, 
\bea
&&G(t,t',z,z') = -2\pi i \sum_\pm\sum_{n=0}^\infty\res\left(I_+,\omega^\pm_n\right)\\
&=& \sum_\pm\sum_{n=0}^\infty\frac{\Gamma\left(\pm2\nu-n\right)}{2n!}P_{\nu-\frac{1}{2}}^{i\omega_n^\pm}(z_<)P_{\nu-\frac{1}{2}}^{i\omega_n^\pm}(z_>)e^{-i\omega_n^\pm(t-t')}\qquad
\eea
where $\omega_n^\pm$ are given in \eqref{PTQNMfreq}. The functions $P_{\nu-\frac{1}{2}}^{i\omega_n^\pm}(z)$ are the QNM radial wavefunctions, and so the residue sum is indeed the QNM mode sum.

One can show that this QNM sum converges in this region using a ratio test. In particular, we use the hypergeometric representation of $P_\alpha^{\mu}(z)$ to get the asymptotic approximation
\be
P_\alpha^{\mu}(z) \sim \frac{1}{\Gamma(1-\mu)}\left(\frac{1+z}{1-z}\right)^{\frac{\mu}{2}},
\label{LegendreAsy}
\ee
for $z>0$ and $|\mu| \to \infty$ with $\arg(\mu)\neq 0$, while for $z<0$ the right hand side of \eqref{LegendreAsy} is modified by $z\to -z$.
Combining with Stirling's formula for the gamma functions, we find\footnote{We focus on the case of $\im{\nu} \neq 0$ for simplicity, i.e. $V_0 > \frac{1}{4}$.}
\be
L_\text{QNM} \equiv \lim_{n\to \infty}\left|\frac{\res\left(I_+,\omega^\pm_{n+1}\right)}{\res\left(I_+,\omega^\pm_n\right)}\right| = e^{-t+t' + |r_*| + r_*'}. \label{PTQNMcrit}
\ee
Recall that we took $r_*' > 0$ without loss of generality, and $r_*$ can take any sign in this region. The definition \eqref{PTreg1def} is equivalent to $t- t' -r_*' > |r_*|$ and thus $L_\text{QNM}<1$ in region I where the QNM sum converges absolutely. On the boundary of the region where $L=1$ the ratio test is inconclusive, however, a numerical investigation of the coefficients indicates that the terms in the series are alternating and monotonically decreasing, implying convergence via the alternating series test. Notice that the convergence criterion fails as soon as we pass from region I into region II, or from region I into region III, i.e. the QNM representation is not valid in regions II or III.

\subsection{Region II}\label{sec:PTreg2}
See row \textbf{b.II} in FIG.\ref{fig:introimage} for reference. In this region, we close the $\widetilde{G}_+$ integral in the UHP and $\widetilde{G}_-$ in the LHP. Thus, there are no QNM contributions, only MMs. A convention choice places our contour slightly above the real axis, and so from $\widetilde{G}_+$ we pick up Matsubara residues for $n>0$ and from $\widetilde{G}_-$ we pick up Matsubara residues for $n \leq 0$. Thus, 
\be
G(t,t',z,z') = 2\pi i \sum_{n=1}^\infty\res\left(I_+,i n\right)-2\pi i \sum_{n=-\infty}^0\res\left(I_-,i n\right). \label{PTmatsum}
\ee
Again, these residues are straightforwardly obtained from \eqref{PTIplus} and \eqref{PTIminus}.

The Matsubara residue sum \eqref{PTmatsum} is a sum over radial functions that are regular in the Euclidean thermal circle associated to the horizon at $z=1$. To see this, we again turn to the Rindler coordinate $z = 1 - \alpha \rho^2$ then
\be
\res\left(I_\pm,i n\right) \propto \rho^{|n|} e^{nt} = \rho^{|n|} e^{in\phi}
\ee
in the near-horizon region $\rho \to 0$, where $\phi = -i t$ is an angular coordinate on the Euclidean disk, $\rho^2d\phi^2 + d\rho^2$. In other words, these modes are regular as $r_*\to\infty$ ($\rho \to 0$), as appropriate for the portion of spacetime covered by region II.

Similarly to the QNM case, one can show that the Matsubara residue sum converges in this region using a ratio test. As before we use \eqref{LegendreAsy} and Stirling's formula, however, there is one associated Legendre function in $\res\left(I_-, in\right)$ for which \eqref{LegendreAsy} does not apply; for this term we first use a reflection formula,
\bea
\frac{\sin\!\big((\alpha-\mu)\pi\big)}{\Gamma(\alpha+\mu+1)}\,P_\alpha^{\mu}(z) &=& \frac{\sin(\alpha\pi)}{\Gamma(\alpha-\mu+1)}\,P_\alpha^{-\mu}(z)\\
&&- \frac{\sin(\mu\pi)}{\Gamma(\alpha-\mu+1)}\,P_\alpha^{-\mu}(-z),\nonumber
\eea
and then use \eqref{LegendreAsy} and Stirling.
The result is,\footnote{Again, we focus on the case of $\im{\nu} \neq 0$ for simplicity.}
\bea
L_\text{MM}^+ \equiv \lim_{n\to \infty}\left|\frac{\res\left(I_+,i(n+1)\right)}{\res\left(I_+, in\right)}\right| &=& e^{t-t' - r_* - r_*'}, \label{PTMMcrit1}\\
L_\text{MM}^-\equiv \lim_{n\to -\infty}\left|\frac{\res\left(I_-,i(n-1)\right)}{\res\left(I_-, in\right)}\right| &=& e^{-t+t' - r_* - r_*'},\;\;\label{PTMMcrit2}
\eea
for which both $L_\text{MM}^+ < 1$ and $L_\text{MM}^- < 1$ in region II. Note that the convergence criterion $L_\text{MM}^+ < 1$ fails as soon as we pass from region II into region I. Passing from region II into region III this Matsubara sum remains finite (so long as $L_\text{MM}^- < 1$ holds) -- numerically we find that this contribution is cancelled by a finite arc integral, i.e. the LHP arc for $I_-$. 

Finally, we note that in both region I and region II, the mode sum can be performed exactly at large $r_*, r_*'$, which gives the same closed form expression upon analytic continuation,
\be
G(t,t',z,z') \sim \frac{1}{2}\,{}_2F_1\left(\frac{1}{2}-\nu, \frac{1}{2}+\nu, 1, -e^{t-t'-r_*-r_*'}\right), \label{PTGasy}
\ee
as $r_*,r_*'\to\infty$.
The spacetime dependence appears only in the form of $e^{t-t'-r_*-r_*'}$ with a logarithmic branch point singularity at $e^{t-t'-r_*-r_*'} = -1$, which is responsible for the radius of convergence of the regional expansions.

\subsection{Region III}\label{sec:PTreg3}
See row \textbf{b.III} in FIG.\ref{fig:introimage} for reference. In this region, both terms are closed in the UHP. The only poles in the UHP correspond to MMs, however, their residue contributions cancel because of the property \eqref{PTsignprop}. Therefore, 
\be
G(t,t',z,z') = 0.
\ee
This is the usual result that the retarded Green's function vanishes outside the forward lightcone, which follows from analyticity of $\widetilde{G}$ in the UHP.

\subsection{Partial sums}
\begin{figure}[h]
\centering
\includegraphics[width=\columnwidth]{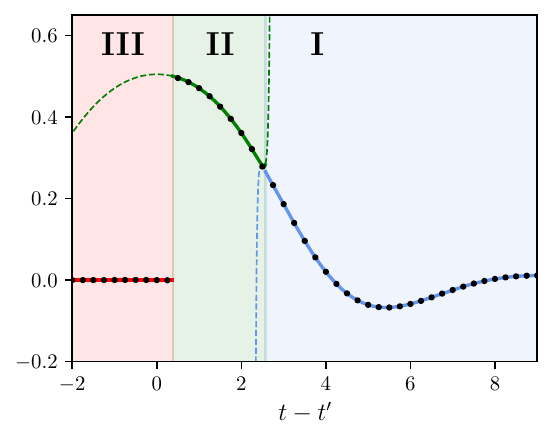}
\caption{The retarded Green's function $G(t,t',z,z')$ for P\"oschl-Teller as a convergent mode sum at fixed $z,z'$. In \textbf{region I} (blue) the Green's function is a convergent sum of QNMs, see section \ref{sec:PTreg1}. In \textbf{region II} (green) it is a convergent sum of MMs, see section \ref{sec:PTreg2}. In \textbf{region III} (red) it is zero, see section \ref{sec:PTreg3}. The lines show sums of the first 30 modes in each sum -- in the region where they apply (solid), and their extension to the regions where they do not (dashed). Black points show the Green's function obtained by direct numerical integration for comparison. The parameters chosen are $z' = 8/10$, $z=9/10$, $\nu = i \sqrt{3}/2$.}
\label{fig:PTpartial}
\end{figure}

As a demonstration of the convergent mode sum in practise, in this section we construct $G(t,t',z,z')$ using a partial sum of modes and compare to a direct numerical integration of \eqref{intro:FT} along the real $\omega$ axis. The result is shown in FIG. \ref{fig:PTpartial}. In the region where a mode sum converges -- QNMs in region I (section \ref{sec:PTreg1}) and MMs in region II (section \ref{sec:PTreg2}) -- the agreement is excellent. Applied in an incorrect region one sees the lack of convergence, as proven from ratio tests in \eqref{PTQNMcrit}, \eqref{PTMMcrit1} and \eqref{PTMMcrit2}.

\section{BTZ} \label{sec:BTZ}
In the last section, we presented the complete mode decomposition of the retarded Green's function for the P\"oschl-Teller toy model as a first analytic example that illustrates the main results of this paper. In this section, we turn our attention to linear perturbations on the (non-rotating) BTZ black hole. We choose to study BTZ since, given that it is an asymptotically AdS$_3$ spacetime, it allows us to demonstrate the universality of our results across different asymptotics, discussing the slightly different details that arise in AdS where needed. Moreover, it is convenient to work with BTZ as its QNM frequencies and wavefunctions are known analytically. 

We note that there are a number of similarities to a previous study of Wightman functions in the Rindler patch of AdS$_2$ \cite{Chen:2023hra}, which we discuss at length in Appendix \ref{sec:AdS2}. 

We consider a massive scalar field $\Phi$ on the (non-rotating) BTZ background \cite{Banados:1992wn}
\bea
ds^2&=&-f(r)dt^2+\frac{1}{f(r)}dr^2+r^2d\varphi^2,\\  
f(r)&=&\frac{1}{L_{\text{AdS}}^2}(r^2-r_h^2),
\eea
fixing for simplicity $L_{\text{AdS}}=1$, $r_h=1$. The radial coordinate $r$ spans from the location of the event horizon, $r=1$, to the conformal timelike boundary of AdS, $r\to+\infty$, while $\varphi$ corresponds to an angular coordinate $\varphi \in [0,2\pi[$. The conformal boundary requires a choice of boundary conditions, which, in the context of the AdS/CFT correspondence \cite{Maldacena:1997re,Witten:1998qj,Gubser:1998bc}, corresponds to a quantisation choice in the dual CFT and is needed for a well-posed evolution of the bulk field $\Phi$. As we shall see, this is one of the principal differences that distinguishes this section from the other examples considered in this paper. 

We study a scalar field dual to a dimension $\Delta$ operator in the CFT, so that $m_\Phi^2=\Delta(\Delta-2)$. In particular, we restrict our attention to cases where $\Delta$ is the largest root of this equation, i.e. $\Delta > 1$. The dynamics of $\Phi$ is governed by the massive Klein-Gordon equation on this background, $(\Box-m^2_\Phi)\Phi=0$. Without loss of generality, we consider the radial problem for $\Phi$ by performing the following decomposition,
\be
\Phi(t,r,\varphi) = \sum_{m=-\infty}^{\infty} \int d\omega \,e^{-i\omega t}\frac{1}{\sqrt{r}}\phi(r)e^{i m \varphi}, \label{BTZ_decomposition}
\ee
where we have introduced a factor of $1/\sqrt{r}$ such that $\phi$ obeys the homogeneous form of \eqref{GreensPDE} in frequency space,
\be
\left(\partial_{r_*}^2 + \omega^2 - V\right)\phi(r) = 0, \label{BTZhom}
\ee
with effective potential
\be
V(r)=\Delta(\Delta-2)f+\frac{m^2f}{r^2}-\frac{f^2}{4r^2}+\frac{ff'}{2r}, \label{BTZ_effV}
\ee
and where $r_\ast$ corresponds to the standard tortoise radial coordinate
\bea
r_\ast &=& \int \frac{dr}{f(r)} \\
&=& \frac{1}{2}\log\left(\frac{r-1}{r+1}\right),
\eea
where we have fixed the integration constant such that $r_*(r\to+\infty)=0$.

In order to find the solutions to \eqref{BTZhom}, it is convenient to change coordinates to $z=(r^2-1)/r^2$ -- where now $z=0$ corresponds to the black hole horizon and $z=1$ to the conformal boundary -- and make the following field redefinition
\be
\phi(r(z)) = z^{-i\omega/2}(1-z)^{\Delta/2-1/4}\psi(z),
\ee
under which \eqref{BTZhom} takes the form of the hypergeometric differential equation \cite{Erdelyi1953}
\be
z(1-z)\frac{d^2 \psi}{dz^2} + \left(c-(a+b+1)z\right)\frac{d\psi}{dz}-ab\,\psi=0, \label{BTZhyperODE}
\ee
with parameters
\bea
a&=&-\frac{im}{2}+\frac{\Delta}{2}-\frac{i\omega}{2}, \label{BTZ_hyper_a}\\
b&=&\frac{im}{2}+\frac{\Delta}{2}-\frac{i\omega}{2}, \label{BTZ_hyper_b}\\
c&=&1-i\omega. \label{BTZ_hyper_c}
\eea
The solutions to \eqref{BTZhyperODE} around $z=0$ are
\bea
\psi_{\text{in}}(z)&=&{}_2F_1(a, b; c; z), \label{BTZpsiin}\\
\psi_{\text{out}}(z)&=&z^{1-c}{}_2F_1(1+a-c, 1+b-c; 2-c; z), \label{BTZpsiout}
\eea
and around $z=1$,
\bea
\psi_{\text{n.}}(z)&=&{}_2F_1(a, b; 1+a+b-c; 1-z), \\
\psi_{\text{n.n.}}(z)&=&(1-z)^{c-a-b} \\
&&\times\, {}_2F_1(c-a, c-b; 1-a-b+c; 1-z). \nonumber
\eea
Noting that \eqref{BTZhyperODE} is second order, these are not all linearly independent solutions, and are related by connection formulae. In what follows, it will be useful to connect the solutions around the boundary to those around the horizon via \cite{Erdelyi1953}
\bea
\psi_{\text{n.}}(z) &=& \cnin \psi_{\text{in}}(z) + \cnout \psi_{\text{out}}(z) \label{BTZn.toinout},\\
\psi_{\text{n.n.}}(z) &=& \cnnin \psi_{\text{in}}(z) + \cnnout \psi_{\text{out}}(z),\label{BTZn.n.toinout}
\eea
with connection coefficients given by \eqref{BTZ_cnin}-\eqref{BTZ_cnnout}.

\begin{figure}[h]
\centering
\includegraphics[width=0.6\columnwidth]{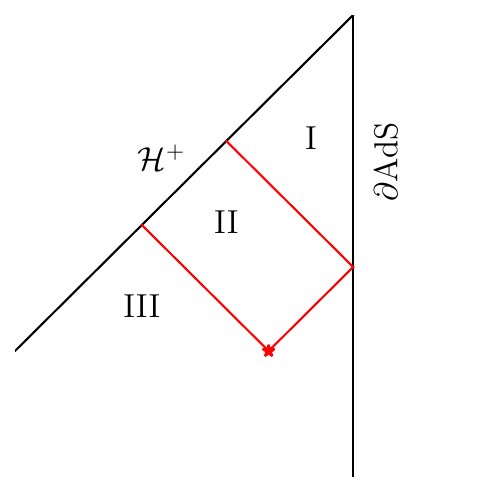}
\caption{Conformal diagram for asymptotically AdS black hole spacetimes showing three regions delineated by lightrays (red lines) emanating from a delta insertion source (red star) and reflecting off the timelike boundary ($\partial$AdS). In regions I and II, the retarded $G$ is constructed from a convergent sum of QNMs and a convergent sum of MMs, respectively. In region III, $G=0$ by causality. For further details, see the analogous construction for asymptotically flat and dS black holes in FIG. \ref{fig:introimage}.}
\label{fig:btzregions}
\end{figure}

Armed with these results, we can construct $\widetilde{G}(\omega,z,z')$ as indicated by \eqref{Gformal}, with the caveat that, due to the presence of the conformal boundary at infinity, boundary conditions are of a different nature there. Namely, for $\Delta>1$, the retarded $G$ requires a Dirichlet zero boundary condition at $z=1$ which is satisfied by the \emph{normalisable} perturbation, $\phi_{\text{n.}}(r(z))$. Physically, this is equivalent to perfectly reflecting boundary conditions at the boundary.\footnote{In the dual CFT, normalisable boundary conditions correspond to an absence of sources for the operator of dimension $\Delta$ dual to the bulk scalar field $\Phi$.} Thus, $\widetilde{G}$ is given by
\be
\widetilde{G}(\omega, z, z') = \frac{\phi_\text{in}(r(z_<))\phi_\text{n.}(r(z_>))}{\mathcal{W}}, \label{BTZGtilde}
\ee
where we adopt the same definition for $z_<,z_>$ as in section \ref{sec:PT}, and the Wronskian reads
\bea
\mathcal{W}&=&-2 i \omega\,\cnout \nonumber\\
&=&2\frac{\Gamma(c)\Gamma(1+a+b-c)}{\Gamma(a)\Gamma(b)}.
\eea
The timelike boundary acts as a reflective barrier where lightrays bounce off, which gives rise to the region subdivision of spacetime shown in FIG. \ref{fig:btzregions} -- the analogous construction to FIG. \ref{fig:introimage} for asymptotically AdS spacetimes. 

In order to see how this separation of spacetime into three regions arises, it is convenient to express $\widetilde{G}$ in terms of the two independent solutions around the horizon \eqref{BTZpsiin}, \eqref{BTZpsiout} using the connection formula \eqref{BTZn.toinout} in \eqref{BTZGtilde}, yielding the following two-term split
\bea
\widetilde{G}(\omega, z, z') &=& \widetilde{G}_+(\omega, z, z')  + \widetilde{G}_-(\omega, z, z'), \\[10pt]
\widetilde{G}_+ &=& i\frac{\cnin \phi_\text{in}(r(z_<))\phi_\text{in}(r(z_>))}{2 \omega\,\cnout}, \\
\widetilde{G}_- &=& i\frac{\phi_\text{in}(r(z_<))\phi_\text{out}(r(z_>))}{2\omega}. 
\eea
\noindent Similarly to section \ref{sec:PT}, we can first study each term at asymptotically large $r_\ast$ and $r_\ast'$, simplifying the analysis greatly. Here, it is natural to look at the near-horizon asymptotics taking $r_\ast,r_\ast' \to -\infty$ ($z,z'\to0^+$) where
\be
\widetilde{G}_\pm (\omega, z, z') \sim \alpha_\pm e^{i\omega \left|r_{*} \pm r_{*}'\right|}, \label{BTZGtildeasy}
\ee
with coefficients
\bea
\alpha_+ &=& i \frac{\cnin 4^{-i\omega}}{2\cnout \omega}, \label{BTZ_alpha+}\\
\alpha_- &=& \frac{i}{2\omega}.
\eea
The integrand in \eqref{intro:FT} thus behaves as $\sim \frac{\alpha_\pm}{2\pi}e^{-i\omega(t-t'-\left|r_\ast \pm r_\ast'\right|)}$ as $r_\ast,r_\ast'\to-\infty$, hinting to close the integration contour in the UHP when $t-t'<\left|r_\ast \pm r_\ast'\right|$ and in the LHP when $t-t'>\left|r_\ast \pm r_\ast'\right|$ for the large $\omega$ arc contribution to vanish. One can prove that this is indeed the case by Jordan's lemma in the same fashion as \ref{sec:PTasyarcs}.\footnote{At first sight, one may think that the $4^{-i\omega}$ factor in \eqref{BTZ_alpha+} could contribute to a constant shift in the exponent $t-t'-\left|r_\ast \pm r_\ast'\right|$, hence changing the location of the lightrays delineating the three spacetime regions. However, a careful analysis of the large $\omega$ asymptotics for this term including the contributions coming from $\cnin/(\omega \,\cnout)$ shows that there is no such shift.} This leads to the following definitions of three spacetime regions, shown in FIG. \ref{fig:btzregions},
\bea
\text{region I:}\quad && t-t' > \left|r_{*} + r_{*}'\right| 
\;\cap\; t-t' > \left| r_{*} - r_{*}'\right|\quad\\
\text{region II:}\quad && t-t' <  \left|r_{*} + r_{*}'\right| 
\;\cap\; t-t' > \left| r_{*} - r_{*}'\right|\\
\text{region III:}\quad && t-t' < \left| r_{*} - r_{*}'\right| 
\eea
Note that, written in terms of the $r_\ast$ coordinate for each spacetime, the region subdivision for BTZ mimics that of the previous P\"oschl-Teller case in section \ref{sec:PT}. Therefore, it has an analogous interpretation, adapted to the AdS asymptotics. In particular, the location of the forward lightcone is again given by $t-t' = |r_{*} - r_{*}'|$, while now $t-t' = |r_{*} + r_{*}'|$ determines the location of the lightray reflected from the timelike boundary of AdS -- infinite potential barrier in \eqref{BTZ_effV} at $r_\ast=0$ -- as opposed to the finite $r_*=0$ potential peak in P\"oschl-Teller. In order to have vanishing arc contributions, hence leading to a convergent mode sum representation for $G$ in \eqref{intro:FT}, we choose different contours for the $\widetilde{G}_+$ and $\widetilde{G}_-$ integrals, separately, in each region. In region I, we close both contours in the LHP. In region II, we close the contour for $\widetilde{G}_+$ in the UHP, and in the LHP for $\widetilde{G}_-$. Finally, in region III, both contours for $\widetilde{G}_+$ and $\widetilde{G}_-$ are closed in the UHP. Next, we shall see that these choices also lead to convergent mode sums at finite $r_\ast, r_\ast'$.

Let us now study the analytic structure in the complex $\omega$ plane of the $\widetilde{G}_+$ and $\widetilde{G}_-$ integrands in \eqref{intro:FT} at finite $r_\ast$, $r_\ast'$. The integrands read, respectively,
\bea
I_+ &=& e^{-i\omega(t-t')}\frac{\Gamma(a)\Gamma(b) \Gamma(1-c)}{4\pi\Gamma(c)\Gamma(1+a-c)\Gamma(1+b-c)} \nonumber\\
&\times& (1-z_<)^{\Delta/2-1/4}(1-z_>)^{\Delta/2-1/4}z_{<}^{-i\omega/2}z_{>}^{-i\omega/2} ,\nonumber \\
&\times& {}_2F_1(a, b; c; z_<) {}_2F_1(a, b; c; z_>) \label{BTZ_Iplus}\\[10pt]
I_- &=& e^{-i\omega(t-t')}\frac{i}{4\pi\omega} \label{BTZ_Iminus}\\
&\times& (1-z_<)^{\Delta/2-1/4}(1-z_>)^{\Delta/2-1/4}z_{<}^{-i\omega/2}z_{>}^{i\omega/2} \nonumber\\
&\times& {}_2F_1(a, b; c; z_<)\,{}_2F_1(1+a-c, 1+b-c; 2-c; z_>) .\nonumber
\eea
$I_+$ has poles at $a=-n$ and $b=-n$ for $n\in\mathbb{Z}_{\geq0}$ coming from the poles of the Gamma functions $\Gamma(a)\Gamma(b)$(zeroes of $\mathcal{W}$), which, using \eqref{BTZ_hyper_a}, \eqref{BTZ_hyper_b}, correspond to the BTZ QNM frequencies \cite{Cardoso:2001hn, Birmingham:2001pj}
\be
\omega_n^\pm = \pm m -i(2n+\Delta), \qquad n\in \mathbb{Z}_{\geq 0}. \label{BTZ_QNFs}
\ee
Also from $\Gamma(1-c)$ in the numerator, there are poles at $1-c=-n$ for $n\in\mathbb{Z}_{\geq0}$, including a simple pole at $\omega=0$ and the UHP MMs for BTZ
\be
\omega_n = in, \qquad n\in \mathbb{Z}_{\geq1}. \label{BTZ_MMs_UHP}
\ee
Finally, $I_+$ also presents poles at $c=-n$ for $n\in\mathbb{Z}_{\geq0}$, where the hypergeometric functions in the last line of \eqref{BTZ_Iplus} are not well-defined (see \ref{app:hypergeometric_residues}). One of these sets of poles is cancelled by a set of zeroes arising from the $1/\Gamma(c)$ factor, but the other set remains and corresponds to the LHP MMs
\be
\omega_n = -in, \qquad n\in \mathbb{Z}_{\geq1}. \label{BTZ_MMs_LHP}
\ee
On the other hand, $I_-$ has a simple pole at $\omega=0$, as well as at all the MMs, \eqref{BTZ_MMs_UHP} and \eqref{BTZ_MMs_LHP}, from the hypergeometric functions in the last line of \eqref{BTZ_Iminus}.

Knowing the analytic structure of the integrands, it is now straightforward to construct the retarded $G$ in each region as a convergent sum of the corresponding modes using the residue theorem as sketched in FIG. \ref{fig:introimage}, adapted to the AdS regions in FIG. \ref{fig:btzregions}. For this purpose, note that, as expected, the property \eqref{PTsignprop} also holds in this case. In the remainder of this section, we make the choice $z<z'$ ($r_\ast<r_\ast'$).

\subsection{Region I}\label{sec:BTZreg1}
In this region, we close both integrals for $I_+$ and $I_-$ in the LHP. The MMs residues from both terms cancel each other by \eqref{PTsignprop}, leaving the QNM residues for $I_+$ as the only contribution to $G$ in \eqref{intro:FT}. Hence,
\bea
&&G(t,t',z,z') = -2\pi i \sum_\pm\sum_{n=0}^\infty\res\left(I_+,\omega^\pm_n\right) \\
&=&-2\pi i \sum_\pm\sum_{n=0}^\infty \frac{(-1)^{n+1}}{2\pi \omega^\pm_n n!}e^{-i\omega^\pm_n(t-t')} \nonumber\\
&\times&\frac{\Gamma(\mp im - n)\Gamma(i\omega^\pm_n)}{\Gamma(n+\Delta)\Gamma(\pm im+n+\Delta)\Gamma(-i\omega^\pm_n)} \nonumber\\
&\times& (1-z)^{\Delta/2-1/4}(1-z')^{\Delta/2-1/4}z^{-i\omega^\pm_n/2}z'^{-i\omega^\pm_n/2} \nonumber \\
&\times& {}_2F_1(\mp im-n, -n; 1-i\omega^\pm_n; z) \nonumber\\
&\times& {}_2F_1(\mp im-n, -n; 1-i\omega^\pm_n; z') \nonumber.
\eea
where $\omega^\pm_n$ are given in \eqref{BTZ_QNFs}. The function $(1-z)^{\Delta/2-1/4}z^{-i\omega^\pm_n/2}{}_2F_1(\mp im-n, -n; 1-i\omega^\pm_n; z)$ is $(1-z)^{-1/4}$ -- this factor comes from (the inverse of) the $1/\sqrt{r}$ field redefinition in \eqref{BTZ_decomposition} -- times the QNM radial wavefunction, thus we have constructed $G$ in this region as a QNM sum.

One can show that this QNM sum converges in this region using a ratio test. In particular, using the following identity for $n\in\mathbb{Z}_{\geq0}$
\bea
&&{}_2F_1(-n,n+\alpha+\beta+1;\alpha+1;x)=\frac{(-1)^n n!}{(\alpha+1)_n} \nonumber\\
&&\times x^n P_n^{(\beta, -2n-\alpha-\beta-1)}\left(\frac{2-x}{x}\right),
\eea
where $P_n^{(\alpha,\beta)}(x)$ are Jacobi polynomials, and using the asymptotic formula \cite{szego}
\be
\frac{P_{n+1}^{(\alpha,\beta)}(x)}{P_{n}^{(\alpha,\beta)}(x)} \sim x + \sqrt{x^2-1} \qquad \text{as}\;\; n\to+\infty,
\ee
we arrive at
\be
L_\text{QNM} \equiv \lim_{n\to \infty}\left|\frac{\res\left(I_+,\omega^\pm_{n+1}\right)}{\res\left(I_+,\omega^\pm_n\right)}\right| = e^{2(-t+t' - r_* - r_*')}. \label{BTZcrit}
\ee
The residue sum converges absolutely, i.e. $L_\text{QNM}<1$, in region I where $t- t' > -r_*-r_*'$. Note that in region II $L_{\text{QNM}}>1$, yielding a divergent sum.

\subsection{Region II}\label{sec:BTZreg2}
In this region there are no QNM contributions, as we close the $\widetilde{G}_+$ integral in the UHP and $\widetilde{G}_-$ in the LHP, such that we do not pick up the QNM poles in the LHP of $\widetilde{G}_+$. Thus, there are only MMs contributions which, following the same convention choice as in section \ref{sec:PT} to place our contours slightly above the real axis, lead to the following residue sum
\be
G(t,t',z,z') = 2\pi i \sum_{n=1}^\infty\res\left(I_+,i n\right)-2\pi i \sum_{n=-\infty}^0\res\left(I_-,i n\right). \label{BTZmatsum}
\ee
These residues are obtained from \eqref{BTZ_Iplus} and \eqref{BTZ_Iminus}, following \ref{app:hypergeometric_residues}.

As expected from the previous section, the Matsubara modes in the residue sum \eqref{BTZmatsum} are also regular functions at the black hole's horizon in Euclidean time. That is, in the new radial coordinate, $z=\rho^2(1-\rho^2/4)$, where spacetime adopts a Rindler form $-\rho^2dt^2+d\rho^2$ in the near-horizon region $\rho\to0^+$, the residues behave as
\be
\res\left(I_\pm,i n\right) \propto \rho^{|n|} e^{nt}
\ee
around $\rho=0$. This was required as region II intersects the future event horizon.

One can again prove convergence of this residue sum in this region by means of a ratio test. Namely, using Stirling's formula for $\Gamma$ functions and the asymptotic formula \cite{watson1918asymptotic}
\bea
&&{}_2F_1(n+\alpha,n+\beta;2n+\gamma,x)\sim(1-x)^{\frac{\gamma-\alpha-\beta-1/2}{2}} \nonumber\\
&&\times\left( \frac{2}{x}(1-\sqrt{1-x})\right)^{2n+\gamma-1}
\eea
as $n\to\infty$, one has
\bea
L_\text{MM}^+ \equiv \lim_{n\to \infty}\left|\frac{\res\left(I_+,i(n+1)\right)}{\res\left(I_+, in\right)}\right| &=& e^{t-t' + r_* + r_*'}, \label{BTZcrit1}\\
L_\text{MM}^-\equiv \lim_{n\to -\infty}\left|\frac{\res\left(I_-,i(n-1)\right)}{\res\left(I_-, in\right)}\right| &=& e^{-(t-t' - r_* - r_*')}, \label{BTZcrit2}
\eea
for which both $L_\text{MM}^+ < 1$ and $L_\text{MM}^- < 1$ in region II. Note that $L_\text{MM}^+ > 1$ in region I, implying a divergent MM sum. However, similarly to P\"oschl-Teller, in region III (as long as $L_\text{MM}^- < 1$ holds) the MM sum gives a finite result, indicating that there must be a finite arc integral contribution to cancel it. Indeed, we numerically test that the latter is given by the LHP arc for $I_-$.

As in the P\"oschl-Teller case, we note that the mode sum can be performed exactly at $r_*, r_*'\to-\infty$, which gives \eqref{BTZ_asyG_regI} in region I and \eqref{BTZ_asyG_regII} in region II. The spacetime dependence appears only in the form of $e^{t-t'+r_\ast+r_\ast'}$ with a logarithmic branch point singularity at $e^{t-t'+r_\ast+r_\ast'} = 1$, which is responsible for the radius of convergence of the regional expansions.

\subsection{Region III}\label{sec:BTZreg3}
In this region both terms are closed in the UHP. The only poles in the UHP correspond to MMs, but their residue contributions cancel due to \eqref{PTsignprop}. Thus, one arrives at 
\be
G(t,t',z,z') = 0.
\ee
This result follows from analyticity of the full $\widetilde{G}$ in the UHP, and is required to respect causality.

\subsection{Partial sums}
\begin{figure}[h]
\centering
\includegraphics[width=\columnwidth]{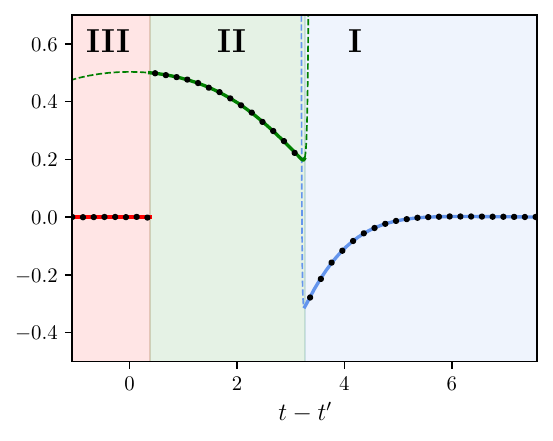}
\caption{The retarded Green's function $G(t,t',z,z')$ for BTZ as a convergent mode sum at fixed $z,z'$. In \textbf{region I} (blue) the Green's function is a convergent sum of QNMs, see section \ref{sec:BTZreg1}. In \textbf{region II} (green) it is a convergent sum of MMs, see section \ref{sec:BTZreg2}. In \textbf{region III} (red) it is zero, see section \ref{sec:BTZreg3}. The lines show sums of the first 50 modes in each sum -- in the region where they apply (solid), and their extension to the regions where they do not (dashed). Black points show the Green's function obtained by direct numerical integration for comparison. The parameters chosen are $z' = 0.2$, $z=0.1$, $\Delta = 3/2$, $m=1$.}
\label{fig:BTZpartial}
\end{figure}

Similarly to the P\"oschl-Teller case, in this section we construct $G(t,t',z,z')$ using a partial sum of modes and compare to a direct numerical integration of \eqref{intro:FT} along the real $\omega$ axis, shown in FIG. \ref{fig:BTZpartial}. Again for BTZ, there is excellent agreement between the mode sums -- following sections \ref{sec:BTZreg1}, \ref{sec:BTZreg2} and \ref{sec:BTZreg3} -- and the numerics in the regions where the former converge. Outside these regions, there is lack of convergence, as expected from the ratio tests \eqref{BTZcrit}, \eqref{BTZcrit1} and \eqref{BTZcrit2}.

\section{Schwarzschild(-de Sitter)} \label{sec:SdS}

In this section, we discuss the case of Schwarzschild-de Sitter black hole. In particular, we focus our attention on the case in which the cosmological constant is parametrically small to make contact with the asymptotically flat Schwarzschild case.\footnote{The extraction of results for the asymptotically flat geometry from the asymptotically de Sitter one was already considered in \cite{Hatsuda:2020sbn}, and subsequently in \cite{Oshita:2021iyn}, where QNMs and excitation factors of Kerr black holes were studied by taking the small cosmological constant regime of Kerr-de Sitter. We further remark that a study on the Kerr–de Sitter black hole Green's function using the Heun function was developed in \cite{Motohashi:2021zyv}, and, in the subsequent work \cite{Motohashi:2024fwt}, a detailed discussion on a consistent way to obtain the Kerr excitation factors was provided, together with a comparison with previous results. The results of \cite{Motohashi:2024fwt} were cross-validated in \cite{Lo:2025njp, Kubota:2025hjk}. See also \cite{Zhou:2025xta} for a study of QNMs in Schwarzschild-de Sitter as $\Lambda \to 0^+$.}

The metric of the four-dimensional Schwarzschild-de Sitter black hole is 
\begin{equation}
ds^2=-f(r)\,dt^2+\frac{dr^2}{f(r)}+r^2\,d\Omega_2^2,
\end{equation}
where 
\begin{equation}
f(r)=1-\frac{2M}{r}-\frac{\Lambda}{3}\,r^2.
\end{equation}
Assuming
\begin{equation}
0<9\,\Lambda\,M^2<1,
\end{equation}
the equation $f(r)=0$ has three real roots. We denote with $R_h$ the event horizon, which is the smallest positive root. The other two roots are then given by 
\begin{equation}
R_{\pm}=\frac{-R_h\pm\sqrt{\frac{12}{\Lambda}-3 R_h^2}}{2},
\end{equation}
where $R_+ > R_h$ denotes the cosmological horizon, and $R_-<0$.

The surface gravity (and temperature) at the horizon and at the cosmological horizon are defined as
\begin{equation}
\begin{aligned}
\kappa_h&=2\pi T_h=\frac{|f'(R_h)|}{2}=\frac{\Lambda  (R_h-R_-) (R_+-R_h)}{6 R_h}\\
\kappa_+&=2\pi T_+=\frac{|f'(R_+)|}{2}=\frac{\Lambda  (R_+-R_h) (R_+-R_-)}{6 R_+}.
\end{aligned}
\end{equation}
Due to the existence of two distinct horizons in the Schwarzschild-de Sitter geometry, each with an associated temperature, there exist two sets of Matsubara frequencies, each being defined as imaginary integer multiples of the corresponding surface gravity.\footnote{There exists a branch of Schwarzschild-de Sitter black hole solutions for which the temperatures $T_+$ and $T_h$ are equal, the so-called lukewarm branch. In this branch of solutions, the two sets of frequencies coincide.}

In the following discussion, we will focus on the region close to the cosmological horizon, and therefore we will need the expression of the Matsubara frequencies at the cosmological horizon:
\begin{equation}\label{Matsubaracosmhor}
\omega_{k,+}^{(M,\pm)}=\pm i\,k\,\kappa_+,\ \ 
k\in\mathbb{Z}_{>0}.
\end{equation}
If, instead, we considered the region close to the event horizon, we would need the corresponding Matsubara frequencies 
\begin{equation}
\omega_{k,h}^{(M,\pm)}=\pm i\,k\,\kappa_h,\ \ k\in\mathbb{Z}_{>0}.
\end{equation}

We consider linear spin $s$ perturbations, where $s=0$ describes conformally coupled scalar perturbation, $s=1$ describes electromagnetic perturbation, and $s=2$ describes vector-type gravitational perturbation.
By considering the decomposition in Fourier modes
\begin{equation}
\Phi(t,r,\theta,\varphi)=\int d\omega\sum_{\ell,m}e^{-i \omega t}\,Y_{\ell m}(\theta,\varphi)\,\frac{\phi(r)}{r},
\end{equation}
the radial problem is
\begin{equation}\label{reggewheeler}
\phi''(r) + \frac{f'(r)}{f(r)}\, \phi'(r) +\frac{\omega^2-V(r)}{f(r)^2}\,\phi(r)=0,
\end{equation}
where
\begin{equation}\label{potentialRW}
V(r)= f(r)\left[\frac{\ell(\ell+1)}{r^2}+(1-s^2)\left(\frac{3R_h-\Lambda R_h^3}{3 r^3}\right)\right].
\end{equation}

From the homogeneous solutions $\phi_{\text{up}}$ satisfying the outgoing boundary condition at the cosmological horizon and $\phi_{\text{in}}$ satisfying the ingoing boundary condition at the horizon, we construct $\widetilde{G}$ as in \eqref{Gformal}, where
\begin{equation}
\begin{aligned}
\mathcal{W} &=\phi_{\mathrm{up}} \frac{d}{d r_*}\phi_{\mathrm{in}} - \phi_{\mathrm{in}}\frac{d}{d r_*} \phi_{\mathrm{up}}\\
&=f(r) \left( \phi_{\mathrm{up}}(r) \frac{d}{d r}\phi_{\mathrm{in}}(r) - \phi_{\mathrm{in}}(r) \frac{d}{d r} \phi_{\mathrm{up}}(r) \right),
\end{aligned}
\end{equation}
and the explicit expression for $r_*$ can be found in \eqref{rstarSdS}.

The differential equation \eqref{reggewheeler} can be rewritten as a Heun equation \cite{heun1888theorie, ronveaux1995heun} in normal form 
\begin{equation}\label{heunnormalform}
\begin{aligned}
&\frac{d^2\,\psi(z)}{d\,z^2} + \biggl[\frac{\frac{1}{4}-a_0^2}{z^2}+\frac{\frac{1}{4}-a_1^2}{(z-1)^2} + \frac{\frac{1}{4}-a_x^2}{(z-x)^2}+\\
&\frac{u}{z(z-x)}- \frac{\frac{1}{2}-a_1^2 -a_x^2 -a_0^2 +a_\infty^2 + u}{z(z-1)} \biggr]\psi(z)=0.
\end{aligned}
\end{equation}
This is obtained by first introducing the new variable $z$
\begin{equation}
z=\frac{R_h}{r}\frac{r-R_-}{R_h-R_-},
\end{equation}
under which $r=0, R_h, R_-$ are mapped to $z=\infty, 1, 0$, and $r=R_+$ is mapped to
\begin{equation}
x=\frac{R_h (R_+-R_-)}{R_+ (R_h-R_-)},
\end{equation}
where for physical parameters $x\in ]0,1[$, and then redefining the wave function as
\begin{equation}
\phi(r(z))=p(z)\,\psi(z),
\end{equation}
where
\begin{equation}
p(z)=\frac{1}{\sqrt{z(1-z)} \sqrt{R_h (R_-+R_+ (z-1))-R_- R_+ z}}.
\end{equation}
The dictionary for the indicial parameters $a_0,a_1,a_x,a_\infty$ and for the accessory parameter $u$ is given in Appendix \ref{appendixgauge}.

The local solutions selected by the boundary conditions are
\begin{equation}\label{selectedsolSdS}
\begin{aligned}
\psi_{\text{in}}(z)&=z^{\gamma /2} (1-z)^{\delta /2} (z-x)^{\epsilon /2}(1-x)^{-\epsilon /2}\\
&\ \text{Heun}\left(1-x,\alpha  \beta -q,\alpha ,\beta ,\delta ,\gamma ,1-z\right),\\
\psi_{\text{up}}(z)&=z^{\gamma /2} (1-z)^{\delta /2}x^{-\gamma /2} (1-x)^{-\delta /2} (z-x)^{\epsilon /2}\\
&\ \text{Heun}\left(\frac{x}{x-1},\frac{q-\alpha  \beta  x}{1-x},\alpha ,\beta ,\epsilon ,\delta ,\frac{z-x}{1-x}\right).
\end{aligned}
\end{equation}
These solutions are written in terms of the canonical Heun functions and of the parameters of the differential equation in the form
\begin{equation}\label{heuncanonical}
\begin{aligned}
&g''(z)+\left(\frac{\gamma}{z}+\frac{\delta}{z-1}+\frac{\epsilon}{z-x}\right)g'(z) \\
&+\frac{\alpha\,\beta\,z-q}{z(z-1)(z-x)}\,g(z)=0,
\end{aligned}
\end{equation}\\[1pt]
and are normalised so that
\begin{equation}
\begin{aligned}
\psi_{\text{in}}(z)&\sim (1-z)^{\delta/2}\left[1+\mathcal{O}(1-z)\right],\quad\text{as}\quad z\to 1,\label{psiinnorm}\\
\psi_{\text{up}}(z)&\sim (z-x)^{\epsilon/2}\left[1+\mathcal{O}(z-x)\right],\quad\text{as}\quad z\to x.
\end{aligned}
\end{equation}
We again postpone the details to Appendix \ref{appendixgauge}.

In terms of the new variable $z$ and wave function $\psi$, we can write
\begin{eqnarray}
\widetilde{G}(\omega,z,z')&=& \,\frac{\frac{3 R_h R_-\,}{\Lambda\,(R_h-R_-)}p(z)p(z')}{\left( \psi_{\mathrm{up}}(z) \frac{d}{d z}\psi_{\mathrm{in}}(z) - \psi_{\mathrm{in}}(z) \frac{d}{d z} \psi_{\mathrm{up}}(z) \right)} \nonumber\\
&&\qquad\times
\begin{cases}
\psi_{\mathrm{in}}(z) \, \psi_{\mathrm{up}}(z'), & r < r' \\
\psi_{\mathrm{in}}(z') \, \psi_{\mathrm{up}}(z), & r > r',
\end{cases}\label{tildeGSdSinz}
\end{eqnarray}
where we used
\begin{equation}
f(r)\,p^2(z)\,\frac{dz(r)}{dr}=\frac{\Lambda\,(R_h-R_-)}{3\,R_h\,R_-}.
\end{equation}
In what follows, we will assume $r>r'$.

The expressions for the connection coefficients relating the ingoing solution at the horizon to the basis of solutions at the cosmological horizon are 
\begin{widetext}
\begin{equation}\label{conncoeffheun}
\begin{aligned}
\conid &=e^{-\frac{1}{2}\partial_{a_{1}}F(x)}\sum_{\sigma=\pm}\frac{\Gamma\left(1-2a_{1}\right)\Gamma\left(-2\sigma a\right)\Gamma\left(1-2\sigma a\right)\Gamma\left(-2a_{x}\right)}{\prod_{\pm}\Gamma\left(\frac{1}{2}- a_{1}-\sigma a\pm a_\infty\right)\Gamma\left(\frac{1}{2}-\sigma a-a_{x}\pm a_0\right)}x^{\sigma a}e^{-\frac{\sigma}{2}\partial_{a}F(x)}x^{-a_{x}}e^{-\frac{1}{2}\partial_{a_{x}}F(x)},\\
\coniu &=e^{-\frac{1}{2}\partial_{a_{1}}F(x)}\sum_{\sigma=\pm}\frac{\Gamma\left(1-2a_{1}\right)\Gamma\left(-2\sigma a\right)\Gamma\left(1-2\sigma a\right)\Gamma\left(2a_{x}\right)}{\prod_{\pm}\Gamma\left(\frac{1}{2}- a_{1}-\sigma a\pm a_\infty\right)\Gamma\left(\frac{1}{2}-\sigma a+a_{x}\pm a_0\right)}x^{\sigma a}e^{-\frac{\sigma}{2}\partial_{a}F(x)}x^{a_{x}}e^{\frac{1}{2}\partial_{a_{x}}F(x)}.
\end{aligned}
\end{equation}
\end{widetext}
These were obtained in \cite{Bonelli:2022ten} by employing a gauge theory language that we briefly describe in Appendix \ref{appendixgauge} (see also \cite{Lisovyy:2022flm}).
In terms of the connection coefficients \eqref{conncoeffheun}, we can rewrite the Wronskian in the denominator of \eqref{tildeGSdSinz} as
\begin{equation}\label{SdSWronskian}
\psi_{\mathrm{up}}(z) \frac{d}{d z}\psi_{\mathrm{in}}(z) - \psi_{\mathrm{in}}(z) \frac{d}{d z} \psi_{\mathrm{up}}(z)=2a_x\,C_{\text{in},\text{down}}.
\end{equation}

\subsection{Region I}

The QNMs are poles of the Green's function \eqref{tildeGSdSinz}, that is, they are given by the zeros of $C_{\text{in},\text{down}}$.
We consider a positive but parametrically small cosmological constant $\Lambda$. In this regime, the set of the QNMs can be seen as the union of the set of poles that reduce to the Schwarzschild QNMs in the $\Lambda\to 0^+$ limit, and a branch of purely imaginary de Sitter modes, coalescing into a branch cut in the $\Lambda\to 0^+$ limit \cite{Arnaudo:2025kit}.

Following the previous sections, region I is defined as the region where the QNM sum 
\begin{equation}\label{QNMsumSdS}
\sum_{\text{QNM}}\res\left[\frac{e^{-i \omega(t-t')}}{2\pi}\frac{3 R_h R_-\,p(z)p(z')}{\Lambda\,(R_h-R_-)}\,\frac{\psi_{\text{in}}(z')\psi_{\text{up}}(z)}{2a_x\,C_{\text{in},\text{down}}},\omega_{\text{QNM}}\right]
\end{equation}
converges. 

As described in Appendix \ref{appendixgauge}, the definition of the connection coefficient $C_{\text{in},\text{down}}$ involves an expansion in instantons, and the QNMs can be found by looking at the zeros of the truncation of this expansion up to some finite order.

Some analytic results for the QNMs within this approach can be found in \cite{Aminov:2023jve}, where a small $R_h$ regime was assumed.

\subsection{Region II}

When the sum \eqref{QNMsumSdS} diverges, we split $\widetilde{G}=\widetilde{G}_++\widetilde{G}_-$ by using the Heun connection formulas, and we analyse the singularity structure of both summands and their convergence properties in region II.

We have
\begin{equation}
\begin{aligned}
\widetilde{G}_+
&=\frac{3 R_h R_-\,}{\Lambda\,(R_h-R_-)}p(z)p(z')\,\frac{C_{\text{in},\text{up}}}{C_{\text{in},\text{down}}}\frac{\psi_{\mathrm{up}}(z)\,\psi_{\mathrm{up}}(z')}{2a_x},\\
\widetilde{G}_-
&=\frac{3 R_h R_-\,}{\Lambda\,(R_h-R_-)}p(z)p(z')\,\frac{\psi_{\mathrm{up}}(z)\,\psi_{\mathrm{down}}(z')}{2a_x},
\end{aligned}
\end{equation}
where $\psi_{\text{down}}$ is normalised as
\begin{equation}
\psi_{\text{down}}(z)\sim (z-x)^{1-\epsilon/2}\left[1+\mathcal{O}(z-x)\right],\quad\text{as}\quad z\to x,
\end{equation}
and its explicit expression can be found in \eqref{secondHeuncosmhor}.

The summand $\widetilde{G}_+$ has poles in QNM frequencies (that are zeros of $C_{\text{in},\text{down}}$),
and Matsubara frequencies of the cosmological horizon that appear in the UHP as poles of the ratio of $\Gamma$ functions $\frac{\Gamma(2a_x)}{\Gamma(-2a_x)}$ in $\frac{C_{\text{in},\text{up}}}{C_{\text{in},\text{down}}}$:
\footnote{We remark that at these values of the parameters the basis of local solutions needs to be changed, to include the logarithmic solutions. However, we will in practise always avoid these points and only compute the residues as limits from the generic cases.}
\begin{equation}\label{MMUHPSdS}
\begin{aligned}
&2a_x=-k,\ k\in\mathbb{Z}_{>0}\Leftrightarrow \omega=\omega_{k,+}^{(M,+)}, 
\end{aligned}
\end{equation}
and in the LHP as poles of the Heun functions in $\psi_{\mathrm{up}}(z),\psi_{\mathrm{up}}(z')$ given by the condition
\begin{equation}\label{MMLHPSdS}
\begin{aligned}
&\epsilon=-k+1,\  k\in\mathbb{Z}_{>0}\Leftrightarrow \omega=\omega_{k,+}^{(M,-)}. 
\end{aligned}
\end{equation}
We describe the origin of the latter singularities \eqref{MMLHPSdS} in Appendix \ref{appendixgauge}. Notice that in the full $\widetilde{G}$ the latter are present only in $\psi_{\mathrm{up}}(z)$ and get simplified by the poles of $\Gamma(-2a_x)$ in the numerator of $C_{\text{in},\text{down}}$. After the splitting, however, the poles remain with multiplicity 1. For this term, we close the integration contour in the UHP and sum over the residues of the poles \eqref{MMUHPSdS}.
We remark that in the confluence limit $\Lambda\to 0^+$, the surface gravity $\kappa_+\to 0^+$, and, as a consequence, the set of poles located at \eqref{MMUHPSdS} coalesce into a branch cut in the $\Lambda\to 0^+$ limit \cite{Arnaudo:2024sen, Arnaudo:2025kit}.

The summand $\widetilde{G}_-$ has poles in the Matsubara frequencies \eqref{MMUHPSdS} in the UHP that are poles of $\psi_{\mathrm{down}}(z')$, and in the Matsubara frequencies \eqref{MMLHPSdS} in the LHP that are poles of $\psi_{\mathrm{up}}(z)$. For this term, we close the integration contour in the LHP and sum over the residues of the poles \eqref{MMLHPSdS}. Taking the integration line slightly above the real axis, we also include the contribution from the $\omega = 0$ pole in this term. 

In general, the residue sums to be considered are
\begin{widetext}
\bea
&&\sum_{k\ge 1}\res\left(\widetilde{G}_+\frac{e^{-i\omega\left(t-t'\right)}}{2\pi},\,\omega=\omega_{k,+}^{(M, +)}\right)=\label{residuesumUHPSdS}\\
&&\qquad\sum_{k\ge 1}\frac{3 R_h R_-}{\Lambda\,(R_h-R_-)}p(z)p(z')\frac{\psi_{\mathrm{up}}(z)\,\psi_{\mathrm{up}}(z')}{2a_x}\,\frac{e^{-i\,\omega\left(t-t'\right)}}{2\pi}\,\frac{\coniu\,\Gamma(-2a_x)}{\conid\,\Gamma(2a_x)}\left(\frac{d a_x}{d \omega}\right)^{-1}\Bigg|_{\omega = \omega_{k,+}^{(M,+)}}\times \frac{(-1)^k}{2\Gamma(k)\Gamma(k+1)}\nonumber
\eea
\end{widetext}
in the UHP, and 
\begin{widetext}
\bea\label{residuesumLHPSdS}
&&\sum_{k\ge 1}\res\left(\widetilde{G}_-\frac{e^{-i\omega(t-t')}}{2\pi},\, \omega = \omega_{k,+}^{(M,-)}\right) = \sum_{k\ge 1} \frac{3 R_h R_-}{\Lambda\,(R_--R_h)}p(z)p(z')\frac{\psi_\text{down}(z')}{4a_x}\frac{e^{-i\omega(t-t')}}{2\pi}\times\\
&&z^{\gamma /2} (1-z)^{\delta /2}x^{-\gamma /2} (1-x)^{-\delta /2} (z-x)^{\epsilon /2} \left(\frac{da_x}{d\omega}\right)^{-1}\res\left[\text{Heun}\left(\frac{x}{x-1},\frac{q-\alpha  \beta  x}{1-x},\alpha ,\beta ,\epsilon ,\delta ,\frac{z-x}{1-x}\right),\epsilon=1-k\right]\Bigg|_{\omega = \omega_{k,+}^{(M,-)}},\nonumber
\eea
\end{widetext}
in the LHP,
where the residue of Heun is given in \eqref{heunresidueepsilon} in terms of three-term recurrence relations.

At $\omega=0$, the pole comes from the factor $1/a_x$, and the residue is given by
\begin{widetext}
\begin{equation}
\begin{aligned}\label{SdSzeroresidue}
\res\left[\widetilde{G}_- \frac{e^{-i\omega(t-t')}}{2\pi},\omega=0\right]=\frac{3 R_h R_-}{\Lambda\,(R_--R_h)}p(z)p(z')\frac{\psi_{\text{up}}(z)\psi_\text{down}(z')}{2}\frac{e^{-i\omega(t-t')}}{2\pi}\left(\frac{da_x}{d\omega}\right)^{-1}\Bigg|_{\omega = 0}.
\end{aligned}
\end{equation}
\end{widetext}

\begin{figure}[h]
\centering
\includegraphics[width=0.9\columnwidth]{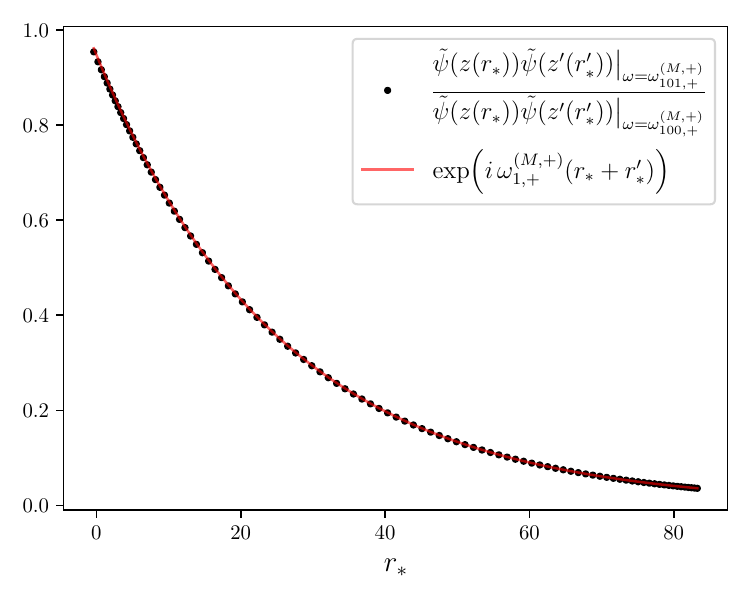}
\caption{
Demonstration that ratios of Heun functions can be replaced by plane waves at large Matsubara mode number, $k$.
The plot range is over values of $r_*$ that correspond to values of $z$ in the interval $]x,1[\,\sim\, ]0.0786,1[$\,. The choices for the parameters were $\ell=s=2$, $R_h=1$, $\Lambda=1/200$, $t'=0$, and $z' = 6/10$, showing that the agreement is good independently on the asymptotic approximation.}
\label{fig:plot_SdS_wave_approx}
\end{figure}

For simplicity, we restrict here to the analysis of the UHP contribution.
We want to find analytically the location in which region II ends and region I begins. To do this, we perform the ratio test for the residue sum \eqref{residuesumUHPSdS}. Because of the difficulty in taking ratios of Heun functions at large parameters, we start by replacing
\begin{equation}\label{replacement}
\tilde{\psi}(z)\equiv \rho\,p(z)\,\psi_{\text{up}}(z)
\end{equation}
with $e^{i\omega r_*(z)}$, where $\rho$ is independent on $z$ and its expression is given in \eqref{rhoconst},
and $\tilde{\psi}(z')$ with $e^{i\omega r'_*(z)}$. 
As we verified in FIG. \ref{fig:plot_SdS_wave_approx}, this provides a good approximation of the wave functions at large $k$, for all values of $r_*,r'_*$.

The ratio test can therefore be applied to the sum 
\begin{widetext}
\begin{equation}
\scalebox{0.88}{$
\begin{aligned}
&\sum_{k\ge 1}\frac{3 R_h R_-}{\Lambda\,(R_h-R_-)}\frac{e^{-i\omega\left(t-t'-r_*-r'_*\right)}}{4\pi}\frac{x^{2a_x}\,e^{\partial_{a_x}F(x)}}{\rho^2}\frac{\sum_{\sigma=\pm}\frac{\Gamma\left(-2\sigma a\right)\Gamma\left(1-2\sigma a\right)}{\prod_{\pm}\Gamma\left(\frac{1}{2}- a_{1}-\sigma a\pm a_\infty\right)\Gamma\left(\frac{1}{2}-\sigma a+a_{x}\pm a_0\right)}x^{\sigma a}e^{-\frac{\sigma}{2}\partial_{a}F(x)}}{\sum_{\sigma=\pm}\frac{\Gamma\left(-2\sigma a\right)\Gamma\left(1-2\sigma a\right)}{\prod_{\pm}\Gamma\left(\frac{1}{2}- a_{1}-\sigma a\pm a_\infty\right)\Gamma\left(\frac{1}{2}-\sigma a-a_{x}\pm a_0\right)}x^{\sigma a}e^{-\frac{\sigma}{2}\partial_{a}F(x)}}\frac{(-1)^{k+1}}{\Gamma(k+1)^2}\left(\frac{d a_x}{d \omega}\right)^{-1}\bigg|_{ \omega=\omega_{k,+}^{(M,+)}}.
\end{aligned}
$}
\end{equation}
\end{widetext}
In the small $\Lambda$ regime, the ratio of connection coefficients can be simplified by considering the leading order in the small $\Lambda$ expansion, that corresponds to the instanton expansion as explained in Appendix \ref{appendixgauge}. In particular, for both sums in the numerator and in the denominator, with the convention specified in Appendix \ref{appendixgauge}, the summand proportional to $x^a$ is subleading with respect to the one proportional to $x^{-a}$. Keeping only the leading terms, and substituting $a$ with $a^{(0)}$ in \eqref{leadingainst}, the residue sum simplifies to
\begin{widetext}
\begin{equation}
\begin{aligned}
\sum_{k\ge 1}\res_k^{(M,+)}\equiv\sum_{k\ge 1}\frac{3 R_h R_-}{\Lambda\,(R_h-R_-)}\frac{e^{-i\omega\left(t-t'-r_*-r'_*\right)}}{4\pi}\frac{x^{2a_x}}{\rho^2}\frac{\prod_{\pm}\Gamma\left(\frac{1}{2}+a^{(0)}-a_{x}\pm a_0\right)}{\prod_{\pm}\Gamma\left(\frac{1}{2}+a^{(0)}+a_{x}\pm a_0\right)}\frac{(-1)^{k+1}}{\Gamma(k+1)^2}\left(\frac{d a_x}{d \omega}\right)^{-1}\bigg|_{ \omega=\omega_{k,+}^{(M,+)}}.
\end{aligned}
\end{equation}
\end{widetext}
Using Stirling approximation for the ratio of $\Gamma$ functions and performing the ratio test, we find
\begin{equation}\label{ratiotest1MMSdS}
\begin{aligned}
&\lim_{k\to\infty}\bigg|\frac{\res_{k+1}^{(M,+)}}{\res_k^{(M,+)}}\bigg|=\\
&1 + \sqrt{\frac{\Lambda}{3}}\left(t-t'-r_*-r_*'+2r_*^\text{bounce}\right) +\mathcal{O}\left(\Lambda\right),
\end{aligned}
\end{equation}
where $r_*^\text{bounce}$ is a constant independent of $t,t',r_*,r_*'$ and corresponds to the radius at which the lightray bounces from the black hole potential. Therefore, we conclude that the residue sum is convergent in the region\footnote{The ratio test was performed by considering the residues of the MMs. In the $\Lambda\to 0^+$ limit, these poles coalesce into a branch cut, and the result in \eqref{ratiotest1MMSdS} loses meaning. However, we expect the result \eqref{ratiotest2MMSdS} to survive the $\Lambda\to 0^+$ limit, by comparison with the region of convergence of the QNM sum.}
\begin{equation}\label{ratiotest2MMSdS}
t<t'+r_*+r'_*-2r_*^\text{bounce}.
\end{equation}
The bounce location at 0-instanton order of approximation is given by
\begin{equation}
\begin{aligned}
&r_*^\text{bounce} \equiv  R_h\left[-1-\frac{\log (2)}{2}-\frac{1}{2} \sqrt{3} \log \left(2-\sqrt{3}\right)\right]
\end{aligned}
\end{equation}
At higher instanton orders, this receives additional corrections. For example, at 1-instanton order, 
\begin{equation}
\begin{aligned}
r_*^\text{bounce} \equiv  R_h\biggl[&-\frac{11}{49}+\frac{1}{2} \log \left(\frac{11}{2}\right)-\frac{3 \log (3)}{2}\\
&+\frac{7 \text{arctanh}\left(\frac{3 \sqrt{3}}{7}\right)}{3 \sqrt{3}}\biggr].
\end{aligned}
\end{equation}
For comparison, we provide $r_*$ at the peak of $V(r)$ in \eqref{peakpotentialSdS2}.

\subsection{Region III}

Due to the analyticity of $\widetilde{G}$ in the UHP, we can close the integration contour in the UHP and obtain $G= 0$ in region III.

\subsection{Partial sums}

To demonstrate the applicability of the mode sums in each region, we perform partial sums and compare to a numerical solution of \eqref{GreensPDE}. The results are shown in FIG. \ref{fig:SdSpartial}.

For region I, a partial sum of the QNM residues \eqref{QNMsumSdS} is shown. The residues can be computed numerically by root-finding for the zeros of $\conid$ and then finding the $\partial_\omega \conid$ there. This can be performed numerically using Heun functions through \eqref{SdSWronskian}, or by utilising the instanton approximation formulae via \eqref{conncoeffheun}. In FIG. \ref{fig:SdSpartial} we have used the former approach (giving the blue curve), avoiding additional approximations, however, the instanton approximation works well too.

For region II, we have partial sums for UHP Matsubara residues \eqref{residuesumUHPSdS}, partial sums for LHP Matsubara residues \eqref{residuesumLHPSdS}, as well as the $\omega = 0$ contribution \eqref{SdSzeroresidue}. In the UHP, we numerically compute residues by root finding in $\frac{\conid}{\coniu}$, using Heun functions, through both \eqref{SdSWronskian} and
\be
\psi_\text{down}(z)\frac{d}{dz}\psi_\text{in}(z) - \psi_\text{in}(z)\frac{d}{dz} \psi_\text{down}(z) = -2a_x \coniu.
\ee
In the LHP we employ the three-term recurrence approach to compute Heun residues at Matsubara frequencies, as discussed around \eqref{residuesumLHPSdS}. These partial sums are shown in green in FIG. \ref{fig:SdSpartial}. Again, one may also use the instanton approximation formulae to compute residue sums through \eqref{conncoeffheun}, with similar results.

The partial modes sums are compared to a direct numerical solution of the  homogeneous form of \eqref{GreensPDE} with initial data $G=0$ and $\partial_t G = \frac{1}{\sqrt{2\pi}\sigma}e^{-\frac{(r_*-r_*')^2}{2\sigma^2}}$ to approximate a delta function initial data, and which reproduces the source term with the correct normalisation. This is shown as the black points in FIG. \ref{fig:SdSpartial}. Clearly, the partial sums of modes gives an excellent construction of the Green's function within their respective regions.

At late times in region I, since $\Lambda > 0$ the falloff is finally dominated by the de Sitter QNMs which are longer-lived than the black hole QNMs, with a spacing along the negative imaginary axis proportional to $\sqrt{\Lambda}$. As $\Lambda \to 0^+$ and more and more residues are included in the sum, one recovers the Price-law tail $t^{-2\ell-3}$ \cite{Arnaudo:2025kit}. 

\section{General initial data}\label{sec:initialdata}
Our main focus has been the Green's function, from which general solutions for the fields themselves $f(t,r_*)$ can be obtained through convolution with initial data $f(t, r_*)|_{t=t_0}$ and $\partial_t f(t, r_*)|_{t=t_0}$(see e.g. \cite{Leaver}),
\bea
f(t,r_*) &=& \int_{-\infty}^\infty dr_*' \bigg[G(t,t',r_*,r_*')\partial_{t'}f(t',r_*') \nonumber\\
&&\qquad- f(t',r_*')\partial_{t'}G(t,t',r_*,r_*')\bigg]\bigg|_{t' = t_0}. \label{GT}
\eea
Through \eqref{GT} we can use our mode decomposition of $G$ to determine the mode decomposition of $f$.

\begin{figure}[h]
\centering
\includegraphics[width=\columnwidth]{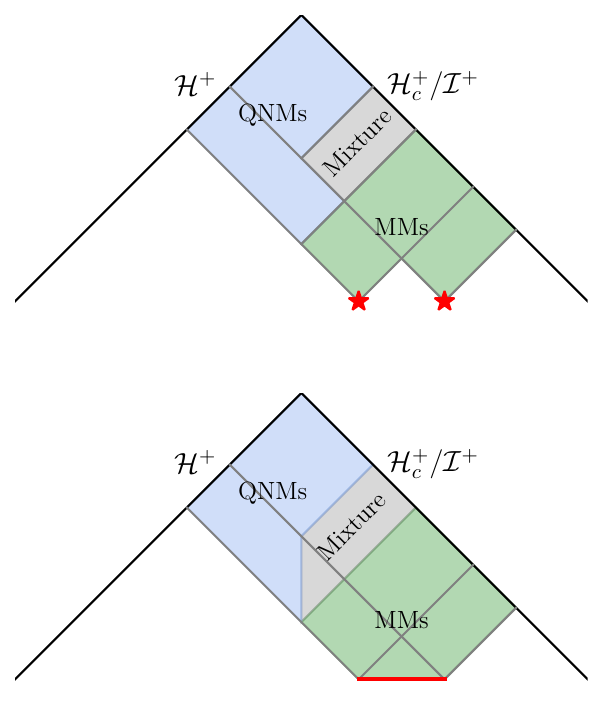}
\caption{Mode expansions of field configurations developed from initial data, following from convergent mode decompositions of $G$. In the \textbf{upper panel} we show the sum of two delta function sources (red stars), leading to a QNM expansion region (blue), a MM expansion region (green) and a region expressed as a sum of both QNMs and MMs (grey). In the \textbf{lower panel} we smear the sources (red line) to produce compact initial data.}
\label{fig:genericdata}
\end{figure}

Consider as a first example, two delta function sources or localised initial data, as in the upper panel of FIG. \ref{fig:genericdata}. Each delta function comes with its own region I and region II, and these then overlap with each other for multiple sources. Thus, there is an overlap region where the Green's function is expressed as a convergent sum over both QNMs and MMs.

As a second example, we smear the source into initial data supported on an interval, as in the lower panel of FIG. \ref{fig:genericdata}. Assuming that either the QNM sum or the MM sum converges on the lightcones themselves (which is indeed the case where we have checked, for the QNM sum in P\"oschl-Teller), this then extends the region where the mixed sum applies, and there is still a window where the QNM expansion applies, consistent with results such as \cite{Beyer:1998nu}. However, extending this initial data to an entire Cauchy surface would then appear to remove the QNM region entirely. This resembles what was described in the context of QNM excitation by infalling matter \cite{NAKAMURA1983403, Berti:2006hb}, where it was argued that the quasinormal ringing is absent from the waveforms in generic infalling matter configurations.

\section{Discussion} \label{sec:discussion}

In this work we showed that retarded black hole Green's functions in the time-domain, $G$, can be expressed as a convergent QNM sum at late times and a convergent sum of Matsubara modes (MMs) at earlier times. We demonstrated convergence of these sums analytically using ratio tests and the asymptotic form of residues of a two-term split of the frequency-space Green's function, $\widetilde{G}$. We also showed in some cases that one expansion can be obtained as the analytic continuation of the other. 

We considered several black hole examples. In some black holes, the frequency space expression $\widetilde{G}$ contains branch point singularities. This occurs in Schwarzschild at $\omega = 0$, for example. Our approach in this work was to resolve the branch point into a set of poles along the negative imaginary axis, by adding a small cosmological constant, $\Lambda >0$. This provides a mode sum description of the physics of Schwarzschild (in particular the power-law tail). At strictly $\Lambda = 0$, region I would contain a QNM branch cut discontinuity integral, and region II would contain MM branch cut discontinuity integrals too. Resolving these discontinuity integrals with a small $\Lambda >0$ gives the same outcome as $\Lambda \to 0^+$ \cite{Arnaudo:2025kit} and is easier to compute.

The results of our analysis have also implications when it comes to fitting ringdown signals with a superposition of QNMs. Indeed, as is evident from the discussion in section \ref{sec:initialdata}, the inclusion of MM contributions is necessary in order to achieve a consistent fit for generic initial data. This also suggests the need to extend recent QNM orthogonality relations \cite{Jafferis:2013qia, Green:2022htq, London:2023aeo, London:2023idh, Arnaudo:2025bnm} to include MMs too.

There are several aspects of our work that could be investigated further.

We focused on static spacetimes. An obvious extension of our study would be the analysis of the Green's function in the background of more general stationary spacetimes such as Kerr.

In the case of BTZ, region II did not intersect the boundary, except at the lightcone singularity. Therefore the holographic response is restricted to region I. Additionally, as the insertion is moved towards the boundary, region II disappears, and thus residues of the CFT retarded Green's function decay exponentially in time at QNM rates. It would be interesting to investigate whether this is always true for holographic theories dual to asymptotically AdS black holes. For example, one could consider designing a bottom-up model where perturbations experience a local maximum in the bulk potential, and whether this affects the regions and convergence properties.

There has been much recent interest in the spectral stability of black hole spacetimes \cite{Nollert:1996rf, Nollert:1998ys, Daghigh:2020jyk, Jaramillo:2020tuu}, which illustrates that small perturbations to black hole potentials can lead to large changes in the QNM frequencies (with respect to a chosen norm). In the context of our present work, an interesting question would be whether the residue sum of the perturbed poles are significantly affected. Relatedly, the existence of QNMs is intimately tied to the open nature of the black hole system with energy flux for perturbations escaping through the future horizon. A natural question is whether other open systems also feature contributions from Matsubara modes, in the same way that they exhibit QNMs, such as perturbations of the viscous Navier-Stokes equations in a pipe \cite{Carballo:2024kbk, Carballo:2025ajx, Besson:2025ghu}, or perturbations of neutron stars.

In \cite{Warnick:2022hnc} a specific class of counterexamples to the completeness of QNMs were formulated. It would be interesting to explicitly construct an example of the type given in \cite{Warnick:2022hnc}, where there is no support on QNMs, using the Matsubara mode contributions.

Given that Matsubara modes are the Fourier modes of the Euclidean thermal circle, a natural question is whether the representation of $G$ in region II as a convergent sum of MMs can be obtained from an analytic continuation of the Euclidean Green's function. In \cite{Jafferis:2013qia}, it was shown for a massive scalar field perturbation on dS$_4$, with mass $m^2\ell_{\text{dS}}^2<9/4$, that its corresponding Euclidean Green's function can be constructed as a sum of QNMs. It would be interesting to investigate whether this result also holds in the systems considered in this paper, as it would lead to a closer understanding of the relations between the convergent mode expansions in each region and the Euclidean Green's function upon analytic continuation.

We constructed retarded Green's function at fixed values for the angular quantum numbers. The analytic continuation of the Green's function in the $\omega$ complex plane provides an expansion of the time-domain Green's function in terms of QNMs \cite{Leaver} and, as explored in this work, of MMs. If we were to also consider the sum over $\ell$, then the analytic continuation in the complex plane of $\ell$, through a Sommerfeld-Watson transform, leads to an expansion of the position-space Green's function in terms of a sum of residues of Regge poles. Such an alternative representation can be useful when studying thermal correlators by analytically continuing the Euclidean Green's function, which is expressed as a Fourier sum of the retarded Green's function evaluated at the Matsubara frequencies, and rewriting it as a sum over the Regge poles in the complex $\ell$ plane. This was used, for example, in \cite{Dodelson:2023nnr} to study the singularity structure of the real-time Wightman function, and to obtain the light-cone singularity as well as the bulk-cone singularities. A natural question is whether the Regge poles decomposition of the Euclidean correlator has access to the decomposition in different regions we studied, and in particular, which relations it shares with the region II decomposition in terms of MMs.
It would also be interesting to analyse relations with caustics \cite{Zenginoglu:2012xe}.

\begin{acknowledgments}
It is a pleasure to thank Nils Andersson, Leor Barack, Simon Caron-Huot, Stephen R. Green, Robin Karlsson, Adrien Kuntz, Jacopo Lestingi, Hayato Motohashi, Christiana Pantelidou,
Frans Pretorius, Laura Sberna, Kostas Skenderis, Claude Warnick and Anil Zenginoglu for useful discussions. PA is supported by the Royal Society grant URF{\textbackslash}R{\textbackslash}231002, ‘Dynamics of holographic field theories’. JC is supported by the Royal Society grant RF{\textbackslash}ERE{\textbackslash}210267. BW is supported by a Royal Society University Research Fellowship and by the STFC consolidated grant ST/X000583/1, ‘New Frontiers In Particle Physics, Cosmology And Gravity’.
\end{acknowledgments}

\appendix
\renewcommand{\addcontentsline}[3]{}

\section{Rindler patch of AdS$_2$}\label{sec:AdS2}
In \cite{Chen:2023hra} the following correlator is considered
\be
G(t,t',r,r') \equiv \left<\Phi(t,r)\partial_{t'} \Phi(t',r')\right>, \label{AdS2corr}
\ee
where $\Phi$ is a massless scalar, in a Rindler patch of AdS$_2$. Unlike the other correlators studied in this paper, \eqref{AdS2corr} is not retarded. Nevertheless, much of the construction outlined in this paper applies (particularly the separation of region I and II). Indeed, as noted in Appendix B of \cite{Chen:2023hra}, $G$ can be expressed as a convergent sum over QNMs after the lightray has bounced from the boundary (i.e. just as in region I of FIG. \ref{fig:btzregions}). In this Appendix, we revisit this illustrative example, and also analyse the analogue of region II in this case. 

The Rindler patch of AdS$_2$ is given by
\be
ds^2 = -\sinh^2{\rho} dt^2 + d\rho^2,
\ee
with the Rindler horizon at $\rho = 0$ and conformal boundary at $\rho \to \infty$. The massless scalar equation is given by
\be
\left(-\partial_t^2 + \partial_{r_*}^2\right)\Phi(t,r) = 0,
\ee
where $r = \tanh(\rho/2)$ and $r_* = \log(r)$, admitting general plane wave solutions, $\Phi(t,r) = e^{-i\omega t}\phi(r)$,
\be
\phi(r) = c_L e^{-i \omega r_*} + c_R e^{i \omega r_*}.
\ee
In this case, to compute $G$, we want to impose Dirichlet boundary conditions at the conformal boundary in both Rindler wedges (in contrast to Dirichlet + ingoing conditions in the case of a retarded correlator). Following \cite{Arnaudo:2025bnm}, the left Rindler wedge is reached through analytic continuation from the right wedge $r_* \to r_* - i \pi$. Thus, the homogeneous solutions that are normalisable in the right and left wedge respectively, are, 
\bea
\phi_{Rn.}(r) &=& e^{-i \omega r_*} - e^{i \omega r_*},\\
\phi_{Ln.}(r) &=& e^{-i \omega (r_*-i \pi)} - e^{i \omega (r_*-i\pi)},
\eea
with Wronskian,
\be
\mathcal{W} = \phi_{Rn.}\partial_{r_*}\phi_{Ln.} - \phi_{Ln.}\partial_{r_*}\phi_{Rn.} = -4 i \omega \sinh(\pi \omega).
\ee
The zeros of $\mathcal{W}$ correspond to $\omega = i n$ with $n\in\mathbb{Z}$. The case $n<0$ corresponds to QNMs, since at these frequencies in ingoing time $v = t + r_*$, we have $\Phi_{Rn.} \equiv e^{-i \omega t}\phi_{Rn.} = e^{nv}\left(1-e^{-2n r_*}\right)$, so that at fixed $v$ with $r_*\to -\infty$ (i.e. on $\mathcal{H}^+$) the mode is regular. Similarly $n>0$ are anti-QNMs (regular on $\mathcal{H}^-$), while $n=0$ is trivial. Note that in this case orthogonality relations between QNMs \cite{Arnaudo:2025bnm}, with a suitable contour deformation by Cauchy, reduce to orthogonality of trigonometric functions on the interval $r_* \in i[0,\pi]$. The frequency space version of \eqref{AdS2corr} is then
\bea
\widetilde{G}(\omega, r, r') &=& \frac{i\omega}{\mathcal{W}}\phi_{Ln.}(r_<)\phi_{Rn.}(r_>),\label{AdS2Gtilde}
\eea
where $r_< = \min(r,r')$, $r_> = \max(r,r')$ and where the extra factor of $i\omega$ comes from the $\partial_{t'}$ in \eqref{AdS2corr} (as can be seen by differentiating \eqref{intro:FT} with respect to $t'$). 

The position space result $G(t,t',r,r')$ can be computed as a convergent mode sum following the techniques outlined in the introduction. A key difference here, however, is that because the Green's function is not retarded, there are four relevant light rays. Correspondingly, we employ a four-term split (instead of the two-term split for retarded functions \eqref{introsplit}),
\bea
\widetilde{G} &=& \widetilde{G}_1 + \widetilde{G}_2 + \widetilde{G}_3 + \widetilde{G}_4,\\
\widetilde{G}_1 &\equiv& -\frac{\text{csch}(\pi \omega)}{4}e^{(-\pi - i r_{*} - i r_{*}')\omega},\\
\widetilde{G}_2 &\equiv& \frac{\text{csch}(\pi \omega)}{4}e^{(\pi  - i |r_{*} -  r_{*}'|)\omega},\\
\widetilde{G}_3 &\equiv& \frac{\text{csch}(\pi \omega)}{4}e^{(-\pi + i |r_{*} -  r_{*}'|)\omega},\\
\widetilde{G}_4 &\equiv& -\frac{\text{csch}(\pi \omega)}{4}e^{(\pi + i r_{*} + i r_{*}')\omega},
\eea
with corresponding Fourier integrand residues, 
\bea
\res\left(\widetilde{G}_1 \frac{e^{-i \omega(t-t')}}{2\pi}, i n\right) &=& - \frac{e^{n(t-t'+r_{*} + r_{*}')}}{8\pi^2},\\
\res\left(\widetilde{G}_2\frac{e^{-i \omega(t-t')}}{2\pi}, i n\right) &=&  \frac{e^{n(t - t'+|r_*-r_*'|)}}{8\pi^2},\\
\res\left(\widetilde{G}_3\frac{e^{-i \omega(t-t')}}{2\pi}, i n\right) &=&  \frac{e^{n(t - t'-|r_* - r_*'|)}}{8\pi^2},\\
\res\left(\widetilde{G}_4\frac{e^{-i \omega(t-t')}}{2\pi}, i n\right) &=& - \frac{e^{n(t-t'-r_* - r_*')}}{8\pi^2}.
\eea
Clearly, then, each residue sum is a geometric series, with a finite radius of convergence. The convergence of each geometric sum is set by lightrays that border the region of interest.
In region I (see FIG. \ref{fig:btzregions}), each of the four integrals are closed in the LHP to obtain a convergent QNM mode sum,
\be
G(t,t',r,r') = -\frac{i}{\pi}\sum_{n=1}^\infty \sinh(n r_{*<})\sinh(n r_{*>}) e^{-n(t-t')}. \label{AdS2regionIsum}
\ee
From region I to region II we cross the lightray $r_{*} + r_{*}' + t-t' = 0$, and so we must close the integral for $\widetilde{G}_1$ in the UHP instead, leading to the following convergent sum,
\bea
G(t,t',r,r') = -\frac{i}{4\pi}-\frac{i}{4\pi}\sum_{n=1}^\infty\Bigg[e^{n(r_*+r_*')}e^{n(t-t')}\qquad\qquad\label{AdS2regionIIsum}\\
\qquad+\left(e^{n|r_*-r_*'|} + e^{-n|r_*-r_*'|} -e^{n(r_*+r_*')}\right)e^{-n(t-t')}\Bigg].\nonumber 
\eea
Note that the modes in this sum are not QNMs and do not obey Dirichlet boundary conditions at $r_* = 0$ -- indeed, this is an unnecessary requirement as region II does not intersect that boundary (see FIG. \ref{fig:btzregions}). Region II \emph{does} intersect $\mathcal{H}^+$, and these modes are all regular there, as required. To see this, we move to ingoing time, $v = t + r_*$, then at fixed $r_*', t', v$ each mode is either constant or decays as $e^{2nr_*}$ as $r_\ast \to -\infty$.

As every region mode sum is a geometric series, including \eqref{AdS2regionIsum} and \eqref{AdS2regionIIsum}, the analytic continuation from one mode sum into all regions is straightforward, and is given by
\bea
G(t,t',r,r') = \frac{i}{4\pi}\bigg(- \frac{1}{1-e^{t-t' + r_* + r_*'}}+ \frac{1}{1-e^{t-t'+|r_*-r_*'|}}\nonumber\\
+ \frac{1}{1-e^{t-t'-|r_*-r_*'|}}  -\frac{1}{1-e^{t-t'-r_*-r_*'}}\bigg)\qquad\qquad\label{AdS2ac}
\eea
as listed in (140) of \cite{Chen:2023hra}. From this one can see each lightcone singularity as the four poles of \eqref{AdS2ac}, and read off the associated convergent mode sum for any chosen region.

\section{P\"oschl-Teller further details} \label{sec:PTapp}
In this Appendix, we provide additional technical details for section \ref{sec:PT}.

The coefficients appearing in the homogeneous solutions \eqref{PThom1}-\eqref{PThom4} are as follows,
\bea
c_d &=& \Gamma(1 + i \omega), \label{PTmodescd}\\
c_u &=& \Gamma(1 - i \omega), \label{PTmodescu}\\
c_i^+ &=& -\omega \cos(\pi \nu) \, \text{csch}(\pi \omega) \, \Gamma(-i \omega), \label{PTmodescip}\\
c_i^- &=& \frac{\pi \omega \,\text{csch}(\pi \omega)\Gamma(-i \omega)}{\Gamma(1/2-\nu-i\omega)\Gamma(1/2+\nu-i\omega)}, \label{PTmodescim}\\
c_o^+ &=& \frac{\pi \omega \,\text{csch}(\pi \omega)\Gamma(i \omega)}{\Gamma(1/2-\nu+i\omega)\Gamma(1/2+\nu+i\omega)},\label{PTmodescop}\\
c_o^-&=& - \omega \cos(\pi \nu) \, \text{csch}(\pi \omega) \, \Gamma(i \omega).\label{PTmodescom}
\eea

The connection coefficients appearing in \eqref{PTcon1}, \eqref{PTcon2} are
\bea
\conid &=& \frac{i\Gamma(1-i\omega)^2}{\omega\Gamma(\frac{1}{2}-\nu - i\omega)\Gamma(\frac{1}{2}+\nu - i\omega)}, \label{PTconid}\\
\coniu &=& -i \cos(\pi \nu)\text{csch}(\pi \omega),\\
\conod &=& i \cos(\pi \nu)\text{csch}(\pi \omega),\\
\conou &=& -\frac{i\Gamma(1+i\omega)^2}{\omega\Gamma(\frac{1}{2}-\nu + i\omega)\Gamma(\frac{1}{2}+\nu + i\omega)}\label{PTconou}.
\eea

\subsection{Arc contributions at asymptotically large radii}\label{sec:PTasyarcs}
In \eqref{PTasy} we gave the expression for the large $r_*, r_*'$ asymptotics of the two-term split of the Green's function, with coefficients,
\bea
\alpha_+ &=&  \frac{\cos(\pi \nu)\Gamma\left(\frac{1}{2}-\nu-i\omega\right)\Gamma\left(\frac{1}{2}+\nu-i\omega\right)\Gamma(i\omega)}{2\pi \Gamma\left(1-i\omega\right)},\label{PTalphaplus}\\
\alpha_- &=& \frac{i}{2\omega}.\label{PTalphaminus}
\eea
We are interested in arc contributions to the Fourier transform integral, which take the form
\be
I_\pm^C = \int_{C} \frac{\alpha_\pm}{2\pi} e^{i\kappa\omega} d\omega,
\ee\\[0.5em]
where $\kappa \equiv -t+t'+\left| r_{*} \pm r_{*}'\right|$ and where $C$ is some portion of an arc of radius $R$ (to be taken to infinity), parameterised by angle $\theta$ with $\theta \in [\theta_1, \theta_2]$ through $\omega = R e^{i \theta}$. The calculation proceeds essentially according to Jordan's lemma, but adapted to the analytic structure of $\alpha_\pm$. We caution that the analytic structure of $\alpha_\pm$ does not reflect the analytic structure of $\widetilde{G}_\pm$ at \emph{finite} $r_*, r_*'$. This highlights the differences between analyticity of the  Green's function and that of the S-matrix \cite{CorreiaUpcoming}. An intermediate result is that
\be
\left|I_\pm^C\right| \leq R M_C \int_{\theta_1}^{\theta_2}e^{-\kappa R \sin\theta}d\theta,
\ee
where $M_C \equiv \max_{\theta \in [\theta_1, \theta_2]}\left|\frac{\alpha_\pm}{2\pi}\right|$.

In the case of $\alpha_-$, there are no poles for the arc to avoid, and so we can use Jordan's lemma. First, we take $\kappa > 0$ in the UHP ($\theta_1 = 0$, $\theta_2 = \pi$),
\begin{widetext}
\be
\left|I_-^C\right| \leq R M_C \int_{0}^{\pi}e^{-\kappa R \sin\theta}d\theta = 2R M_C\int_{0}^{\frac{\pi}{2}}e^{-\kappa R \sin\theta}d\theta \leq 2R M_C\int_{0}^{\frac{\pi}{2}}e^{-\kappa R \frac{2\theta}{\pi}}d\theta = \frac{\pi M_C}{\kappa}\left(1 - e^{-\kappa R}\right)
\ee
\end{widetext}
with $M_C = \frac{1}{4\pi R}$,
thus in this case $I_-^C\to 0$ as $R\to \infty$. A similar analysis shows that when $\kappa < 0$ closing in the LHP yields a vanishing arc contribution as $R\to \infty$.

\begin{figure}[h]
\centering
\includegraphics[width=0.6\columnwidth]{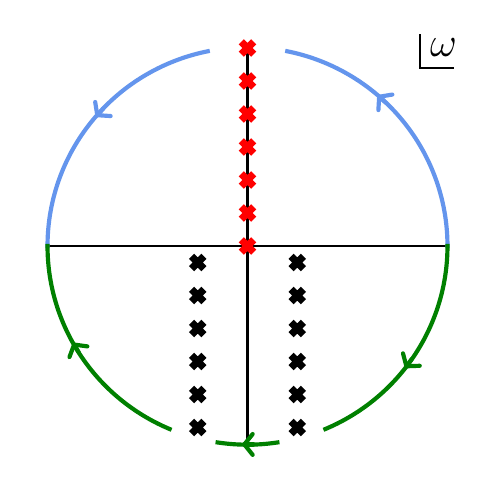}
\caption{Analytic structure of $\alpha_+(\omega)$ and definition of the arc contributions evaluated in section \ref{sec:PTasyarcs}.}
\label{fig:PTasyarcs}
\end{figure}

In the case of $\alpha_+$ there are lines of poles, see FIG.\ref{fig:PTasyarcs}. These poles originate in the gamma functions of \eqref{PTalphaplus}, and away from these lines of poles we can approximate $\alpha_+$ at large $R$ using Stirling's formula to obtain,
\be
\left|\frac{\alpha_+}{2\pi}\right| \sim \frac{e^{-\pi R\left|\cos\theta\right|} |\cos(\pi\nu)|}{2\pi R}. \label{PTalphapluslargeR}
\ee
Thus, with the understanding that the arc contributions should have infinitesimal pieces removed along these lines of poles (as in FIG. \eqref{fig:PTasyarcs} -- this is where vertical segments are attached to pick up mode sums), the same result holds. That is, when $\kappa > 0$ we close in the UHP to obtain, as above,
\be
|I_+^C| \leq  \frac{\pi M_C}{\kappa}\left(1- e^{-\kappa R}\right)
\ee
where the maximum of \eqref{PTalphapluslargeR} occurs at $\theta = \pi/2$, so that $M_C=\frac{|\cos(\pi\nu)|}{2\pi R}$. Hence, as $R\to \infty$, we recover $I_+^C = 0$. A similar derivation holds when $\kappa < 0$ closing in the LHP.

\section{BTZ further details}\label{appendixBTZ}
In this Appendix, we provide additional technical details for section \ref{sec:PT}.

The connection coefficients appearing in \eqref{BTZn.toinout}, \eqref{BTZn.n.toinout} are
\bea
\cnin &=& \frac{\Gamma(1-c)\Gamma(a+b-c+1)}{\Gamma(a-c+1)\Gamma(b-c+1)}, \label{BTZ_cnin}\\
\cnout &=& \frac{\Gamma(c-1)\Gamma(a+b-c+1)}{\Gamma(a)\Gamma(b)}, \label{BTZ_cnout}\\
\cnnin &=& \frac{\Gamma(1-c)\Gamma(c-a-b+1)}{\Gamma(1-a)\Gamma(1-b)}, \label{BTZ_cnnin}\\
\cnnout &=& \frac{\Gamma(c-1)\Gamma(c-a-b+1)}{\Gamma(c-a)\Gamma(c-b)}. \label{BTZ_cnnout}
\eea

The closed form expression at $r_\ast,r_\ast'\to-\infty$ (asymptotically close to the horizon) for the mode sum in region I reads
\begin{widetext}
\bea
&G(t,t',z,z')\sim\frac{\pi \, \mathrm{csch}(\pi m)}{m \, \Gamma(\Delta)} \,
 2^{-2 i m - 2 \Delta} \,
 e^{-(t - t' + r_\ast + r_\ast')(i m + \Delta)} \nonumber\\
&\times \Bigg[
 \frac{2^{4 i m} \, e^{2 i m (t - t' + r_\ast + r_\ast')}}{
   \Gamma(- i m)\,\Gamma(1 + i m - \Delta)} \; {}_4F_{3}\!\left(
   \begin{array}{c}
   -\tfrac{i m}{2} + \tfrac{\Delta}{2},\;
   -\tfrac{i m}{2} + \tfrac{\Delta}{2}, \\[1pt]
   \tfrac{1}{2} - \tfrac{i m}{2} + \tfrac{\Delta}{2},\;
   \tfrac{1}{2} - \tfrac{i m}{2} + \tfrac{\Delta}{2}
   \end{array};
   \begin{array}{c}
   1 - i m,\; \Delta,\; -i m + \Delta
   \end{array};
   e^{-2(t - t' + r_\ast + r_\ast')}
   \right) \nonumber\\
&\qquad + \frac{1}{\Gamma(i m)\,\Gamma(1 - i m - \Delta)} \; {}_4F_{3}\!\left(
   \begin{array}{c}
   \tfrac{i m}{2} + \tfrac{\Delta}{2},\;
   \tfrac{i m}{2} + \tfrac{\Delta}{2}, \\[1pt]
   \tfrac{1}{2} + \tfrac{i m}{2} + \tfrac{\Delta}{2},\;
   \tfrac{1}{2} + \tfrac{i m}{2} + \tfrac{\Delta}{2}
   \end{array};
   \begin{array}{c}
   1 + i m,\; \Delta,\; i m + \Delta
   \end{array};
   e^{-2(t - t' + r_\ast + r_\ast')}
   \right)
\Bigg]. \label{BTZ_asyG_regI}
\eea
\end{widetext}
Similarly, the closed form expression at $r_\ast,r_\ast'\to-\infty$ for the mode sum in region II reads
\begin{widetext}
\begin{equation}
\begin{aligned}
&&G(t,t',z,z')\sim\frac{1}{2}\Bigg[
 {}_4F_{3}\!\left(
   \begin{array}{c}
   1 - \tfrac{i m}{2} - \tfrac{\Delta}{2},\;
   1 + \tfrac{i m}{2} - \tfrac{\Delta}{2},\\[4pt]
   -\tfrac{i m}{2} + \tfrac{\Delta}{2},\;
   \tfrac{i m}{2} + \tfrac{\Delta}{2}
   \end{array};
   \begin{array}{c}
   \tfrac{1}{2},\; \tfrac{1}{2},\; 1
   \end{array};
   e^{2(t - t' + r_\ast + r_\ast')}
 \right) \nonumber\\
&&\qquad\qquad\qquad
 - e^{\,t - t' + r_\ast + r_\ast'}\bigl(m^{2}+(-1+\Delta)^{2}\bigr)\,
 {}_4F_{3}\!\left(
   \begin{array}{c}
   \tfrac{3}{2} - \tfrac{i m}{2} - \tfrac{\Delta}{2},\;
   \tfrac{3}{2} + \tfrac{i m}{2} - \tfrac{\Delta}{2},\\[4pt]
   \tfrac{1}{2} - \tfrac{i m}{2} + \tfrac{\Delta}{2},\;
   \tfrac{1}{2} + \tfrac{i m}{2} + \tfrac{\Delta}{2}
   \end{array};
   \begin{array}{c}
   1,\; \tfrac{3}{2},\; \tfrac{3}{2}
   \end{array};
   e^{2(t - t' + r_\ast + r_\ast')}
 \right)
\Bigg]. \label{BTZ_asyG_regII}
\end{aligned}
\end{equation}
\end{widetext}


\section{Schwarzschild-de Sitter details, Heun equation, and gauge theory}\label{appendixgauge}

This Appendix is devoted to make explicit the notations used in \ref{sec:SdS} and some more technical details. The Appendix is divided in four subsections.

\subsection{Longer formulas}

We define the tortoise coordinate for the Schwarzschild-de Sitter black hole as
\begin{widetext}
\begin{equation}\label{rstarSdS}
\begin{aligned}
r_*=\,&-\frac{3 \left[R_h (R_--R_+) \log \left(\frac{r}{R_h}-1\right)+R_- (R_+-R_h) \log \left(\frac{r}{R_h}-\frac{R_-}{R_h}\right)+R_+ (R_h-R_-) \log \left(\frac{R_+}{R_h}-\frac{r}{R_h}\right)\right]}{\Lambda  (R_h-R_-) (R_h-R_+) (R_--R_+)}\\
&+\frac{1}{2} R_h \left[\log\left(\frac{3}{\Lambda\,R_h^2}\right)-1\right].
\end{aligned}
\end{equation}
\end{widetext}
satisfying $\frac{dr_*}{dr} = \frac{1}{f(r)}$, and such that, in the $\Lambda\to 0^+$ limit, it reduces to the Schwarzschild tortoise coordinate
\begin{equation}
r_*^{(\text{S})}=r+R_h\,\log \left(\frac{r}{R_h}-1\right).
\end{equation}

The local solution satisfying ingoing boundary condition at the cosmological horizon is
\begin{widetext}
\begin{equation}\label{secondHeuncosmhor}
\begin{aligned}
&\psi_{\text{down}}(z)=z^{\gamma /2} (1-z)^{\delta /2}x^{-\gamma /2} (1-x)^{-\delta /2} (z-x)^{1-\epsilon /2}\\
&\times\text{Heun}\biggl(\frac{x}{x-1},\frac{-q+x\,(\beta-\gamma-\delta)\,(\alpha-\gamma-\delta)+\gamma\,(\alpha+\beta-\gamma-\delta)}{x-1},-\alpha+\gamma+\delta,-\beta+\gamma+\delta,2-\epsilon,\delta,\frac{z-x}{1-x}\biggr).
\end{aligned}
\end{equation}
\end{widetext}

The full expression for $\rho$ appearing in \eqref{replacement} is
\begin{widetext}
\begin{equation}\label{rhoconst}
\begin{aligned}
\rho=\,&\frac{\left(\frac{3}{e}\right)^{\frac{i R_h \omega }{2}} \Lambda ^{-\frac{1}{2} i R_h \omega } R_h^{-i \omega   R_h}}{\sqrt{\frac{1}{R_h R_--R_- R_+}+\frac{1}{R_h (R_+-R_-)}}} \left(\frac{R_h^2 R_-}{R_+^2 (R_--R_h)}\right)^{\frac{3 i R_+ \omega }{\Lambda  (R_h-R_+) (R_--R_+)}} \left(\frac{R_+-R_-}{R_h}\right)^{\frac{3 i R_- \omega }{\Lambda  (R_h-R_-) (R_--R_+)}} \left(\frac{R_+}{R_h}-1\right)^{-\frac{3 i R_h \omega }{\Lambda  (R_h-R_-) (R_h-R_+)}}.
\end{aligned}
\end{equation}
\end{widetext}

The peak of the potential in the $r_*$ variable reads
\begin{widetext}
\begin{equation}\label{peakpotentialSdS2}
\begin{aligned}
r_*^{(\text{peak})}=&R_h\biggl[\frac{3 \ell (\ell+1)+3 s^2-3+\sqrt{-14 \ell (\ell+1) s^2+\ell (\ell+1) (9 \ell (\ell+1)+14)+9 s^4-18 s^2+9}}{4 \ell (\ell+1)}+\\
&+\log \left(\sqrt{-14 \ell (\ell+1) s^2+\ell (\ell+1) (9 \ell (\ell+1)+14)+9 s^4-18 s^2+9}-\ell (\ell+1)+3 s^2-3\right)-\log (4 \ell (\ell+1))\biggr].
\end{aligned}
\end{equation}
\end{widetext}

\subsection{Heun equation}

In this second subsection, we write the full dictionaries for the parameters of the Heun equations. We begin with the dictionary of \eqref{heunnormalform}:
\bea
a_0&=\frac{3 i R_- \omega }{\Lambda  (R_--R_h) (R_--R_+)},
\eea
\bea
a_1&=\frac{3 i R_h \omega }{\Lambda  (R_h-R_-) (R_+-R_h)},
\eea
\bea
a_x&=\frac{3 i R_+ \omega }{\Lambda  (R_+-R_h) (R_+-R_-)},
\eea
\bea
a_{\infty}&=s,
\eea
\bea
u&=&\frac{3 \ell(\ell+1)+3(1-s^2)}{\Lambda  \left(R_h^2-R_+^2\right)}-\frac{2 R_h^2+2 R_h R_+-2 R_+^2 s^2+R_+^2}{2 R_h^2-2 R_+^2}\nonumber\\
&&\ +\frac{18 R_+^2 \omega ^2 \left(-2 R_h^2-2 R_h R_++R_+^2\right)}{\Lambda ^2 (R_h-R_+)^3 (R_h+R_+) (R_h+2 R_+)^2}.
\eea

The redefinition of the wave function to pass from the Heun equation as in \eqref{heunnormalform} to the one in \eqref{heuncanonical} is
\begin{equation}
g(z)=z^{-\gamma/2}(1-z)^{-\delta/2}(z-x)^{-\epsilon/2}\psi(z),
\end{equation} 
and the dictionary for the parameters is
\begin{equation}
\begin{aligned}
\alpha&=1-a_0-a_1-a_x+a_{\infty},\\
\beta&=1-a_0-a_1-a_x-a_{\infty},\\
\gamma&=1-2a_0,\\
\delta&=1-2a_1,\\
\epsilon&=1-2a_x,\\
q&=\frac{1}{2}+x\left(a_0^2+a_1^2+a_x^2-a_{\infty}^2\right)-a_x-a_1\,x+\\
&\quad a_0\left[2a_x-1+x\left(2a_1-1\right)\right]+\left(1-x\right)u.
\end{aligned}
\end{equation}

\subsection{Heun poles}

In this subsection, we describe the origin of poles in the Heun functions when the parameters assume nonpositive integer values. We begin the discussion by considering the standard Heun function
\begin{equation}\label{Heunin0}
\text{Heun}(x,q,\alpha,\beta,\gamma,\delta,z),
\end{equation}
which has simple poles when $\gamma=-n$, $n\in\mathbb{Z}_{\ge 0}$. The Heun function \eqref{Heunin0} admits a Taylor series representation around $z=0$,
\begin{equation}
\text{Heun}(x,q,\alpha,\beta,\gamma,\delta,z)=\sum_{r\ge 0}c_r\,z^r,
\end{equation}
where the coefficients $c_r$ satisfy a three-term recurrence relation defined as
\begin{equation}
\begin{aligned}
&x\,\gamma\,c_1-q\,c_0=0,\\
&C_j\,c_{j+1}-B_j\,c_j+A_j\,c_{j-1}=0,\quad j\ge 1,
\end{aligned}
\end{equation}
where
\begin{equation}
\begin{aligned}
A_j&=(j-1+\alpha)(j-1+\beta),\\
B_j&=q+j\,[(j-1+\gamma)(1+x)+x\,\delta+\epsilon],\\
C_j&=x\,(j+1)(j+\gamma).
\end{aligned}
\end{equation}
The normalization is fixed by $c_0=1$. When $\gamma=0$, the pole appears from the condition $x\,\gamma\,c_1-q\,c_0=0$ (for generic values of $q$). When $\gamma=-n$, $n\ge 1$, the pole appears in the definition of $c_{n+1}$ since $C_n=0$ (again, for generic values of the other parameters).

For the analysis of the Matsubara frequencies as poles of $\widetilde{G}_-(\omega,z,z')$ in the Schwarzschild-de Sitter case, we are interested in the poles of the Heun function appearing in $\psi_{\text{up}}(z)$ \eqref{selectedsolSdS}:
\begin{equation}
\begin{aligned}
\text{Heun}\left(\frac{x}{x-1},\frac{q-\alpha  \beta  x}{1-x},\alpha ,\beta ,\epsilon ,\delta ,\frac{z-x}{1-x}\right).
\end{aligned}
\end{equation}
This Heun function admits the Taylor series representation 
\begin{equation}
\sum_{r\ge 0}d_r\left(\frac{z-x}{1-x}\right)^r,
\end{equation}
with
\begin{equation}
d_0=1,\quad x\,\epsilon\,d_1+\left(q-\alpha  \beta  x\right)=0,
\end{equation}
and 
\begin{equation}
\tilde{C}_j\,d_{j+1}-\tilde{B}_j\,d_j+\tilde{A}_j\,d_{j-1}=0,\quad j\ge 1,
\end{equation}
where
\begin{equation}
\begin{aligned}
\tilde{A}_j&=(j-1+\alpha)(j-1+\beta),\\
\tilde{B}_j&=\frac{q-\alpha  \beta  x}{1-x}+j\left[(j-1+\epsilon)\left(1+\frac{x}{x-1}\right)+\frac{x}{x-1}\,\delta+\gamma\right],\\
\tilde{C}_j&=\frac{x}{x-1}(j+1)(j+\epsilon).
\end{aligned}
\end{equation}
Therefore, this Heun function has simple poles when $\epsilon=-n$, $n\in\mathbb{Z}_{\ge 0}$.

To evaluate the residues at these poles, we first describe the procedure for the simpler case of hypergemetric functions and then apply the same reasoning to the Heun context.

\subsubsection{Hypergeometric residues} \label{app:hypergeometric_residues}

The hypergeometric function ${}_2F_1(a,b;c;z)$ has poles when $c=-n$, $n\in\mathbb{Z}_{\ge 0}$. To compute the residue of the pole, we start by considering the Taylor series representation
\begin{equation}
{}_2F_1(a,b;c;z)=\sum_{r\ge 0}\frac{(a)_r\,(b)_r}{(c)_r\,r!}\,z^r,
\end{equation}
in which the presence of poles at nonpositive integer values of $c$ is made explicit by the Pochhammer symbol.

From this representation, it is apparent that the pole structure can be absorbed in the $\Gamma$ function's one, by introducing the regularized version of the hypergeometric function:
\begin{equation}
\begin{aligned}
&\res\left[\sum_{r\ge 0}\frac{(a)_r\,(b)_r}{(c)_r\,r!}\,z^r,c=-n\right]=\\
&\res\left[\frac{\Gamma(c)}{\Gamma(c)}\sum_{r\ge 0}\frac{(a)_r\,(b)_r}{(c)_r\,r!}\,z^r,c=-n\right]=\\
&\res\left[\Gamma(c),c=-n\right]\lim_{c\to -n}\frac{1}{\Gamma(c)}\sum_{r\ge 0}\frac{(a)_r\,(b)_r}{(c)_r\,r!}\,z^r=\\
&\frac{(-1)^n}{n!}\frac {(a)_{n+1}(b)_{n+1}}{(n+1)!}z^{n+1}{}_{2}F_{1}(a+n+1,b+n+1;n+2;z),
\end{aligned}
\end{equation}
where we used Equation 15.1.2 in \cite{abramowitz+stegun}.

The final result can be explained as follows. When $c\in\mathbb{Z}$ the two standard solutions around $z=0$ become linearly dependent. Therefore, the regularized version ${}_2F_1(a,b;c;z)/\Gamma(c)$ of the hypergeometric function ${}_2F_1(a,b;c;z)$ must be equal to the evaluation at $c=-n$ of the other solution $z^{1-c}\,{}_2F_1(1+a-c,1+b-c;2-c;z)$, up to a constant factor. This constant can be obtained by considering the Taylor series representation of ${}_2F_1(a,b;c;z)$, and by taking the limit $\lim_{c\to -n}(c+n){}_2F_1(a,b;c;z)$. Indeed, the first non-zero term is the one with $r=n+1$, whose coefficient is $\frac{(a)_{n+1}(b)_{n+1}}{(n+1)!}$.

\subsubsection{Heun residues}

As in the hypergeometric case, we claim that the residues of the Heun function $\text{Heun}(x,q,\alpha,\beta,\gamma,\delta,z)$ at $\gamma=-n$, with $n\in\mathbb{Z}_{\ge 0}$ are of the form
\begin{equation}
\begin{aligned}
&\res\left[\text{Heun}(x,q,\alpha,\beta,\gamma,\delta,z),\gamma=-n\right]=\\
&\text{const}\,z^{1-\gamma}\,\text{Heun}(x,q+(x \delta+\epsilon)(1-\gamma),\\
&\alpha+1-\gamma,\beta+1-\gamma,2-\gamma,\delta,z)\left.\right|_{\gamma=-n},
\end{aligned} \label{resHeun}
\end{equation}
meaning they are proportional to the evaluation of the independent local solution at $\gamma=-n$, up to a constant which is the residue of the first coefficient $c_{n+1}$ having a pole at $\gamma=-n$:
\begin{equation}
\text{const}=\res\left[c_{n+1},\gamma=-n\right]=\frac{B_n\,c_n-A_n\,c_{n-1}}{x\,(n+1)}.
\end{equation}
We show this by computing
\begin{equation}
\lim_{\gamma\to -n}(\gamma+n)\,\text{Heun}(x,q,\alpha,\beta,\gamma,\delta,z).
\end{equation}
By using the Taylor series representation and the three-term recurrence relation for the Heun function, we have that the finite sum $\sum_{r=0}^nc_r\,z^r$ does not contribute to the residue:
\begin{equation}
\lim_{\gamma\to -n}(\gamma+n)\,\sum_{r=0}^nc_r\,z^r=0.
\end{equation}
The first term being non-zero is given by
\begin{equation}
\lim_{\gamma\to -n}(\gamma+n)\,c_{n+1}=\frac{B_n\,c_n-A_n\,c_{n-1}}{C_n}=\frac{B_n\,c_n-A_n\,c_{n-1}}{x(n+1)}.
\end{equation}
Then, the pole is propagated by the recurrence relation and the overall residue is given by
\begin{equation}
\lim_{\gamma\to -n}(\gamma+n)\,\sum_{r=0}^\infty\hat{c}_r\,z^{r+n+1}=\lim_{\gamma\to -n}(\gamma+n)\,z^{n+1}\,\sum_{r=0}^\infty\hat{c}_r\,z^{r},
\end{equation}
where the new coefficients $\hat{c}_r$ satisfy the following recursion relation:
\begin{equation}\label{newthreeterm1}
\begin{aligned}
&\hat{c}_0=\frac{B_n\,c_n-A_n\,c_{n-1}}{x(n+1)},\quad \hat{c}_1=\frac{B_{n+1}\,\hat{c}_0}{x(n+2)},\\
&\hat{c}_{j+1}=\frac{\hat{B}_j\,\hat{c}_j-\hat{A}_j\,\hat{c}_{j-1}}{\hat{C}_j},\quad j\ge 1,
\end{aligned}
\end{equation}
with
\begin{equation}\label{newthreeterm2}
\begin{aligned}
\hat{A}_j&=A_{j+n+1},\\
\hat{B}_j&=B_{j+n+1},\\
\hat{C}_j&=x(j+n+2)(j+1).
\end{aligned}
\end{equation}
The functions in \eqref{newthreeterm2} are precisely the ones defining the three-term recurrence relation satisfied by the coefficients of the Taylor series representation of the Heun function
\begin{equation}\label{secondHeunin0}
\text{Heun}(x,q+(x \delta+\epsilon)(1-\gamma),\alpha+1-\gamma,\beta+1-\gamma,2-\gamma,\delta,z)\left.\right|_{\gamma=-n}.
\end{equation}
The claim is proved taking into account the overall factor of $\hat{c}_0$, due to the normalisation of the Heun function \eqref{secondHeunin0}.

For our analysis in Section \ref{sec:SdS}, the following result will be used:
\begin{equation}\label{heunresidueepsilon}
\begin{aligned}
&\res\left[\text{Heun}\left(\frac{x}{x-1},\frac{q-\alpha  \beta  x}{1-x},\alpha ,\beta ,\epsilon ,\delta ,\frac{z-x}{1-x}\right),\epsilon=-n\right]=\\
&\frac{\tilde{B}_n\,d_n-\tilde{A}_n\,d_{n-1}}{\frac{x}{x-1}(n+1)}\,\left(\frac{z-x}{1-x}\right)^{n+1}\times\\
&\text{Heun}\biggl(\frac{x}{x-1},\frac{q-(\beta-\gamma-\delta)(\alpha-\gamma-\delta)x+\gamma(1+n)}{1-x},\\
&\ \ -\alpha+\gamma+\delta,-\beta+\gamma+\delta,n+2,\delta,\frac{z-x}{1-x}\biggr).
\end{aligned}
\end{equation}

\subsection{Gauge theory details}

In this subsection, we introduce the conventions used for the gauge theory quantities and the Nekrasov-Shatashvili (NS) functions \cite{Nekrasov:2009rc}.

The connection between black hole perturbation theory and Seiberg-Witten curves was established in \cite{Aminov:2020yma} and was then developed in many subsequent works \cite{Bianchi:2021xpr, Bonelli:2021uvf, Bianchi:2021mft, daCunha:2022ewy, Fioravanti:2021dce, Dodelson:2022yvn, Jia:2024zes, Arnaudo:2025kof}.

The expressions for the connection coefficients of the Heun differential equation were obtained in \cite{Bonelli:2022ten} from the connection formulae for the semiclassical Virasoro conformal blocks with a degenerate insertion in Liouville conformal field theory. More precisely, correlation functions with a degenerate insertion satisfy a partial differential equation, known as Belavin–Polyakov–Zamolodchikov equation \cite{Belavin:1984vu}, that in the semiclassical limit becomes an ordinary differential equation with regular singularities at the positions of the insertions. Thanks to the crossing symmetry property of the correlation functions, it is possible to extract the connection coefficients that relate different conformal block expansions for the same correlator, obtained by performing operator product expansions between the degenerate insertion and one of the primary ones.
Finally, these quantities admit explicit and combinatorial expressions provided by the Alday-Gaiotto-Tachikawa correspondence \cite{Alday:2009aq}, which makes a dictionary between quantities in 2-dimensional Lioville CFT and 4-dimensional $\mathcal{N}=2$ supersymmetric gauge theory. 
The relevant theory for the Heun equation is the $SU(2)$ theory with $N_f=4$ fundamental hypermultiplets.

We introduce the main contributions crucial for defining the instanton partition function appearing in the connection formulas. We denote with $\vec{Y}=(Y_1,Y_2)$ a pair of Young diagrams, with $\vec{a}=(a_1,a_2)$ the vacuum expectation value of the scalar in the vector multiplet, and with $\epsilon_1,\epsilon_2$ the parameters characterizing the $\Omega$-background. We define the hypermultiplet and vector contributions as
\begin{equation}\small
\begin{aligned}
z_{\text{hyp}} \left( \vec{a}, \vec{Y}, m \right)&=\prod_{k= 1,2} \prod_{(i,j) \in Y_k} \left[ a_k + m + \epsilon_1 \left( i - \frac{1}{2} \right) + \epsilon_2 \left( j - \frac{1}{2} \right) \right],\\
z_{\text{vec}} \left( \vec{a}, \vec{Y} \right)&=\prod_{i,j=1}^2\prod_{s\in Y_i}\frac{1}{a_i-a_j-\epsilon_1L_{Y_j}(s)+\epsilon_2(A_{Y_i}(s)+1)}\\
&\quad\times\prod_{t\in Y_j}\frac{1}{-a_j+a_i+\epsilon_1(L_{Y_i}(t)+1)-\epsilon_2A_{Y_j}(s)}.
\end{aligned}
\end{equation}
We adopt the conventions $\epsilon_1=1$ and $\vec{a}=(a,-a)$.
By denoting with $m_1,m_2,m_3,m_4$ the masses of the four hypermultiplets, the monodromy parameters $a_0,a_x,a_1,a_{\infty}$ are defined as
\begin{equation}\label{gaugemasses}
\begin{aligned}
m_1&=-a_x-a_0,\quad
&m_2=-a_x+a_0,\\
m_3&=a_{\infty}+a_1,\quad
&m_4=-a_{\infty}+a_1.
\end{aligned}
\end{equation}
The position of the fourth singularity of the Heun equation, denoted with $x$, is the instanton counting parameter $x=e^{2\pi i\tau}$, where $\tau$ is related to the gauge coupling by 
\begin{equation}
\tau=\frac{\theta}{2\pi}+i\frac{4\pi}{g_{\rm YM}^2}.
\end{equation}
The instanton part of the NS free energy is then given as a power series in $x$ by
\begin{widetext}
\begin{equation}
F(x)=\lim_{\epsilon_2\to 0}\epsilon_2\log\Biggl[(1-x)^{-2\epsilon_2^{-1}\left(\frac{1}{2}+a_1\right)\left(\frac{1}{2}+a_x\right)}\sum_{\vec{Y}}x^{|\vec{Y}|}z_{\text{vec}} \left( \vec{a}, \vec{Y} \right)\prod_{i=1}^4z_{\text{hyp}} \left( \vec{a}, \vec{Y}, m_i \right)\Biggr].
\end{equation}
\end{widetext}

The gauge parameter $a$ parametrizes the composite monodromy around the points $z=0$ and $z=x$, and is expressed as a series expansion in the instanton counting parameter $x$, obtained by inverting the Matone relation \cite{Matone:1995rx}
\begin{equation}
u =-\frac{1}{4} - a^2 + a_x^2 + a_0^2 + x \partial_x F(x),
\end{equation}
where the parameter $u$ appearing in the differential equation \eqref{heunnormalform} is the complex moduli parametrizing the corresponding Seiberg-Witten curve.
The instanton expansion of $a$ reads 
\begin{widetext}
\begin{equation} 
a=\pm\left\{\sqrt{-\frac{1}{4}-u+a_x^2+a_0^2}+\frac{\bigl(\frac{1}{2}+u-a_x^2-a_0^2-a_1^2+a_{\infty}^2\Bigr)\Bigl(\frac{1}{2}+u-2a_x^2\Bigr)}{2(1+2u-2a_x^2-2a_0^2)\sqrt{-\frac{1}{4}-u+a_x^2+a_0^2}}x+\mathcal{O}(x^2)\right\}.
\end{equation} 
\end{widetext}
As can be seen, $a$ is in principle defined up to a sign. We always choose the branch for which $a$ has a positive real part. We will denote with $a^{(0)}$ the leading order of $a$ in the instanton expansion:
\begin{equation}\label{leadingainst}
a^{(0)}=\sqrt{-\frac{1}{4}-u+a_x^2+a_0^2}.
\end{equation}


\begin{thebibliography}{95}%
\makeatletter
\providecommand \@ifxundefined [1]{%
 \@ifx{#1\undefined}
}%
\providecommand \@ifnum [1]{%
 \ifnum #1\expandafter \@firstoftwo
 \else \expandafter \@secondoftwo
 \fi
}%
\providecommand \@ifx [1]{%
 \ifx #1\expandafter \@firstoftwo
 \else \expandafter \@secondoftwo
 \fi
}%
\providecommand \natexlab [1]{#1}%
\providecommand \enquote  [1]{``#1''}%
\providecommand \bibnamefont  [1]{#1}%
\providecommand \bibfnamefont [1]{#1}%
\providecommand \citenamefont [1]{#1}%
\providecommand \href@noop [0]{\@secondoftwo}%
\providecommand \href [0]{\begingroup \@sanitize@url \@href}%
\providecommand \@href[1]{\@@startlink{#1}\@@href}%
\providecommand \@@href[1]{\endgroup#1\@@endlink}%
\providecommand \@sanitize@url [0]{\catcode `\\12\catcode `\$12\catcode `\&12\catcode `\#12\catcode `\^12\catcode `\_12\catcode `\%12\relax}%
\providecommand \@@startlink[1]{}%
\providecommand \@@endlink[0]{}%
\providecommand \url  [0]{\begingroup\@sanitize@url \@url }%
\providecommand \@url [1]{\endgroup\@href {#1}{\urlprefix }}%
\providecommand \urlprefix  [0]{URL }%
\providecommand \Eprint [0]{\href }%
\providecommand \doibase [0]{https://doi.org/}%
\providecommand \selectlanguage [0]{\@gobble}%
\providecommand \bibinfo  [0]{\@secondoftwo}%
\providecommand \bibfield  [0]{\@secondoftwo}%
\providecommand \translation [1]{[#1]}%
\providecommand \BibitemOpen [0]{}%
\providecommand \bibitemStop [0]{}%
\providecommand \bibitemNoStop [0]{.\EOS\space}%
\providecommand \EOS [0]{\spacefactor3000\relax}%
\providecommand \BibitemShut  [1]{\csname bibitem#1\endcsname}%
\let\auto@bib@innerbib\@empty
\bibitem [{\citenamefont {Abbott}\ \emph {et~al.}(2016{\natexlab{a}})\citenamefont {Abbott} \emph {et~al.}}]{LIGOScientific:2016aoc}%
  \BibitemOpen
  \bibfield  {author} {\bibinfo {author} {\bibfnamefont {B.~P.}\ \bibnamefont {Abbott}} \emph {et~al.} (\bibinfo {collaboration} {LIGO Scientific, Virgo}),\ }\bibfield  {title} {\bibinfo {title} {{Observation of Gravitational Waves from a Binary Black Hole Merger}},\ }\href {https://doi.org/10.1103/PhysRevLett.116.061102} {\bibfield  {journal} {\bibinfo  {journal} {Phys. Rev. Lett.}\ }\textbf {\bibinfo {volume} {116}},\ \bibinfo {pages} {061102} (\bibinfo {year} {2016}{\natexlab{a}})},\ \Eprint {https://arxiv.org/abs/1602.03837} {arXiv:1602.03837 [gr-qc]} \BibitemShut {NoStop}%
\bibitem [{\citenamefont {Abbott}\ \emph {et~al.}(2016{\natexlab{b}})\citenamefont {Abbott} \emph {et~al.}}]{GWcollab}%
  \BibitemOpen
  \bibfield  {author} {\bibinfo {author} {\bibfnamefont {B.~P.}\ \bibnamefont {Abbott}} \emph {et~al.} (\bibinfo {collaboration} {KAGRA, LIGO Scientific, Virgo}),\ }\bibfield  {title} {\bibinfo {title} {{Prospects for observing and localizing gravitational-wave transients with Advanced LIGO, Advanced Virgo and KAGRA}},\ }\href {https://doi.org/10.1007/s41114-020-00026-9} {\bibfield  {journal} {\bibinfo  {journal} {Living Rev. Rel.}\ }\textbf {\bibinfo {volume} {19}},\ \bibinfo {pages} {1} (\bibinfo {year} {2016}{\natexlab{b}})},\ \Eprint {https://arxiv.org/abs/1304.0670} {arXiv:1304.0670 [gr-qc]} \BibitemShut {NoStop}%
\bibitem [{\citenamefont {Abac}\ \emph {et~al.}(2025)\citenamefont {Abac} \emph {et~al.}}]{LIGOScientific:2025slb}%
  \BibitemOpen
  \bibfield  {author} {\bibinfo {author} {\bibfnamefont {A.~G.}\ \bibnamefont {Abac}} \emph {et~al.} (\bibinfo {collaboration} {LIGO Scientific, VIRGO, KAGRA}),\ }\bibfield  {title} {\bibinfo {title} {{GWTC-4.0: Updating the Gravitational-Wave Transient Catalog with Observations from the First Part of the Fourth LIGO-Virgo-KAGRA Observing Run}},\ }\href@noop {} {\  (\bibinfo {year} {2025})},\ \Eprint {https://arxiv.org/abs/2508.18082} {arXiv:2508.18082 [gr-qc]} \BibitemShut {NoStop}%
\bibitem [{\citenamefont {Regge}\ and\ \citenamefont {Wheeler}(1957)}]{Regge_Wheeler}%
  \BibitemOpen
  \bibfield  {author} {\bibinfo {author} {\bibfnamefont {T.}~\bibnamefont {Regge}}\ and\ \bibinfo {author} {\bibfnamefont {J.~A.}\ \bibnamefont {Wheeler}},\ }\bibfield  {title} {\bibinfo {title} {{Stability of a Schwarzschild Singularity}},\ }\href {https://doi.org/10.1103/PhysRev.108.1063} {\bibfield  {journal} {\bibinfo  {journal} {Phys. Rev.}\ }\textbf {\bibinfo {volume} {108}},\ \bibinfo {pages} {1063} (\bibinfo {year} {1957})}\BibitemShut {NoStop}%
\bibitem [{\citenamefont {Zerilli}(1970)}]{Zerilli}%
  \BibitemOpen
  \bibfield  {author} {\bibinfo {author} {\bibfnamefont {F.~J.}\ \bibnamefont {Zerilli}},\ }\bibfield  {title} {\bibinfo {title} {{Effective Potential for Even-Parity Regge-Wheeler Gravitational Perturbation Equations}},\ }\href {https://doi.org/10.1103/PhysRevLett.24.737} {\bibfield  {journal} {\bibinfo  {journal} {Phys. Rev. Lett.}\ }\textbf {\bibinfo {volume} {24}},\ \bibinfo {pages} {737} (\bibinfo {year} {1970})}\BibitemShut {NoStop}%
\bibitem [{\citenamefont {Berti}\ \emph {et~al.}(2025)\citenamefont {Berti} \emph {et~al.}}]{Berti:2025hly}%
  \BibitemOpen
  \bibfield  {author} {\bibinfo {author} {\bibfnamefont {E.}~\bibnamefont {Berti}} \emph {et~al.},\ }\bibfield  {title} {\bibinfo {title} {{Black hole spectroscopy: from theory to experiment}},\ }\href@noop {} {\  (\bibinfo {year} {2025})},\ \Eprint {https://arxiv.org/abs/2505.23895} {arXiv:2505.23895 [gr-qc]} \BibitemShut {NoStop}%
\bibitem [{LIG(2025)}]{LIGOScientific:2025obp}%
  \BibitemOpen
  \bibfield  {title} {\bibinfo {title} {{Black Hole Spectroscopy and Tests of General Relativity with GW250114}},\ }\href@noop {} {\  (\bibinfo {year} {2025})},\ \Eprint {https://arxiv.org/abs/2509.08099} {arXiv:2509.08099 [gr-qc]} \BibitemShut {NoStop}%
\bibitem [{\citenamefont {Arnaudo}\ and\ \citenamefont {Withers}(2025{\natexlab{a}})}]{Arnaudo:2025kit}%
  \BibitemOpen
  \bibfield  {author} {\bibinfo {author} {\bibfnamefont {P.}~\bibnamefont {Arnaudo}}\ and\ \bibinfo {author} {\bibfnamefont {B.}~\bibnamefont {Withers}},\ }\bibfield  {title} {\bibinfo {title} {{Price's law from quasinormal modes}},\ }\href@noop {} {\  (\bibinfo {year} {2025}{\natexlab{a}})},\ \Eprint {https://arxiv.org/abs/2511.17703} {arXiv:2511.17703 [gr-qc]} \BibitemShut {NoStop}%
\bibitem [{\citenamefont {Arnaudo}\ and\ \citenamefont {Withers}(2025{\natexlab{b}})}]{Arnaudo:2024sen}%
  \BibitemOpen
  \bibfield  {author} {\bibinfo {author} {\bibfnamefont {P.}~\bibnamefont {Arnaudo}}\ and\ \bibinfo {author} {\bibfnamefont {B.}~\bibnamefont {Withers}},\ }\bibfield  {title} {\bibinfo {title} {{Exact low-temperature Green{\textquoteright}s functions in AdS/CFT: From the Heun equation to the confluent Heun equation}},\ }\href {https://doi.org/10.1103/8n3f-2d33} {\bibfield  {journal} {\bibinfo  {journal} {Phys. Rev. D}\ }\textbf {\bibinfo {volume} {111}},\ \bibinfo {pages} {L121903} (\bibinfo {year} {2025}{\natexlab{b}})},\ \Eprint {https://arxiv.org/abs/2412.01923} {arXiv:2412.01923 [hep-th]} \BibitemShut {NoStop}%
\bibitem [{\citenamefont {Besson}\ and\ \citenamefont {Jaramillo}(2025)}]{Besson:2024adi}%
  \BibitemOpen
  \bibfield  {author} {\bibinfo {author} {\bibfnamefont {J.}~\bibnamefont {Besson}}\ and\ \bibinfo {author} {\bibfnamefont {J.~L.}\ \bibnamefont {Jaramillo}},\ }\bibfield  {title} {\bibinfo {title} {{Quasi-normal mode expansions of black hole perturbations: a hyperboloidal Keldysh{\textquoteright}s approach}},\ }\href {https://doi.org/10.1007/s10714-025-03438-6} {\bibfield  {journal} {\bibinfo  {journal} {Gen. Rel. Grav.}\ }\textbf {\bibinfo {volume} {57}},\ \bibinfo {pages} {110} (\bibinfo {year} {2025})},\ \Eprint {https://arxiv.org/abs/2412.02793} {arXiv:2412.02793 [gr-qc]} \BibitemShut {NoStop}%
\bibitem [{\citenamefont {Leaver}(1986)}]{Leaver}%
  \BibitemOpen
  \bibfield  {author} {\bibinfo {author} {\bibfnamefont {E.~W.}\ \bibnamefont {Leaver}},\ }\bibfield  {title} {\bibinfo {title} {Spectral decomposition of the perturbation response of the schwarzschild geometry},\ }\href {https://doi.org/10.1103/PhysRevD.34.384} {\bibfield  {journal} {\bibinfo  {journal} {Phys. Rev. D}\ }\textbf {\bibinfo {volume} {34}},\ \bibinfo {pages} {384} (\bibinfo {year} {1986})}\BibitemShut {NoStop}%
\bibitem [{\citenamefont {Nollert}\ and\ \citenamefont {Schmidt}(1992)}]{PhysRevD.45.2617}%
  \BibitemOpen
  \bibfield  {author} {\bibinfo {author} {\bibfnamefont {H.-P.}\ \bibnamefont {Nollert}}\ and\ \bibinfo {author} {\bibfnamefont {B.~G.}\ \bibnamefont {Schmidt}},\ }\bibfield  {title} {\bibinfo {title} {Quasinormal modes of schwarzschild black holes: Defined and calculated via laplace transformation},\ }\href {https://doi.org/10.1103/PhysRevD.45.2617} {\bibfield  {journal} {\bibinfo  {journal} {Phys. Rev. D}\ }\textbf {\bibinfo {volume} {45}},\ \bibinfo {pages} {2617} (\bibinfo {year} {1992})}\BibitemShut {NoStop}%
\bibitem [{\citenamefont {Andersson}(1998)}]{AnderssonBookSection}%
  \BibitemOpen
  \bibfield  {author} {\bibinfo {author} {\bibfnamefont {N.}~\bibnamefont {Andersson}},\ }\bibfield  {title} {\bibinfo {title} {Black hole perturbations},\ }in\ \href {https://eprints.soton.ac.uk/29422/} {\emph {\bibinfo {booktitle} {Physics of Black Holes}}},\ \bibinfo {editor} {edited by\ \bibinfo {editor} {\bibfnamefont {I.}~\bibnamefont {Nivikov}}\ and\ \bibinfo {editor} {\bibfnamefont {V.}~\bibnamefont {Frolov}}}\ (\bibinfo  {publisher} {Kluwer Academic Publishers},\ \bibinfo {year} {1998})\ pp.\ \bibinfo {pages} {855--864}\BibitemShut {NoStop}%
\bibitem [{\citenamefont {Andersson}(1997)}]{Andersson:1996cm}%
  \BibitemOpen
  \bibfield  {author} {\bibinfo {author} {\bibfnamefont {N.}~\bibnamefont {Andersson}},\ }\bibfield  {title} {\bibinfo {title} {{Evolving test fields in a black hole geometry}},\ }\href {https://doi.org/10.1103/PhysRevD.55.468} {\bibfield  {journal} {\bibinfo  {journal} {Phys. Rev. D}\ }\textbf {\bibinfo {volume} {55}},\ \bibinfo {pages} {468} (\bibinfo {year} {1997})},\ \Eprint {https://arxiv.org/abs/gr-qc/9607064} {arXiv:gr-qc/9607064} \BibitemShut {NoStop}%
\bibitem [{\citenamefont {Dolan}\ and\ \citenamefont {Ottewill}(2011)}]{Dolan:2011fh}%
  \BibitemOpen
  \bibfield  {author} {\bibinfo {author} {\bibfnamefont {S.~R.}\ \bibnamefont {Dolan}}\ and\ \bibinfo {author} {\bibfnamefont {A.~C.}\ \bibnamefont {Ottewill}},\ }\bibfield  {title} {\bibinfo {title} {{Wave Propagation and Quasinormal Mode Excitation on Schwarzschild Spacetime}},\ }\href {https://doi.org/10.1103/PhysRevD.84.104002} {\bibfield  {journal} {\bibinfo  {journal} {Phys. Rev. D}\ }\textbf {\bibinfo {volume} {84}},\ \bibinfo {pages} {104002} (\bibinfo {year} {2011})},\ \Eprint {https://arxiv.org/abs/1106.4318} {arXiv:1106.4318 [gr-qc]} \BibitemShut {NoStop}%
\bibitem [{\citenamefont {Warnick}(2022)}]{Warnick:2022hnc}%
  \BibitemOpen
  \bibfield  {author} {\bibinfo {author} {\bibfnamefont {C.}~\bibnamefont {Warnick}},\ }\bibfield  {title} {\bibinfo {title} {{(In)completeness of Quasinormal Modes}},\ }\href {https://doi.org/10.5506/APhysPolBSupp.15.1-A13} {\bibfield  {journal} {\bibinfo  {journal} {Acta Phys. Polon. Supp.}\ }\textbf {\bibinfo {volume} {15}},\ \bibinfo {pages} {1} (\bibinfo {year} {2022})}\BibitemShut {NoStop}%
\bibitem [{\citenamefont {Sun}\ and\ \citenamefont {Price}(1988)}]{SunPrice88}%
  \BibitemOpen
  \bibfield  {author} {\bibinfo {author} {\bibfnamefont {Y.}~\bibnamefont {Sun}}\ and\ \bibinfo {author} {\bibfnamefont {R.~H.}\ \bibnamefont {Price}},\ }\bibfield  {title} {\bibinfo {title} {Excitation of quasinormal ringing of a schwarzschild black hole},\ }\href {https://doi.org/10.1103/PhysRevD.38.1040} {\bibfield  {journal} {\bibinfo  {journal} {Phys. Rev. D}\ }\textbf {\bibinfo {volume} {38}},\ \bibinfo {pages} {1040} (\bibinfo {year} {1988})}\BibitemShut {NoStop}%
\bibitem [{\citenamefont {Bachelot}\ and\ \citenamefont {Motet-Bachelot}(1993)}]{Bachelot1993}%
  \BibitemOpen
  \bibfield  {author} {\bibinfo {author} {\bibfnamefont {A.}~\bibnamefont {Bachelot}}\ and\ \bibinfo {author} {\bibfnamefont {A.}~\bibnamefont {Motet-Bachelot}},\ }\bibfield  {title} {\bibinfo {title} {Les résonances d'un trou noir de schwarzschild},\ }\href {http://eudml.org/doc/76614} {\bibfield  {journal} {\bibinfo  {journal} {Annales de l'I.H.P. Physique théorique}\ }\textbf {\bibinfo {volume} {59}},\ \bibinfo {pages} {3} (\bibinfo {year} {1993})}\BibitemShut {NoStop}%
\bibitem [{\citenamefont {Ching}\ \emph {et~al.}(1995)\citenamefont {Ching}, \citenamefont {Leung}, \citenamefont {Suen},\ and\ \citenamefont {Young}}]{Ching:1993gt}%
  \BibitemOpen
  \bibfield  {author} {\bibinfo {author} {\bibfnamefont {E.~S.~C.}\ \bibnamefont {Ching}}, \bibinfo {author} {\bibfnamefont {P.~T.}\ \bibnamefont {Leung}}, \bibinfo {author} {\bibfnamefont {W.~M.}\ \bibnamefont {Suen}},\ and\ \bibinfo {author} {\bibfnamefont {K.}~\bibnamefont {Young}},\ }\bibfield  {title} {\bibinfo {title} {{Quasinormal mode expansion for linearized waves in gravitational system}},\ }\href {https://doi.org/10.1103/PhysRevLett.74.4588} {\bibfield  {journal} {\bibinfo  {journal} {Phys. Rev. Lett.}\ }\textbf {\bibinfo {volume} {74}},\ \bibinfo {pages} {4588} (\bibinfo {year} {1995})},\ \Eprint {https://arxiv.org/abs/gr-qc/9408043} {arXiv:gr-qc/9408043} \BibitemShut {NoStop}%
\bibitem [{\citenamefont {Ching}\ \emph {et~al.}(1996)\citenamefont {Ching}, \citenamefont {Leung}, \citenamefont {Suen},\ and\ \citenamefont {Young}}]{Ching:1995rt}%
  \BibitemOpen
  \bibfield  {author} {\bibinfo {author} {\bibfnamefont {E.~S.~C.}\ \bibnamefont {Ching}}, \bibinfo {author} {\bibfnamefont {P.~T.}\ \bibnamefont {Leung}}, \bibinfo {author} {\bibfnamefont {W.~M.}\ \bibnamefont {Suen}},\ and\ \bibinfo {author} {\bibfnamefont {K.}~\bibnamefont {Young}},\ }\bibfield  {title} {\bibinfo {title} {{Wave propagation in gravitational systems: Completeness of quasinormal modes}},\ }\href {https://doi.org/10.1103/PhysRevD.54.3778} {\bibfield  {journal} {\bibinfo  {journal} {Phys. Rev. D}\ }\textbf {\bibinfo {volume} {54}},\ \bibinfo {pages} {3778} (\bibinfo {year} {1996})},\ \Eprint {https://arxiv.org/abs/gr-qc/9507034} {arXiv:gr-qc/9507034} \BibitemShut {NoStop}%
\bibitem [{\citenamefont {S\'a~Barreto}\ and\ \citenamefont {Zworski}(1997)}]{SaBarretoZworski1997}%
  \BibitemOpen
  \bibfield  {author} {\bibinfo {author} {\bibfnamefont {A.}~\bibnamefont {S\'a~Barreto}}\ and\ \bibinfo {author} {\bibfnamefont {M.}~\bibnamefont {Zworski}},\ }\bibfield  {title} {\bibinfo {title} {{Distribution of resonances for spherical black holes}},\ }\href {https://doi.org/10.4310/MRL.1997.v4.n1.a10} {\bibfield  {journal} {\bibinfo  {journal} {Math. Res. Lett.}\ }\textbf {\bibinfo {volume} {4}},\ \bibinfo {pages} {103} (\bibinfo {year} {1997})}\BibitemShut {NoStop}%
\bibitem [{\citenamefont {Beyer}(1999)}]{Beyer:1998nu}%
  \BibitemOpen
  \bibfield  {author} {\bibinfo {author} {\bibfnamefont {H.~R.}\ \bibnamefont {Beyer}},\ }\bibfield  {title} {\bibinfo {title} {{On the completeness of the quasinormal modes of the Poschl-Teller potential}},\ }\href {https://doi.org/10.1007/s002200050651} {\bibfield  {journal} {\bibinfo  {journal} {Commun. Math. Phys.}\ }\textbf {\bibinfo {volume} {204}},\ \bibinfo {pages} {397} (\bibinfo {year} {1999})},\ \Eprint {https://arxiv.org/abs/gr-qc/9803034} {arXiv:gr-qc/9803034} \BibitemShut {NoStop}%
\bibitem [{\citenamefont {Szpak}(2004)}]{Szpak:2004sf}%
  \BibitemOpen
  \bibfield  {author} {\bibinfo {author} {\bibfnamefont {N.}~\bibnamefont {Szpak}},\ }\bibfield  {title} {\bibinfo {title} {{Quasinormal mode expansion and the exact solution of the Cauchy problem for wave equations}},\ }\href@noop {} {\  (\bibinfo {year} {2004})},\ \Eprint {https://arxiv.org/abs/gr-qc/0411050} {arXiv:gr-qc/0411050} \BibitemShut {NoStop}%
\bibitem [{\citenamefont {Berti}\ and\ \citenamefont {Cardoso}(2006)}]{Berti:2006wq}%
  \BibitemOpen
  \bibfield  {author} {\bibinfo {author} {\bibfnamefont {E.}~\bibnamefont {Berti}}\ and\ \bibinfo {author} {\bibfnamefont {V.}~\bibnamefont {Cardoso}},\ }\bibfield  {title} {\bibinfo {title} {{Quasinormal ringing of Kerr black holes. I. The Excitation factors}},\ }\href {https://doi.org/10.1103/PhysRevD.74.104020} {\bibfield  {journal} {\bibinfo  {journal} {Phys. Rev. D}\ }\textbf {\bibinfo {volume} {74}},\ \bibinfo {pages} {104020} (\bibinfo {year} {2006})},\ \Eprint {https://arxiv.org/abs/gr-qc/0605118} {arXiv:gr-qc/0605118} \BibitemShut {NoStop}%
\bibitem [{\citenamefont {Bony}\ and\ \citenamefont {H{\"a}fner}(2008)}]{bony2008decay}%
  \BibitemOpen
  \bibfield  {author} {\bibinfo {author} {\bibfnamefont {J.-F.}\ \bibnamefont {Bony}}\ and\ \bibinfo {author} {\bibfnamefont {D.}~\bibnamefont {H{\"a}fner}},\ }\bibfield  {title} {\bibinfo {title} {Decay and non-decay of the local energy for the wave equation on the de sitter--schwarzschild metric},\ }\href {https://doi.org/10.1007/s00220-008-0553-y} {\bibfield  {journal} {\bibinfo  {journal} {Communications in Mathematical Physics}\ }\textbf {\bibinfo {volume} {282}},\ \bibinfo {pages} {697} (\bibinfo {year} {2008})},\ \Eprint {https://arxiv.org/abs/0706.0350} {arXiv:0706.0350} \BibitemShut {NoStop}%
\bibitem [{\citenamefont {Dyatlov}(2011)}]{Dyatlov:2010hq}%
  \BibitemOpen
  \bibfield  {author} {\bibinfo {author} {\bibfnamefont {S.}~\bibnamefont {Dyatlov}},\ }\bibfield  {title} {\bibinfo {title} {{Quasi-normal modes and exponential energy decay for the Kerr-de Sitter black hole}},\ }\href {https://doi.org/10.1007/s00220-011-1286-x} {\bibfield  {journal} {\bibinfo  {journal} {Commun. Math. Phys.}\ }\textbf {\bibinfo {volume} {306}},\ \bibinfo {pages} {119} (\bibinfo {year} {2011})},\ \Eprint {https://arxiv.org/abs/1003.6128} {arXiv:1003.6128 [math.AP]} \BibitemShut {NoStop}%
\bibitem [{\citenamefont {Ansorg}\ and\ \citenamefont {Panosso~Macedo}(2016)}]{Ansorg:2016ztf}%
  \BibitemOpen
  \bibfield  {author} {\bibinfo {author} {\bibfnamefont {M.}~\bibnamefont {Ansorg}}\ and\ \bibinfo {author} {\bibfnamefont {R.}~\bibnamefont {Panosso~Macedo}},\ }\bibfield  {title} {\bibinfo {title} {{Spectral decomposition of black-hole perturbations on hyperboloidal slices}},\ }\href {https://doi.org/10.1103/PhysRevD.93.124016} {\bibfield  {journal} {\bibinfo  {journal} {Phys. Rev. D}\ }\textbf {\bibinfo {volume} {93}},\ \bibinfo {pages} {124016} (\bibinfo {year} {2016})},\ \Eprint {https://arxiv.org/abs/1604.02261} {arXiv:1604.02261 [gr-qc]} \BibitemShut {NoStop}%
\bibitem [{\citenamefont {Panosso~Macedo}\ \emph {et~al.}(2018)\citenamefont {Panosso~Macedo}, \citenamefont {Jaramillo},\ and\ \citenamefont {Ansorg}}]{PanossoMacedo:2018hab}%
  \BibitemOpen
  \bibfield  {author} {\bibinfo {author} {\bibfnamefont {R.}~\bibnamefont {Panosso~Macedo}}, \bibinfo {author} {\bibfnamefont {J.~L.}\ \bibnamefont {Jaramillo}},\ and\ \bibinfo {author} {\bibfnamefont {M.}~\bibnamefont {Ansorg}},\ }\bibfield  {title} {\bibinfo {title} {{Hyperboloidal slicing approach to quasi-normal mode expansions: the Reissner-Nordstr{\"o}m case}},\ }\href {https://doi.org/10.1103/PhysRevD.98.124005} {\bibfield  {journal} {\bibinfo  {journal} {Phys. Rev. D}\ }\textbf {\bibinfo {volume} {98}},\ \bibinfo {pages} {124005} (\bibinfo {year} {2018})},\ \Eprint {https://arxiv.org/abs/1809.02837} {arXiv:1809.02837 [gr-qc]} \BibitemShut {NoStop}%
\bibitem [{\citenamefont {Chen}\ \emph {et~al.}(2024)\citenamefont {Chen}, \citenamefont {Ivo},\ and\ \citenamefont {Maldacena}}]{Chen:2023hra}%
  \BibitemOpen
  \bibfield  {author} {\bibinfo {author} {\bibfnamefont {Y.}~\bibnamefont {Chen}}, \bibinfo {author} {\bibfnamefont {V.}~\bibnamefont {Ivo}},\ and\ \bibinfo {author} {\bibfnamefont {J.}~\bibnamefont {Maldacena}},\ }\bibfield  {title} {\bibinfo {title} {{Comments on the double cone wormhole}},\ }\href {https://doi.org/10.1007/JHEP04(2024)124} {\bibfield  {journal} {\bibinfo  {journal} {JHEP}\ }\textbf {\bibinfo {volume} {04}},\ \bibinfo {pages} {124}},\ \Eprint {https://arxiv.org/abs/2310.11617} {arXiv:2310.11617 [hep-th]} \BibitemShut {NoStop}%
\bibitem [{\citenamefont {Nollert}(1999)}]{HansPeterNollert_1999}%
  \BibitemOpen
  \bibfield  {author} {\bibinfo {author} {\bibfnamefont {H.-P.}\ \bibnamefont {Nollert}},\ }\bibfield  {title} {\bibinfo {title} {Quasinormal modes: the characteristic `sound' of black holes and neutron stars},\ }\href {https://doi.org/10.1088/0264-9381/16/12/201} {\bibfield  {journal} {\bibinfo  {journal} {Classical and Quantum Gravity}\ }\textbf {\bibinfo {volume} {16}},\ \bibinfo {pages} {R159} (\bibinfo {year} {1999})}\BibitemShut {NoStop}%
\bibitem [{\citenamefont {Kokkotas}\ and\ \citenamefont {Schmidt}(1999)}]{Kokkotas:1999bd}%
  \BibitemOpen
  \bibfield  {author} {\bibinfo {author} {\bibfnamefont {K.~D.}\ \bibnamefont {Kokkotas}}\ and\ \bibinfo {author} {\bibfnamefont {B.~G.}\ \bibnamefont {Schmidt}},\ }\bibfield  {title} {\bibinfo {title} {{Quasinormal modes of stars and black holes}},\ }\href {https://doi.org/10.12942/lrr-1999-2} {\bibfield  {journal} {\bibinfo  {journal} {Living Rev. Rel.}\ }\textbf {\bibinfo {volume} {2}},\ \bibinfo {pages} {2} (\bibinfo {year} {1999})},\ \Eprint {https://arxiv.org/abs/gr-qc/9909058} {arXiv:gr-qc/9909058} \BibitemShut {NoStop}%
\bibitem [{\citenamefont {Berti}\ \emph {et~al.}(2009)\citenamefont {Berti}, \citenamefont {Cardoso},\ and\ \citenamefont {Starinets}}]{Berti:2009kk}%
  \BibitemOpen
  \bibfield  {author} {\bibinfo {author} {\bibfnamefont {E.}~\bibnamefont {Berti}}, \bibinfo {author} {\bibfnamefont {V.}~\bibnamefont {Cardoso}},\ and\ \bibinfo {author} {\bibfnamefont {A.~O.}\ \bibnamefont {Starinets}},\ }\bibfield  {title} {\bibinfo {title} {{Quasinormal modes of black holes and black branes}},\ }\href {https://doi.org/10.1088/0264-9381/26/16/163001} {\bibfield  {journal} {\bibinfo  {journal} {Class. Quant. Grav.}\ }\textbf {\bibinfo {volume} {26}},\ \bibinfo {pages} {163001} (\bibinfo {year} {2009})},\ \Eprint {https://arxiv.org/abs/0905.2975} {arXiv:0905.2975 [gr-qc]} \BibitemShut {NoStop}%
\bibitem [{\citenamefont {Konoplya}\ and\ \citenamefont {Zhidenko}(2011)}]{Konoplya:2011qq}%
  \BibitemOpen
  \bibfield  {author} {\bibinfo {author} {\bibfnamefont {R.~A.}\ \bibnamefont {Konoplya}}\ and\ \bibinfo {author} {\bibfnamefont {A.}~\bibnamefont {Zhidenko}},\ }\bibfield  {title} {\bibinfo {title} {{Quasinormal modes of black holes: From astrophysics to string theory}},\ }\href {https://doi.org/10.1103/RevModPhys.83.793} {\bibfield  {journal} {\bibinfo  {journal} {Rev. Mod. Phys.}\ }\textbf {\bibinfo {volume} {83}},\ \bibinfo {pages} {793} (\bibinfo {year} {2011})},\ \Eprint {https://arxiv.org/abs/1102.4014} {arXiv:1102.4014 [gr-qc]} \BibitemShut {NoStop}%
\bibitem [{\citenamefont {Jafferis}\ \emph {et~al.}(2015)\citenamefont {Jafferis}, \citenamefont {Lupsasca}, \citenamefont {Lysov}, \citenamefont {Ng},\ and\ \citenamefont {Strominger}}]{Jafferis:2013qia}%
  \BibitemOpen
  \bibfield  {author} {\bibinfo {author} {\bibfnamefont {D.~L.}\ \bibnamefont {Jafferis}}, \bibinfo {author} {\bibfnamefont {A.}~\bibnamefont {Lupsasca}}, \bibinfo {author} {\bibfnamefont {V.}~\bibnamefont {Lysov}}, \bibinfo {author} {\bibfnamefont {G.~S.}\ \bibnamefont {Ng}},\ and\ \bibinfo {author} {\bibfnamefont {A.}~\bibnamefont {Strominger}},\ }\bibfield  {title} {\bibinfo {title} {{Quasinormal quantization in de Sitter spacetime}},\ }\href {https://doi.org/10.1007/JHEP01(2015)004} {\bibfield  {journal} {\bibinfo  {journal} {JHEP}\ }\textbf {\bibinfo {volume} {01}},\ \bibinfo {pages} {004}},\ \Eprint {https://arxiv.org/abs/1305.5523} {arXiv:1305.5523 [hep-th]} \BibitemShut {NoStop}%
\bibitem [{\citenamefont {Green}\ \emph {et~al.}(2023)\citenamefont {Green}, \citenamefont {Hollands}, \citenamefont {Sberna}, \citenamefont {Toomani},\ and\ \citenamefont {Zimmerman}}]{Green:2022htq}%
  \BibitemOpen
  \bibfield  {author} {\bibinfo {author} {\bibfnamefont {S.~R.}\ \bibnamefont {Green}}, \bibinfo {author} {\bibfnamefont {S.}~\bibnamefont {Hollands}}, \bibinfo {author} {\bibfnamefont {L.}~\bibnamefont {Sberna}}, \bibinfo {author} {\bibfnamefont {V.}~\bibnamefont {Toomani}},\ and\ \bibinfo {author} {\bibfnamefont {P.}~\bibnamefont {Zimmerman}},\ }\bibfield  {title} {\bibinfo {title} {{Conserved currents for a Kerr black hole and orthogonality of quasinormal modes}},\ }\href {https://doi.org/10.1103/PhysRevD.107.064030} {\bibfield  {journal} {\bibinfo  {journal} {Phys. Rev. D}\ }\textbf {\bibinfo {volume} {107}},\ \bibinfo {pages} {064030} (\bibinfo {year} {2023})},\ \Eprint {https://arxiv.org/abs/2210.15935} {arXiv:2210.15935 [gr-qc]} \BibitemShut {NoStop}%
\bibitem [{\citenamefont {London}(2023)}]{London:2023aeo}%
  \BibitemOpen
  \bibfield  {author} {\bibinfo {author} {\bibfnamefont {L.~T.}\ \bibnamefont {London}},\ }\bibfield  {title} {\bibinfo {title} {{A radial scalar product for Kerr quasinormal modes}},\ }\href@noop {} {\  (\bibinfo {year} {2023})},\ \Eprint {https://arxiv.org/abs/2312.17678} {arXiv:2312.17678 [gr-qc]} \BibitemShut {NoStop}%
\bibitem [{\citenamefont {London}\ and\ \citenamefont {Gurevich}(2023)}]{London:2023idh}%
  \BibitemOpen
  \bibfield  {author} {\bibinfo {author} {\bibfnamefont {L.}~\bibnamefont {London}}\ and\ \bibinfo {author} {\bibfnamefont {M.}~\bibnamefont {Gurevich}},\ }\bibfield  {title} {\bibinfo {title} {{Natural polynomials for Kerr quasi-normal modes}},\ }\href@noop {} {\  (\bibinfo {year} {2023})},\ \Eprint {https://arxiv.org/abs/2312.17680} {arXiv:2312.17680 [gr-qc]} \BibitemShut {NoStop}%
\bibitem [{\citenamefont {Arnaudo}\ \emph {et~al.}(2025{\natexlab{a}})\citenamefont {Arnaudo}, \citenamefont {Carballo},\ and\ \citenamefont {Withers}}]{Arnaudo:2025bnm}%
  \BibitemOpen
  \bibfield  {author} {\bibinfo {author} {\bibfnamefont {P.}~\bibnamefont {Arnaudo}}, \bibinfo {author} {\bibfnamefont {J.}~\bibnamefont {Carballo}},\ and\ \bibinfo {author} {\bibfnamefont {B.}~\bibnamefont {Withers}},\ }\bibfield  {title} {\bibinfo {title} {{QNM orthogonality relations for AdS black holes}},\ }\href@noop {} {\  (\bibinfo {year} {2025}{\natexlab{a}})},\ \Eprint {https://arxiv.org/abs/2505.04696} {arXiv:2505.04696 [hep-th]} \BibitemShut {NoStop}%
\bibitem [{\citenamefont {Finster}\ \emph {et~al.}(2003)\citenamefont {Finster}, \citenamefont {Kamran}, \citenamefont {Smoller},\ and\ \citenamefont {Yau}}]{Finster:2000jz}%
  \BibitemOpen
  \bibfield  {author} {\bibinfo {author} {\bibfnamefont {F.}~\bibnamefont {Finster}}, \bibinfo {author} {\bibfnamefont {N.}~\bibnamefont {Kamran}}, \bibinfo {author} {\bibfnamefont {J.}~\bibnamefont {Smoller}},\ and\ \bibinfo {author} {\bibfnamefont {S.-T.}\ \bibnamefont {Yau}},\ }\bibfield  {title} {\bibinfo {title} {{The Long time dynamics of Dirac particles in the Kerr-Newman black hole geometry}},\ }\href {https://doi.org/10.4310/ATMP.2003.v7.n1.a2} {\bibfield  {journal} {\bibinfo  {journal} {Adv. Theor. Math. Phys.}\ }\textbf {\bibinfo {volume} {7}},\ \bibinfo {pages} {25} (\bibinfo {year} {2003})},\ \Eprint {https://arxiv.org/abs/gr-qc/0005088} {arXiv:gr-qc/0005088} \BibitemShut {NoStop}%
\bibitem [{\citenamefont {Finster}\ \emph {et~al.}(2005)\citenamefont {Finster}, \citenamefont {Kamran}, \citenamefont {Smoller},\ and\ \citenamefont {Yau}}]{Finster:2003fu}%
  \BibitemOpen
  \bibfield  {author} {\bibinfo {author} {\bibfnamefont {F.}~\bibnamefont {Finster}}, \bibinfo {author} {\bibfnamefont {N.}~\bibnamefont {Kamran}}, \bibinfo {author} {\bibfnamefont {J.}~\bibnamefont {Smoller}},\ and\ \bibinfo {author} {\bibfnamefont {S.-T.}\ \bibnamefont {Yau}},\ }\bibfield  {title} {\bibinfo {title} {{An Integral spectral representation of the propagator for the wave equation in the Kerr geometry}},\ }\href {https://doi.org/10.1007/s00220-005-1390-x} {\bibfield  {journal} {\bibinfo  {journal} {Commun. Math. Phys.}\ }\textbf {\bibinfo {volume} {260}},\ \bibinfo {pages} {257} (\bibinfo {year} {2005})},\ \Eprint {https://arxiv.org/abs/gr-qc/0310024} {arXiv:gr-qc/0310024} \BibitemShut {NoStop}%
\bibitem [{\citenamefont {Mino}\ and\ \citenamefont {Brink}(2008)}]{Mino:2008at}%
  \BibitemOpen
  \bibfield  {author} {\bibinfo {author} {\bibfnamefont {Y.}~\bibnamefont {Mino}}\ and\ \bibinfo {author} {\bibfnamefont {J.}~\bibnamefont {Brink}},\ }\bibfield  {title} {\bibinfo {title} {{Gravitational Radiation from Plunging Orbits: Perturbative Study}},\ }\href {https://doi.org/10.1103/PhysRevD.78.124015} {\bibfield  {journal} {\bibinfo  {journal} {Phys. Rev. D}\ }\textbf {\bibinfo {volume} {78}},\ \bibinfo {pages} {124015} (\bibinfo {year} {2008})},\ \Eprint {https://arxiv.org/abs/0809.2814} {arXiv:0809.2814 [gr-qc]} \BibitemShut {NoStop}%
\bibitem [{\citenamefont {Zimmerman}\ and\ \citenamefont {Chen}(2011)}]{Zimmerman:2011dx}%
  \BibitemOpen
  \bibfield  {author} {\bibinfo {author} {\bibfnamefont {A.}~\bibnamefont {Zimmerman}}\ and\ \bibinfo {author} {\bibfnamefont {Y.}~\bibnamefont {Chen}},\ }\bibfield  {title} {\bibinfo {title} {{New Generic Ringdown Frequencies at the Birth of a Kerr Black Hole}},\ }\href {https://doi.org/10.1103/PhysRevD.84.084012} {\bibfield  {journal} {\bibinfo  {journal} {Phys. Rev. D}\ }\textbf {\bibinfo {volume} {84}},\ \bibinfo {pages} {084012} (\bibinfo {year} {2011})},\ \Eprint {https://arxiv.org/abs/1106.0782} {arXiv:1106.0782 [gr-qc]} \BibitemShut {NoStop}%
\bibitem [{\citenamefont {De~Amicis}\ \emph {et~al.}(2025)\citenamefont {De~Amicis}, \citenamefont {Cannizzaro}, \citenamefont {Carullo},\ and\ \citenamefont {Sberna}}]{DeAmicis:2025xuh}%
  \BibitemOpen
  \bibfield  {author} {\bibinfo {author} {\bibfnamefont {M.}~\bibnamefont {De~Amicis}}, \bibinfo {author} {\bibfnamefont {E.}~\bibnamefont {Cannizzaro}}, \bibinfo {author} {\bibfnamefont {G.}~\bibnamefont {Carullo}},\ and\ \bibinfo {author} {\bibfnamefont {L.}~\bibnamefont {Sberna}},\ }\bibfield  {title} {\bibinfo {title} {{Dynamical quasinormal mode excitation}},\ }\href@noop {} {\  (\bibinfo {year} {2025})},\ \Eprint {https://arxiv.org/abs/2506.21668} {arXiv:2506.21668 [gr-qc]} \BibitemShut {NoStop}%
\bibitem [{\citenamefont {Oshita}\ \emph {et~al.}(2025)\citenamefont {Oshita}, \citenamefont {Ma}, \citenamefont {Chen},\ and\ \citenamefont {Yang}}]{Oshita:2025qmn}%
  \BibitemOpen
  \bibfield  {author} {\bibinfo {author} {\bibfnamefont {N.}~\bibnamefont {Oshita}}, \bibinfo {author} {\bibfnamefont {S.}~\bibnamefont {Ma}}, \bibinfo {author} {\bibfnamefont {Y.}~\bibnamefont {Chen}},\ and\ \bibinfo {author} {\bibfnamefont {H.}~\bibnamefont {Yang}},\ }\bibfield  {title} {\bibinfo {title} {{Probing Direct Waves in Black Hole Ringdowns}},\ }\href@noop {} {\  (\bibinfo {year} {2025})},\ \Eprint {https://arxiv.org/abs/2509.09165} {arXiv:2509.09165 [gr-qc]} \BibitemShut {NoStop}%
\bibitem [{\citenamefont {Okuzumi}\ \emph {et~al.}(2008)\citenamefont {Okuzumi}, \citenamefont {Ioka},\ and\ \citenamefont {Sakagami}}]{Okuzumi:2008ej}%
  \BibitemOpen
  \bibfield  {author} {\bibinfo {author} {\bibfnamefont {S.}~\bibnamefont {Okuzumi}}, \bibinfo {author} {\bibfnamefont {K.}~\bibnamefont {Ioka}},\ and\ \bibinfo {author} {\bibfnamefont {M.-a.}\ \bibnamefont {Sakagami}},\ }\bibfield  {title} {\bibinfo {title} {{Possible Discovery of Nonlinear Tail and Quasinormal Modes in Black Hole Ringdown}},\ }\href {https://doi.org/10.1103/PhysRevD.77.124018} {\bibfield  {journal} {\bibinfo  {journal} {Phys. Rev. D}\ }\textbf {\bibinfo {volume} {77}},\ \bibinfo {pages} {124018} (\bibinfo {year} {2008})},\ \Eprint {https://arxiv.org/abs/0803.0501} {arXiv:0803.0501 [gr-qc]} \BibitemShut {NoStop}%
\bibitem [{\citenamefont {Lagos}\ and\ \citenamefont {Hui}(2023)}]{Lagos:2022otp}%
  \BibitemOpen
  \bibfield  {author} {\bibinfo {author} {\bibfnamefont {M.}~\bibnamefont {Lagos}}\ and\ \bibinfo {author} {\bibfnamefont {L.}~\bibnamefont {Hui}},\ }\bibfield  {title} {\bibinfo {title} {{Generation and propagation of nonlinear quasinormal modes of a Schwarzschild black hole}},\ }\href {https://doi.org/10.1103/PhysRevD.107.044040} {\bibfield  {journal} {\bibinfo  {journal} {Phys. Rev. D}\ }\textbf {\bibinfo {volume} {107}},\ \bibinfo {pages} {044040} (\bibinfo {year} {2023})},\ \Eprint {https://arxiv.org/abs/2208.07379} {arXiv:2208.07379 [gr-qc]} \BibitemShut {NoStop}%
\bibitem [{\citenamefont {Chavda}\ \emph {et~al.}(2025)\citenamefont {Chavda}, \citenamefont {Lagos},\ and\ \citenamefont {Hui}}]{Chavda:2024awq}%
  \BibitemOpen
  \bibfield  {author} {\bibinfo {author} {\bibfnamefont {A.}~\bibnamefont {Chavda}}, \bibinfo {author} {\bibfnamefont {M.}~\bibnamefont {Lagos}},\ and\ \bibinfo {author} {\bibfnamefont {L.}~\bibnamefont {Hui}},\ }\bibfield  {title} {\bibinfo {title} {{The impact of initial conditions on quasi-normal modes}},\ }\href {https://doi.org/10.1088/1475-7516/2025/07/084} {\bibfield  {journal} {\bibinfo  {journal} {JCAP}\ }\textbf {\bibinfo {volume} {07}},\ \bibinfo {pages} {084}},\ \Eprint {https://arxiv.org/abs/2412.03435} {arXiv:2412.03435 [gr-qc]} \BibitemShut {NoStop}%
\bibitem [{\citenamefont {Ling}\ \emph {et~al.}(2025)\citenamefont {Ling}, \citenamefont {Shah},\ and\ \citenamefont {Wong}}]{Ling:2025wfv}%
  \BibitemOpen
  \bibfield  {author} {\bibinfo {author} {\bibfnamefont {S.}~\bibnamefont {Ling}}, \bibinfo {author} {\bibfnamefont {S.}~\bibnamefont {Shah}},\ and\ \bibinfo {author} {\bibfnamefont {S.~S.~C.}\ \bibnamefont {Wong}},\ }\bibfield  {title} {\bibinfo {title} {{Dynamical nonlinear tails in the Schwarzschild black hole ringdown}},\ }\href {https://doi.org/10.1103/22lc-62gj} {\bibfield  {journal} {\bibinfo  {journal} {Phys. Rev. D}\ }\textbf {\bibinfo {volume} {112}},\ \bibinfo {pages} {024008} (\bibinfo {year} {2025})},\ \Eprint {https://arxiv.org/abs/2503.19967} {arXiv:2503.19967 [gr-qc]} \BibitemShut {NoStop}%
\bibitem [{\citenamefont {Banados}\ \emph {et~al.}(1992)\citenamefont {Banados}, \citenamefont {Teitelboim},\ and\ \citenamefont {Zanelli}}]{Banados:1992wn}%
  \BibitemOpen
  \bibfield  {author} {\bibinfo {author} {\bibfnamefont {M.}~\bibnamefont {Banados}}, \bibinfo {author} {\bibfnamefont {C.}~\bibnamefont {Teitelboim}},\ and\ \bibinfo {author} {\bibfnamefont {J.}~\bibnamefont {Zanelli}},\ }\bibfield  {title} {\bibinfo {title} {{The Black hole in three-dimensional space-time}},\ }\href {https://doi.org/10.1103/PhysRevLett.69.1849} {\bibfield  {journal} {\bibinfo  {journal} {Phys. Rev. Lett.}\ }\textbf {\bibinfo {volume} {69}},\ \bibinfo {pages} {1849} (\bibinfo {year} {1992})},\ \Eprint {https://arxiv.org/abs/hep-th/9204099} {arXiv:hep-th/9204099} \BibitemShut {NoStop}%
\bibitem [{\citenamefont {Maldacena}(1998)}]{Maldacena:1997re}%
  \BibitemOpen
  \bibfield  {author} {\bibinfo {author} {\bibfnamefont {J.~M.}\ \bibnamefont {Maldacena}},\ }\bibfield  {title} {\bibinfo {title} {{The Large $N$ limit of superconformal field theories and supergravity}},\ }\href {https://doi.org/10.4310/ATMP.1998.v2.n2.a1} {\bibfield  {journal} {\bibinfo  {journal} {Adv. Theor. Math. Phys.}\ }\textbf {\bibinfo {volume} {2}},\ \bibinfo {pages} {231} (\bibinfo {year} {1998})},\ \Eprint {https://arxiv.org/abs/hep-th/9711200} {arXiv:hep-th/9711200} \BibitemShut {NoStop}%
\bibitem [{\citenamefont {Witten}(1998)}]{Witten:1998qj}%
  \BibitemOpen
  \bibfield  {author} {\bibinfo {author} {\bibfnamefont {E.}~\bibnamefont {Witten}},\ }\bibfield  {title} {\bibinfo {title} {{Anti de Sitter space and holography}},\ }\href {https://doi.org/10.4310/ATMP.1998.v2.n2.a2} {\bibfield  {journal} {\bibinfo  {journal} {Adv. Theor. Math. Phys.}\ }\textbf {\bibinfo {volume} {2}},\ \bibinfo {pages} {253} (\bibinfo {year} {1998})},\ \Eprint {https://arxiv.org/abs/hep-th/9802150} {arXiv:hep-th/9802150} \BibitemShut {NoStop}%
\bibitem [{\citenamefont {Gubser}\ \emph {et~al.}(1998)\citenamefont {Gubser}, \citenamefont {Klebanov},\ and\ \citenamefont {Polyakov}}]{Gubser:1998bc}%
  \BibitemOpen
  \bibfield  {author} {\bibinfo {author} {\bibfnamefont {S.~S.}\ \bibnamefont {Gubser}}, \bibinfo {author} {\bibfnamefont {I.~R.}\ \bibnamefont {Klebanov}},\ and\ \bibinfo {author} {\bibfnamefont {A.~M.}\ \bibnamefont {Polyakov}},\ }\bibfield  {title} {\bibinfo {title} {{Gauge theory correlators from noncritical string theory}},\ }\href {https://doi.org/10.1016/S0370-2693(98)00377-3} {\bibfield  {journal} {\bibinfo  {journal} {Phys. Lett. B}\ }\textbf {\bibinfo {volume} {428}},\ \bibinfo {pages} {105} (\bibinfo {year} {1998})},\ \Eprint {https://arxiv.org/abs/hep-th/9802109} {arXiv:hep-th/9802109} \BibitemShut {NoStop}%
\bibitem [{\citenamefont {Erdélyi}\ \emph {et~al.}(1953)\citenamefont {Erdélyi}, \citenamefont {Magnus}, \citenamefont {Oberhettinger},\ and\ \citenamefont {Tricomi}}]{Erdelyi1953}%
  \BibitemOpen
  \bibfield  {author} {\bibinfo {author} {\bibfnamefont {A.}~\bibnamefont {Erdélyi}}, \bibinfo {author} {\bibfnamefont {W.}~\bibnamefont {Magnus}}, \bibinfo {author} {\bibfnamefont {F.}~\bibnamefont {Oberhettinger}},\ and\ \bibinfo {author} {\bibfnamefont {F.~G.}\ \bibnamefont {Tricomi}},\ }\href@noop {} {\emph {\bibinfo {title} {Higher Transcendental Functions. Vol. I}}},\ Bateman Manuscript Project\ (\bibinfo  {publisher} {McGraw--Hill},\ \bibinfo {address} {New York},\ \bibinfo {year} {1953})\BibitemShut {NoStop}%
\bibitem [{\citenamefont {Cardoso}\ and\ \citenamefont {Lemos}(2001)}]{Cardoso:2001hn}%
  \BibitemOpen
  \bibfield  {author} {\bibinfo {author} {\bibfnamefont {V.}~\bibnamefont {Cardoso}}\ and\ \bibinfo {author} {\bibfnamefont {J.~P.~S.}\ \bibnamefont {Lemos}},\ }\bibfield  {title} {\bibinfo {title} {{Scalar, electromagnetic and Weyl perturbations of BTZ black holes: Quasinormal modes}},\ }\href {https://doi.org/10.1103/PhysRevD.63.124015} {\bibfield  {journal} {\bibinfo  {journal} {Phys. Rev. D}\ }\textbf {\bibinfo {volume} {63}},\ \bibinfo {pages} {124015} (\bibinfo {year} {2001})},\ \Eprint {https://arxiv.org/abs/gr-qc/0101052} {arXiv:gr-qc/0101052} \BibitemShut {NoStop}%
\bibitem [{\citenamefont {Birmingham}\ \emph {et~al.}(2002)\citenamefont {Birmingham}, \citenamefont {Sachs},\ and\ \citenamefont {Solodukhin}}]{Birmingham:2001pj}%
  \BibitemOpen
  \bibfield  {author} {\bibinfo {author} {\bibfnamefont {D.}~\bibnamefont {Birmingham}}, \bibinfo {author} {\bibfnamefont {I.}~\bibnamefont {Sachs}},\ and\ \bibinfo {author} {\bibfnamefont {S.~N.}\ \bibnamefont {Solodukhin}},\ }\bibfield  {title} {\bibinfo {title} {{Conformal field theory interpretation of black hole quasinormal modes}},\ }\href {https://doi.org/10.1103/PhysRevLett.88.151301} {\bibfield  {journal} {\bibinfo  {journal} {Phys. Rev. Lett.}\ }\textbf {\bibinfo {volume} {88}},\ \bibinfo {pages} {151301} (\bibinfo {year} {2002})},\ \Eprint {https://arxiv.org/abs/hep-th/0112055} {arXiv:hep-th/0112055} \BibitemShut {NoStop}%
\bibitem [{\citenamefont {Szegő}(1975)}]{szego}%
  \BibitemOpen
  \bibfield  {author} {\bibinfo {author} {\bibfnamefont {G.}~\bibnamefont {Szegő}},\ }\href@noop {} {\emph {\bibinfo {title} {Orthogonal polynomials}}},\ \bibinfo {edition} {4th}\ ed.,\ Colloquium publications (American Mathematical Society), v. 23\ (\bibinfo  {publisher} {American Mathematical Society},\ \bibinfo {year} {1939-1975})\BibitemShut {NoStop}%
\bibitem [{\citenamefont {Watson}(1918)}]{watson1918asymptotic}%
  \BibitemOpen
  \bibfield  {author} {\bibinfo {author} {\bibfnamefont {G.~N.}\ \bibnamefont {Watson}},\ }\bibfield  {title} {\bibinfo {title} {Asymptotic expansions of hypergeometric functions},\ }\href@noop {} {\bibfield  {journal} {\bibinfo  {journal} {Transactions of the Cambridge Philosophical Society}\ }\textbf {\bibinfo {volume} {22}},\ \bibinfo {pages} {277} (\bibinfo {year} {1918})}\BibitemShut {NoStop}%
\bibitem [{\citenamefont {Hatsuda}(2020)}]{Hatsuda:2020sbn}%
  \BibitemOpen
  \bibfield  {author} {\bibinfo {author} {\bibfnamefont {Y.}~\bibnamefont {Hatsuda}},\ }\bibfield  {title} {\bibinfo {title} {{Quasinormal modes of Kerr-de Sitter black holes via the Heun function}},\ }\href {https://doi.org/10.1088/1361-6382/abc82e} {\bibfield  {journal} {\bibinfo  {journal} {Class. Quant. Grav.}\ }\textbf {\bibinfo {volume} {38}},\ \bibinfo {pages} {025015} (\bibinfo {year} {2020})},\ \Eprint {https://arxiv.org/abs/2006.08957} {arXiv:2006.08957 [gr-qc]} \BibitemShut {NoStop}%
\bibitem [{\citenamefont {Oshita}(2021)}]{Oshita:2021iyn}%
  \BibitemOpen
  \bibfield  {author} {\bibinfo {author} {\bibfnamefont {N.}~\bibnamefont {Oshita}},\ }\bibfield  {title} {\bibinfo {title} {{Ease of excitation of black hole ringing: Quantifying the importance of overtones by the excitation factors}},\ }\href {https://doi.org/10.1103/PhysRevD.104.124032} {\bibfield  {journal} {\bibinfo  {journal} {Phys. Rev. D}\ }\textbf {\bibinfo {volume} {104}},\ \bibinfo {pages} {124032} (\bibinfo {year} {2021})},\ \Eprint {https://arxiv.org/abs/2109.09757} {arXiv:2109.09757 [gr-qc]} \BibitemShut {NoStop}%
\bibitem [{\citenamefont {Motohashi}\ and\ \citenamefont {Noda}(2021)}]{Motohashi:2021zyv}%
  \BibitemOpen
  \bibfield  {author} {\bibinfo {author} {\bibfnamefont {H.}~\bibnamefont {Motohashi}}\ and\ \bibinfo {author} {\bibfnamefont {S.}~\bibnamefont {Noda}},\ }\bibfield  {title} {\bibinfo {title} {{Exact solution for wave scattering from black holes: Formulation}},\ }\href {https://doi.org/10.1093/ptep/ptac020} {\bibfield  {journal} {\bibinfo  {journal} {PTEP}\ }\textbf {\bibinfo {volume} {2021}},\ \bibinfo {pages} {083E03} (\bibinfo {year} {2021})},\ \Eprint {https://arxiv.org/abs/2103.10802} {arXiv:2103.10802 [gr-qc]} \BibitemShut {NoStop}%
\bibitem [{\citenamefont {Motohashi}(2025)}]{Motohashi:2024fwt}%
  \BibitemOpen
  \bibfield  {author} {\bibinfo {author} {\bibfnamefont {H.}~\bibnamefont {Motohashi}},\ }\bibfield  {title} {\bibinfo {title} {{Resonant Excitation of Quasinormal Modes of Black Holes}},\ }\href {https://doi.org/10.1103/PhysRevLett.134.141401} {\bibfield  {journal} {\bibinfo  {journal} {Phys. Rev. Lett.}\ }\textbf {\bibinfo {volume} {134}},\ \bibinfo {pages} {141401} (\bibinfo {year} {2025})},\ \Eprint {https://arxiv.org/abs/2407.15191} {arXiv:2407.15191 [gr-qc]} \BibitemShut {NoStop}%
\bibitem [{\citenamefont {Lo}\ \emph {et~al.}(2025)\citenamefont {Lo}, \citenamefont {Sabani},\ and\ \citenamefont {Cardoso}}]{Lo:2025njp}%
  \BibitemOpen
  \bibfield  {author} {\bibinfo {author} {\bibfnamefont {R.~K.~L.}\ \bibnamefont {Lo}}, \bibinfo {author} {\bibfnamefont {L.}~\bibnamefont {Sabani}},\ and\ \bibinfo {author} {\bibfnamefont {V.}~\bibnamefont {Cardoso}},\ }\bibfield  {title} {\bibinfo {title} {{Quasinormal modes and excitation factors of Kerr black holes}},\ }\href {https://doi.org/10.1103/PhysRevD.111.124002} {\bibfield  {journal} {\bibinfo  {journal} {Phys. Rev. D}\ }\textbf {\bibinfo {volume} {111}},\ \bibinfo {pages} {124002} (\bibinfo {year} {2025})},\ \Eprint {https://arxiv.org/abs/2504.00084} {arXiv:2504.00084 [gr-qc]} \BibitemShut {NoStop}%
\bibitem [{\citenamefont {Kubota}\ and\ \citenamefont {Motohashi}(2025)}]{Kubota:2025hjk}%
  \BibitemOpen
  \bibfield  {author} {\bibinfo {author} {\bibfnamefont {K.-i.}\ \bibnamefont {Kubota}}\ and\ \bibinfo {author} {\bibfnamefont {H.}~\bibnamefont {Motohashi}},\ }\bibfield  {title} {\bibinfo {title} {{Resonance in black hole ringdown: Benchmarking quasinormal mode excitation and extraction}},\ }\href@noop {} {\  (\bibinfo {year} {2025})},\ \Eprint {https://arxiv.org/abs/2509.06411} {arXiv:2509.06411 [gr-qc]} \BibitemShut {NoStop}%
\bibitem [{\citenamefont {Zhou}\ and\ \citenamefont {Panosso~Macedo}(2025)}]{Zhou:2025xta}%
  \BibitemOpen
  \bibfield  {author} {\bibinfo {author} {\bibfnamefont {Y.}~\bibnamefont {Zhou}}\ and\ \bibinfo {author} {\bibfnamefont {R.}~\bibnamefont {Panosso~Macedo}},\ }\bibfield  {title} {\bibinfo {title} {{Limiting geometry and spectral instability in Schwarzschild{\textendash}de Sitter spacetimes}},\ }\href {https://doi.org/10.1103/n948-jnfh} {\bibfield  {journal} {\bibinfo  {journal} {Phys. Rev. D}\ }\textbf {\bibinfo {volume} {112}},\ \bibinfo {pages} {084063} (\bibinfo {year} {2025})},\ \Eprint {https://arxiv.org/abs/2507.05370} {arXiv:2507.05370 [gr-qc]} \BibitemShut {NoStop}%
\bibitem [{\citenamefont {Heun}(1888)}]{heun1888theorie}%
  \BibitemOpen
  \bibfield  {author} {\bibinfo {author} {\bibfnamefont {K.}~\bibnamefont {Heun}},\ }\bibfield  {title} {\bibinfo {title} {Zur theorie der riemann'schen functionen zweiter ordnung mit vier verzweigungspunkten},\ }\href@noop {} {\bibfield  {journal} {\bibinfo  {journal} {Mathematische Annalen}\ }\textbf {\bibinfo {volume} {33}},\ \bibinfo {pages} {161} (\bibinfo {year} {1888})}\BibitemShut {NoStop}%
\bibitem [{\citenamefont {Ronveaux}\ and\ \citenamefont {Arscott}(1995)}]{ronveaux1995heun}%
  \BibitemOpen
  \bibfield  {author} {\bibinfo {author} {\bibfnamefont {A.}~\bibnamefont {Ronveaux}}\ and\ \bibinfo {author} {\bibfnamefont {F.}~\bibnamefont {Arscott}},\ }\href {https://books.google.ch/books?id=5p65FD8caCgC} {\emph {\bibinfo {title} {Heun's Differential Equations}}},\ Oxford science publications\ (\bibinfo  {publisher} {Oxford University Press},\ \bibinfo {year} {1995})\BibitemShut {NoStop}%
\bibitem [{\citenamefont {Bonelli}\ \emph {et~al.}(2023)\citenamefont {Bonelli}, \citenamefont {Iossa}, \citenamefont {Panea~Lichtig},\ and\ \citenamefont {Tanzini}}]{Bonelli:2022ten}%
  \BibitemOpen
  \bibfield  {author} {\bibinfo {author} {\bibfnamefont {G.}~\bibnamefont {Bonelli}}, \bibinfo {author} {\bibfnamefont {C.}~\bibnamefont {Iossa}}, \bibinfo {author} {\bibfnamefont {D.}~\bibnamefont {Panea~Lichtig}},\ and\ \bibinfo {author} {\bibfnamefont {A.}~\bibnamefont {Tanzini}},\ }\bibfield  {title} {\bibinfo {title} {{Irregular Liouville Correlators and Connection Formulae for Heun Functions}},\ }\href {https://doi.org/10.1007/s00220-022-04497-5} {\bibfield  {journal} {\bibinfo  {journal} {Commun. Math. Phys.}\ }\textbf {\bibinfo {volume} {397}},\ \bibinfo {pages} {635} (\bibinfo {year} {2023})},\ \Eprint {https://arxiv.org/abs/2201.04491} {arXiv:2201.04491 [hep-th]} \BibitemShut {NoStop}%
\bibitem [{\citenamefont {Lisovyy}\ and\ \citenamefont {Naidiuk}(2022)}]{Lisovyy:2022flm}%
  \BibitemOpen
  \bibfield  {author} {\bibinfo {author} {\bibfnamefont {O.}~\bibnamefont {Lisovyy}}\ and\ \bibinfo {author} {\bibfnamefont {A.}~\bibnamefont {Naidiuk}},\ }\bibfield  {title} {\bibinfo {title} {{Perturbative connection formulas for Heun equations}},\ }\href {https://doi.org/10.1088/1751-8121/ac9ba7} {\bibfield  {journal} {\bibinfo  {journal} {J. Phys. A}\ }\textbf {\bibinfo {volume} {55}},\ \bibinfo {pages} {434005} (\bibinfo {year} {2022})},\ \Eprint {https://arxiv.org/abs/2208.01604} {arXiv:2208.01604 [math-ph]} \BibitemShut {NoStop}%
\bibitem [{\citenamefont {Aminov}\ \emph {et~al.}(2023)\citenamefont {Aminov}, \citenamefont {Arnaudo}, \citenamefont {Bonelli}, \citenamefont {Grassi},\ and\ \citenamefont {Tanzini}}]{Aminov:2023jve}%
  \BibitemOpen
  \bibfield  {author} {\bibinfo {author} {\bibfnamefont {G.}~\bibnamefont {Aminov}}, \bibinfo {author} {\bibfnamefont {P.}~\bibnamefont {Arnaudo}}, \bibinfo {author} {\bibfnamefont {G.}~\bibnamefont {Bonelli}}, \bibinfo {author} {\bibfnamefont {A.}~\bibnamefont {Grassi}},\ and\ \bibinfo {author} {\bibfnamefont {A.}~\bibnamefont {Tanzini}},\ }\bibfield  {title} {\bibinfo {title} {{Black hole perturbation theory and multiple polylogarithms}},\ }\href {https://doi.org/10.1007/JHEP11(2023)059} {\bibfield  {journal} {\bibinfo  {journal} {JHEP}\ }\textbf {\bibinfo {volume} {11}},\ \bibinfo {pages} {059}},\ \Eprint {https://arxiv.org/abs/2307.10141} {arXiv:2307.10141 [hep-th]} \BibitemShut {NoStop}%
\bibitem [{\citenamefont {Nakamura}\ and\ \citenamefont {ichi Oohara}(1983)}]{NAKAMURA1983403}%
  \BibitemOpen
  \bibfield  {author} {\bibinfo {author} {\bibfnamefont {T.}~\bibnamefont {Nakamura}}\ and\ \bibinfo {author} {\bibfnamefont {K.}~\bibnamefont {ichi Oohara}},\ }\bibfield  {title} {\bibinfo {title} {Gravitational radiation emitted by n particles in circular orbits},\ }\href {https://doi.org/https://doi.org/10.1016/0375-9601(83)90248-7} {\bibfield  {journal} {\bibinfo  {journal} {Physics Letters A}\ }\textbf {\bibinfo {volume} {98}},\ \bibinfo {pages} {403} (\bibinfo {year} {1983})}\BibitemShut {NoStop}%
\bibitem [{\citenamefont {Berti}\ \emph {et~al.}(2006)\citenamefont {Berti}, \citenamefont {Cardoso},\ and\ \citenamefont {Will}}]{Berti:2006hb}%
  \BibitemOpen
  \bibfield  {author} {\bibinfo {author} {\bibfnamefont {E.}~\bibnamefont {Berti}}, \bibinfo {author} {\bibfnamefont {V.}~\bibnamefont {Cardoso}},\ and\ \bibinfo {author} {\bibfnamefont {C.~M.}\ \bibnamefont {Will}},\ }\bibfield  {title} {\bibinfo {title} {{Considerations on the excitation of black hole quasinormal modes}},\ }\href {https://doi.org/10.1063/1.2348047} {\bibfield  {journal} {\bibinfo  {journal} {AIP Conf. Proc.}\ }\textbf {\bibinfo {volume} {848}},\ \bibinfo {pages} {687} (\bibinfo {year} {2006})},\ \Eprint {https://arxiv.org/abs/gr-qc/0601077} {arXiv:gr-qc/0601077} \BibitemShut {NoStop}%
\bibitem [{\citenamefont {Nollert}(1996)}]{Nollert:1996rf}%
  \BibitemOpen
  \bibfield  {author} {\bibinfo {author} {\bibfnamefont {H.-P.}\ \bibnamefont {Nollert}},\ }\bibfield  {title} {\bibinfo {title} {{About the significance of quasinormal modes of black holes}},\ }\href {https://doi.org/10.1103/PhysRevD.53.4397} {\bibfield  {journal} {\bibinfo  {journal} {Phys. Rev. D}\ }\textbf {\bibinfo {volume} {53}},\ \bibinfo {pages} {4397} (\bibinfo {year} {1996})},\ \Eprint {https://arxiv.org/abs/gr-qc/9602032} {arXiv:gr-qc/9602032} \BibitemShut {NoStop}%
\bibitem [{\citenamefont {Nollert}\ and\ \citenamefont {Price}(1999)}]{Nollert:1998ys}%
  \BibitemOpen
  \bibfield  {author} {\bibinfo {author} {\bibfnamefont {H.-P.}\ \bibnamefont {Nollert}}\ and\ \bibinfo {author} {\bibfnamefont {R.~H.}\ \bibnamefont {Price}},\ }\bibfield  {title} {\bibinfo {title} {{Quantifying excitations of quasinormal mode systems}},\ }\href {https://doi.org/10.1063/1.532698} {\bibfield  {journal} {\bibinfo  {journal} {J. Math. Phys.}\ }\textbf {\bibinfo {volume} {40}},\ \bibinfo {pages} {980} (\bibinfo {year} {1999})},\ \Eprint {https://arxiv.org/abs/gr-qc/9810074} {arXiv:gr-qc/9810074} \BibitemShut {NoStop}%
\bibitem [{\citenamefont {Daghigh}\ \emph {et~al.}(2020)\citenamefont {Daghigh}, \citenamefont {Green},\ and\ \citenamefont {Morey}}]{Daghigh:2020jyk}%
  \BibitemOpen
  \bibfield  {author} {\bibinfo {author} {\bibfnamefont {R.~G.}\ \bibnamefont {Daghigh}}, \bibinfo {author} {\bibfnamefont {M.~D.}\ \bibnamefont {Green}},\ and\ \bibinfo {author} {\bibfnamefont {J.~C.}\ \bibnamefont {Morey}},\ }\bibfield  {title} {\bibinfo {title} {{Significance of Black Hole Quasinormal Modes: A Closer Look}},\ }\href {https://doi.org/10.1103/PhysRevD.101.104009} {\bibfield  {journal} {\bibinfo  {journal} {Phys. Rev. D}\ }\textbf {\bibinfo {volume} {101}},\ \bibinfo {pages} {104009} (\bibinfo {year} {2020})},\ \Eprint {https://arxiv.org/abs/2002.07251} {arXiv:2002.07251 [gr-qc]} \BibitemShut {NoStop}%
\bibitem [{\citenamefont {Jaramillo}\ \emph {et~al.}(2021)\citenamefont {Jaramillo}, \citenamefont {Panosso~Macedo},\ and\ \citenamefont {Al~Sheikh}}]{Jaramillo:2020tuu}%
  \BibitemOpen
  \bibfield  {author} {\bibinfo {author} {\bibfnamefont {J.~L.}\ \bibnamefont {Jaramillo}}, \bibinfo {author} {\bibfnamefont {R.}~\bibnamefont {Panosso~Macedo}},\ and\ \bibinfo {author} {\bibfnamefont {L.}~\bibnamefont {Al~Sheikh}},\ }\bibfield  {title} {\bibinfo {title} {{Pseudospectrum and Black Hole Quasinormal Mode Instability}},\ }\href {https://doi.org/10.1103/PhysRevX.11.031003} {\bibfield  {journal} {\bibinfo  {journal} {Phys. Rev. X}\ }\textbf {\bibinfo {volume} {11}},\ \bibinfo {pages} {031003} (\bibinfo {year} {2021})},\ \Eprint {https://arxiv.org/abs/2004.06434} {arXiv:2004.06434 [gr-qc]} \BibitemShut {NoStop}%
\bibitem [{\citenamefont {Carballo}\ and\ \citenamefont {Withers}(2024)}]{Carballo:2024kbk}%
  \BibitemOpen
  \bibfield  {author} {\bibinfo {author} {\bibfnamefont {J.}~\bibnamefont {Carballo}}\ and\ \bibinfo {author} {\bibfnamefont {B.}~\bibnamefont {Withers}},\ }\bibfield  {title} {\bibinfo {title} {{Transient dynamics of quasinormal mode sums}},\ }\href {https://doi.org/10.1007/JHEP10(2024)084} {\bibfield  {journal} {\bibinfo  {journal} {JHEP}\ }\textbf {\bibinfo {volume} {10}},\ \bibinfo {pages} {084}},\ \Eprint {https://arxiv.org/abs/2406.06685} {arXiv:2406.06685 [hep-th]} \BibitemShut {NoStop}%
\bibitem [{\citenamefont {Carballo}\ \emph {et~al.}(2025)\citenamefont {Carballo}, \citenamefont {Pantelidou},\ and\ \citenamefont {Withers}}]{Carballo:2025ajx}%
  \BibitemOpen
  \bibfield  {author} {\bibinfo {author} {\bibfnamefont {J.}~\bibnamefont {Carballo}}, \bibinfo {author} {\bibfnamefont {C.}~\bibnamefont {Pantelidou}},\ and\ \bibinfo {author} {\bibfnamefont {B.}~\bibnamefont {Withers}},\ }\bibfield  {title} {\bibinfo {title} {{Non-modal effects in black hole perturbation theory: Transient Superradiance}},\ }\href@noop {} {\  (\bibinfo {year} {2025})},\ \Eprint {https://arxiv.org/abs/2503.05871} {arXiv:2503.05871 [gr-qc]} \BibitemShut {NoStop}%
\bibitem [{\citenamefont {Besson}\ \emph {et~al.}(2025)\citenamefont {Besson}, \citenamefont {Carballo}, \citenamefont {Pantelidou},\ and\ \citenamefont {Withers}}]{Besson:2025ghu}%
  \BibitemOpen
  \bibfield  {author} {\bibinfo {author} {\bibfnamefont {J.}~\bibnamefont {Besson}}, \bibinfo {author} {\bibfnamefont {J.}~\bibnamefont {Carballo}}, \bibinfo {author} {\bibfnamefont {C.}~\bibnamefont {Pantelidou}},\ and\ \bibinfo {author} {\bibfnamefont {B.}~\bibnamefont {Withers}},\ }\bibfield  {title} {\bibinfo {title} {{Transients in black hole perturbation theory}},\ }\href {https://doi.org/10.3389/fphy.2025.1638583} {\bibfield  {journal} {\bibinfo  {journal} {Front. in Phys.}\ }\textbf {\bibinfo {volume} {13}},\ \bibinfo {pages} {1638583} (\bibinfo {year} {2025})},\ \Eprint {https://arxiv.org/abs/2507.16493} {arXiv:2507.16493 [gr-qc]} \BibitemShut {NoStop}%
\bibitem [{\citenamefont {Dodelson}\ \emph {et~al.}(2024)\citenamefont {Dodelson}, \citenamefont {Iossa}, \citenamefont {Karlsson}, \citenamefont {Lupsasca},\ and\ \citenamefont {Zhiboedov}}]{Dodelson:2023nnr}%
  \BibitemOpen
  \bibfield  {author} {\bibinfo {author} {\bibfnamefont {M.}~\bibnamefont {Dodelson}}, \bibinfo {author} {\bibfnamefont {C.}~\bibnamefont {Iossa}}, \bibinfo {author} {\bibfnamefont {R.}~\bibnamefont {Karlsson}}, \bibinfo {author} {\bibfnamefont {A.}~\bibnamefont {Lupsasca}},\ and\ \bibinfo {author} {\bibfnamefont {A.}~\bibnamefont {Zhiboedov}},\ }\bibfield  {title} {\bibinfo {title} {{Black hole bulk-cone singularities}},\ }\href {https://doi.org/10.1007/JHEP07(2024)046} {\bibfield  {journal} {\bibinfo  {journal} {JHEP}\ }\textbf {\bibinfo {volume} {07}},\ \bibinfo {pages} {046}},\ \Eprint {https://arxiv.org/abs/2310.15236} {arXiv:2310.15236 [hep-th]} \BibitemShut {NoStop}%
\bibitem [{\citenamefont {Zenginoglu}\ and\ \citenamefont {Galley}(2012)}]{Zenginoglu:2012xe}%
  \BibitemOpen
  \bibfield  {author} {\bibinfo {author} {\bibfnamefont {A.}~\bibnamefont {Zenginoglu}}\ and\ \bibinfo {author} {\bibfnamefont {C.~R.}\ \bibnamefont {Galley}},\ }\bibfield  {title} {\bibinfo {title} {{Caustic echoes from a Schwarzschild black hole}},\ }\href {https://doi.org/10.1103/PhysRevD.86.064030} {\bibfield  {journal} {\bibinfo  {journal} {Phys. Rev. D}\ }\textbf {\bibinfo {volume} {86}},\ \bibinfo {pages} {064030} (\bibinfo {year} {2012})},\ \Eprint {https://arxiv.org/abs/1206.1109} {arXiv:1206.1109 [gr-qc]} \BibitemShut {NoStop}%
\bibitem [{\citenamefont {Correia}\ \emph {et~al.}()\citenamefont {Correia}, \citenamefont {Gopalka}, \citenamefont {Isabella},\ and\ \citenamefont {Wolz}}]{CorreiaUpcoming}%
  \BibitemOpen
  \bibfield  {author} {\bibinfo {author} {\bibfnamefont {M.}~\bibnamefont {Correia}}, \bibinfo {author} {\bibfnamefont {T.}~\bibnamefont {Gopalka}}, \bibinfo {author} {\bibfnamefont {G.}~\bibnamefont {Isabella}},\ and\ \bibinfo {author} {\bibfnamefont {A.}~\bibnamefont {Wolz}},\ }\bibfield  {title} {\bibinfo {title} {{Analyticity of the Black Hole S-Matrix}},\ }\href@noop {} {\bibinfo  {journal} {(to appear)}\ }\BibitemShut {NoStop}%
\bibitem [{\citenamefont {Abramowitz}\ and\ \citenamefont {Stegun}(1964)}]{abramowitz+stegun}%
  \BibitemOpen
\bibfield  {journal} {  }\bibfield  {author} {\bibinfo {author} {\bibfnamefont {M.}~\bibnamefont {Abramowitz}}\ and\ \bibinfo {author} {\bibfnamefont {I.~A.}\ \bibnamefont {Stegun}},\ }\href@noop {} {\emph {\bibinfo {title} {Handbook of Mathematical Functions with Formulas, Graphs, and Mathematical Tables}}},\ \bibinfo {edition} {ninth dover printing, tenth gpo printing}\ ed.\ (\bibinfo  {publisher} {Dover},\ \bibinfo {address} {New York},\ \bibinfo {year} {1964})\BibitemShut {NoStop}%
\bibitem [{\citenamefont {Nekrasov}\ and\ \citenamefont {Shatashvili}(2010)}]{Nekrasov:2009rc}%
  \BibitemOpen
  \bibfield  {author} {\bibinfo {author} {\bibfnamefont {N.~A.}\ \bibnamefont {Nekrasov}}\ and\ \bibinfo {author} {\bibfnamefont {S.~L.}\ \bibnamefont {Shatashvili}},\ }\bibfield  {title} {\bibinfo {title} {{Quantization of Integrable Systems and Four Dimensional Gauge Theories}},\ }in\ \href {https://doi.org/10.1142/9789814304634_0015} {\emph {\bibinfo {booktitle} {{16th International Congress on Mathematical Physics}}}}\ (\bibinfo {year} {2010})\ pp.\ \bibinfo {pages} {265--289},\ \Eprint {https://arxiv.org/abs/0908.4052} {arXiv:0908.4052 [hep-th]} \BibitemShut {NoStop}%
\bibitem [{\citenamefont {Aminov}\ \emph {et~al.}(2022)\citenamefont {Aminov}, \citenamefont {Grassi},\ and\ \citenamefont {Hatsuda}}]{Aminov:2020yma}%
  \BibitemOpen
  \bibfield  {author} {\bibinfo {author} {\bibfnamefont {G.}~\bibnamefont {Aminov}}, \bibinfo {author} {\bibfnamefont {A.}~\bibnamefont {Grassi}},\ and\ \bibinfo {author} {\bibfnamefont {Y.}~\bibnamefont {Hatsuda}},\ }\bibfield  {title} {\bibinfo {title} {{Black Hole Quasinormal Modes and Seiberg{\textendash}Witten Theory}},\ }\href {https://doi.org/10.1007/s00023-021-01137-x} {\bibfield  {journal} {\bibinfo  {journal} {Annales Henri Poincare}\ }\textbf {\bibinfo {volume} {23}},\ \bibinfo {pages} {1951} (\bibinfo {year} {2022})},\ \Eprint {https://arxiv.org/abs/2006.06111} {arXiv:2006.06111 [hep-th]} \BibitemShut {NoStop}%
\bibitem [{\citenamefont {Bianchi}\ \emph {et~al.}(2022{\natexlab{a}})\citenamefont {Bianchi}, \citenamefont {Consoli}, \citenamefont {Grillo},\ and\ \citenamefont {Morales}}]{Bianchi:2021xpr}%
  \BibitemOpen
  \bibfield  {author} {\bibinfo {author} {\bibfnamefont {M.}~\bibnamefont {Bianchi}}, \bibinfo {author} {\bibfnamefont {D.}~\bibnamefont {Consoli}}, \bibinfo {author} {\bibfnamefont {A.}~\bibnamefont {Grillo}},\ and\ \bibinfo {author} {\bibfnamefont {J.~F.}\ \bibnamefont {Morales}},\ }\bibfield  {title} {\bibinfo {title} {{QNMs of branes, BHs and fuzzballs from quantum SW geometries}},\ }\href {https://doi.org/10.1016/j.physletb.2021.136837} {\bibfield  {journal} {\bibinfo  {journal} {Phys. Lett. B}\ }\textbf {\bibinfo {volume} {824}},\ \bibinfo {pages} {136837} (\bibinfo {year} {2022}{\natexlab{a}})},\ \Eprint {https://arxiv.org/abs/2105.04245} {arXiv:2105.04245 [hep-th]} \BibitemShut {NoStop}%
\bibitem [{\citenamefont {Bonelli}\ \emph {et~al.}(2022)\citenamefont {Bonelli}, \citenamefont {Iossa}, \citenamefont {Lichtig},\ and\ \citenamefont {Tanzini}}]{Bonelli:2021uvf}%
  \BibitemOpen
  \bibfield  {author} {\bibinfo {author} {\bibfnamefont {G.}~\bibnamefont {Bonelli}}, \bibinfo {author} {\bibfnamefont {C.}~\bibnamefont {Iossa}}, \bibinfo {author} {\bibfnamefont {D.~P.}\ \bibnamefont {Lichtig}},\ and\ \bibinfo {author} {\bibfnamefont {A.}~\bibnamefont {Tanzini}},\ }\bibfield  {title} {\bibinfo {title} {{Exact solution of Kerr black hole perturbations via CFT2 and instanton counting: Greybody factor, quasinormal modes, and Love numbers}},\ }\href {https://doi.org/10.1103/PhysRevD.105.044047} {\bibfield  {journal} {\bibinfo  {journal} {Phys. Rev. D}\ }\textbf {\bibinfo {volume} {105}},\ \bibinfo {pages} {044047} (\bibinfo {year} {2022})},\ \Eprint {https://arxiv.org/abs/2105.04483} {arXiv:2105.04483 [hep-th]} \BibitemShut {NoStop}%
\bibitem [{\citenamefont {Bianchi}\ \emph {et~al.}(2022{\natexlab{b}})\citenamefont {Bianchi}, \citenamefont {Consoli}, \citenamefont {Grillo},\ and\ \citenamefont {Morales}}]{Bianchi:2021mft}%
  \BibitemOpen
  \bibfield  {author} {\bibinfo {author} {\bibfnamefont {M.}~\bibnamefont {Bianchi}}, \bibinfo {author} {\bibfnamefont {D.}~\bibnamefont {Consoli}}, \bibinfo {author} {\bibfnamefont {A.}~\bibnamefont {Grillo}},\ and\ \bibinfo {author} {\bibfnamefont {J.~F.}\ \bibnamefont {Morales}},\ }\bibfield  {title} {\bibinfo {title} {{More on the SW-QNM correspondence}},\ }\href {https://doi.org/10.1007/JHEP01(2022)024} {\bibfield  {journal} {\bibinfo  {journal} {JHEP}\ }\textbf {\bibinfo {volume} {01}},\ \bibinfo {pages} {024}},\ \Eprint {https://arxiv.org/abs/2109.09804} {arXiv:2109.09804 [hep-th]} \BibitemShut {NoStop}%
\bibitem [{\citenamefont {da~Cunha}\ and\ \citenamefont {Cavalcante}(2024)}]{daCunha:2022ewy}%
  \BibitemOpen
  \bibfield  {author} {\bibinfo {author} {\bibfnamefont {B.~C.}\ \bibnamefont {da~Cunha}}\ and\ \bibinfo {author} {\bibfnamefont {J.~P.}\ \bibnamefont {Cavalcante}},\ }\bibfield  {title} {\bibinfo {title} {{Expansions for semiclassical conformal blocks}},\ }\href {https://doi.org/10.1007/JHEP08(2024)110} {\bibfield  {journal} {\bibinfo  {journal} {JHEP}\ }\textbf {\bibinfo {volume} {08}},\ \bibinfo {pages} {110}},\ \Eprint {https://arxiv.org/abs/2211.03551} {arXiv:2211.03551 [hep-th]} \BibitemShut {NoStop}%
\bibitem [{\citenamefont {Fioravanti}\ and\ \citenamefont {Gregori}(2021)}]{Fioravanti:2021dce}%
  \BibitemOpen
  \bibfield  {author} {\bibinfo {author} {\bibfnamefont {D.}~\bibnamefont {Fioravanti}}\ and\ \bibinfo {author} {\bibfnamefont {D.}~\bibnamefont {Gregori}},\ }\bibfield  {title} {\bibinfo {title} {{A new method for exact results on Quasinormal Modes of Black Holes}},\ }\href@noop {} {\  (\bibinfo {year} {2021})},\ \Eprint {https://arxiv.org/abs/2112.11434} {arXiv:2112.11434 [hep-th]} \BibitemShut {NoStop}%
\bibitem [{\citenamefont {Dodelson}\ \emph {et~al.}(2023)\citenamefont {Dodelson}, \citenamefont {Grassi}, \citenamefont {Iossa}, \citenamefont {Panea~Lichtig},\ and\ \citenamefont {Zhiboedov}}]{Dodelson:2022yvn}%
  \BibitemOpen
  \bibfield  {author} {\bibinfo {author} {\bibfnamefont {M.}~\bibnamefont {Dodelson}}, \bibinfo {author} {\bibfnamefont {A.}~\bibnamefont {Grassi}}, \bibinfo {author} {\bibfnamefont {C.}~\bibnamefont {Iossa}}, \bibinfo {author} {\bibfnamefont {D.}~\bibnamefont {Panea~Lichtig}},\ and\ \bibinfo {author} {\bibfnamefont {A.}~\bibnamefont {Zhiboedov}},\ }\bibfield  {title} {\bibinfo {title} {{Holographic thermal correlators from supersymmetric instantons}},\ }\href {https://doi.org/10.21468/SciPostPhys.14.5.116} {\bibfield  {journal} {\bibinfo  {journal} {SciPost Phys.}\ }\textbf {\bibinfo {volume} {14}},\ \bibinfo {pages} {116} (\bibinfo {year} {2023})},\ \Eprint {https://arxiv.org/abs/2206.07720} {arXiv:2206.07720 [hep-th]} \BibitemShut {NoStop}%
\bibitem [{\citenamefont {Jia}\ and\ \citenamefont {Rangamani}(2024)}]{Jia:2024zes}%
  \BibitemOpen
  \bibfield  {author} {\bibinfo {author} {\bibfnamefont {H.~F.}\ \bibnamefont {Jia}}\ and\ \bibinfo {author} {\bibfnamefont {M.}~\bibnamefont {Rangamani}},\ }\bibfield  {title} {\bibinfo {title} {{Holographic thermal correlators and quasinormal modes from semiclassical Virasoro blocks}},\ }\href {https://doi.org/10.1007/JHEP12(2024)047} {\bibfield  {journal} {\bibinfo  {journal} {JHEP}\ }\textbf {\bibinfo {volume} {12}},\ \bibinfo {pages} {047}},\ \Eprint {https://arxiv.org/abs/2408.05208} {arXiv:2408.05208 [hep-th]} \BibitemShut {NoStop}%
\bibitem [{\citenamefont {Arnaudo}\ \emph {et~al.}(2025{\natexlab{b}})\citenamefont {Arnaudo}, \citenamefont {Grassi},\ and\ \citenamefont {Hao}}]{Arnaudo:2025kof}%
  \BibitemOpen
  \bibfield  {author} {\bibinfo {author} {\bibfnamefont {P.}~\bibnamefont {Arnaudo}}, \bibinfo {author} {\bibfnamefont {A.}~\bibnamefont {Grassi}},\ and\ \bibinfo {author} {\bibfnamefont {Q.}~\bibnamefont {Hao}},\ }\bibfield  {title} {\bibinfo {title} {{On quivers, spectral networks and black holes}},\ }\href@noop {} {\  (\bibinfo {year} {2025}{\natexlab{b}})},\ \Eprint {https://arxiv.org/abs/2502.01526} {arXiv:2502.01526 [hep-th]} \BibitemShut {NoStop}%
\bibitem [{\citenamefont {Belavin}\ \emph {et~al.}(1984)\citenamefont {Belavin}, \citenamefont {Polyakov},\ and\ \citenamefont {Zamolodchikov}}]{Belavin:1984vu}%
  \BibitemOpen
  \bibfield  {author} {\bibinfo {author} {\bibfnamefont {A.~A.}\ \bibnamefont {Belavin}}, \bibinfo {author} {\bibfnamefont {A.~M.}\ \bibnamefont {Polyakov}},\ and\ \bibinfo {author} {\bibfnamefont {A.~B.}\ \bibnamefont {Zamolodchikov}},\ }\bibfield  {title} {\bibinfo {title} {{Infinite Conformal Symmetry in Two-Dimensional Quantum Field Theory}},\ }\href {https://doi.org/10.1016/0550-3213(84)90052-X} {\bibfield  {journal} {\bibinfo  {journal} {Nucl. Phys. B}\ }\textbf {\bibinfo {volume} {241}},\ \bibinfo {pages} {333} (\bibinfo {year} {1984})}\BibitemShut {NoStop}%
\bibitem [{\citenamefont {Alday}\ \emph {et~al.}(2010)\citenamefont {Alday}, \citenamefont {Gaiotto},\ and\ \citenamefont {Tachikawa}}]{Alday:2009aq}%
  \BibitemOpen
  \bibfield  {author} {\bibinfo {author} {\bibfnamefont {L.~F.}\ \bibnamefont {Alday}}, \bibinfo {author} {\bibfnamefont {D.}~\bibnamefont {Gaiotto}},\ and\ \bibinfo {author} {\bibfnamefont {Y.}~\bibnamefont {Tachikawa}},\ }\bibfield  {title} {\bibinfo {title} {{Liouville Correlation Functions from Four-dimensional Gauge Theories}},\ }\href {https://doi.org/10.1007/s11005-010-0369-5} {\bibfield  {journal} {\bibinfo  {journal} {Lett. Math. Phys.}\ }\textbf {\bibinfo {volume} {91}},\ \bibinfo {pages} {167} (\bibinfo {year} {2010})},\ \Eprint {https://arxiv.org/abs/0906.3219} {arXiv:0906.3219 [hep-th]} \BibitemShut {NoStop}%
\bibitem [{\citenamefont {Matone}(1995)}]{Matone:1995rx}%
  \BibitemOpen
  \bibfield  {author} {\bibinfo {author} {\bibfnamefont {M.}~\bibnamefont {Matone}},\ }\bibfield  {title} {\bibinfo {title} {{Instantons and recursion relations in N=2 SUSY gauge theory}},\ }\href {https://doi.org/10.1016/0370-2693(95)00920-G} {\bibfield  {journal} {\bibinfo  {journal} {Phys. Lett. B}\ }\textbf {\bibinfo {volume} {357}},\ \bibinfo {pages} {342} (\bibinfo {year} {1995})},\ \Eprint {https://arxiv.org/abs/hep-th/9506102} {arXiv:hep-th/9506102} \BibitemShut {NoStop}%
\end{thebibliography}

%

\end{document}